\documentclass{lmcs}
\pdfoutput=1

\usepackage{lastpage}
\lmcsdoi{22}{4}{12}
\lmcsheading{}{\pageref{LastPage}}{}{}%
{Jan.~24,~2025}{Nov.~03,~2026}{}


\usepackage{hyperref}
\usepackage{doi}
\usepackage{graphicx}
\usepackage{graphics}
\usepackage{mathrsfs}
\usepackage{mathtools}
\usepackage{amsfonts}
\usepackage{float}
\usepackage{amssymb}
\usepackage{stmaryrd}
\SetSymbolFont{stmry}{bold}{U}{stmry}{m}{n}
\usepackage{bm}
\usepackage{dsfont}
\usepackage{xspace}
\usepackage[normalem]{ulem}

\usepackage[dvipsnames,svgnames]{xcolor}
\definecolor{webgreen}{rgb}{0,.5,0}
\definecolor{webbrown}{rgb}{.6,0,0}
\definecolor{opcolor}{rgb}{0.4,0,0}
\colorlet{ptcolor}{blue}
\colorlet{mscolor}{webgreen}

\colorlet{globalcol}{yellow!30}
\colorlet{localcol}{green!20}
\colorlet{bluecol}{blue!18}
\colorlet{dglobalcol}{yellow!60!black!50}
\colorlet{dlocalcol}{green!60!black!50}
\colorlet{dbluecol}{blue!70!black!40}
\AtBeginDocument{\hypersetup{%
 linkcolor   = NavyBlue,
 citecolor   = NavyBlue,
 urlcolor    = RubineRed
}}

\usepackage{src/reo-tikz}
\tikzstyle{box}=[rectangle, inner sep=5pt, draw=black]
\tikzstyle{boxin}=[box,fill=black!10]
\tikzstyle{st}=[inner sep=2pt,initial text={~},draw=black,minimum width=5mm,circle]
\tikzstyle{link}=[dotted,very thick]

\tikzstyle{ndistz}=[node distance=10mm]

\usepackage{listings}
\usepackage[nameinlink,capitalise]{cleveref}
\Crefname{section}{Sect.}{Sects.}
\Crefname{thm}{Example}{Examples}
\usepackage{url,cite}
\usepackage{booktabs}
\usepackage[inline]{enumitem}
\usepackage[linesnumbered,lined,commentsnumbered,noend,vlined,ruled]{algorithm2e} 
\usepackage{longtable}
\usepackage{multirow}
\usepackage{footnote}
\usepackage{listings}
\usepackage{bold-extra}
\usepackage{relsize}
\usepackage[T1]{fontenc}
\usepackage[utf8]{inputenc}
\usepackage{tikz-cd}
\usepackage{tikz}
\usetikzlibrary{arrows,calc,shapes.arrows,shapes.multipart,shapes.geometric,shapes.misc,positioning, shadows, automata,fit,decorations.markings,decorations.pathmorphing,backgrounds}
\usepackage{fontawesome}
\usepackage{caption}
\usepackage{subcaption}
\makeatletter
\providecommand{\leftsquigarrow}{%
  \mathrel{\mathpalette\reflect@squig\relax}%
}
\newcommand{\reflect@squig}[2]{%
  \reflectbox{$\m@th#1\rightsquigarrow$}%
}
\makeatother
\usepackage{thm-restate}

\usepackage{wrapfig}

\usepackage{pifont}

\newcounter{rmrk}

\crefalias{rmrk}{remark}

\newcommand{\syn}[3][m]{\ensuremath{\pts{#2}\,{\to}\,\pts{#3}\,{:}\,\msg{#1}}\xspace}
\newcommand{\snd}[3][m]{\ensuremath{\pts{#2}\,\pts{#3}\mkern2mu!\mkern2mu\msg{#1}}\xspace}

\newcommand{\alone}[2][m]{\ensuremath{\pts{#2}\,{:}\,\msg{#1}}\xspace}

\newcommand{\myparagraph}[1]{\medskip\noindent\textbf{#1} }

\newcommand{\wrap}[1]{\begin{tabular}{@{}c@{}}#1\end{tabular}}
\newcommand{\lwrap}[1]{\begin{tabular}{@{}l@{}}#1\end{tabular}}
\newcommand{\mwrap}[1]{\begin{array}{@{}c@{}}#1\end{array}}

\definecolor{dblue}{RGB}{52, 81, 105}
\definecolor{lblue}{RGB}{218, 218, 247}
\definecolor{webgreen}{rgb}{0,.5,0}
\definecolor{webbrown}{rgb}{.6,0,0}
\definecolor{myred}{rgb}{.7,0,0}
\definecolor{mypurple}{rgb}{.5,0,0.6}
\definecolor{mypurple2}{HTML}{800080}
\definecolor{dslcomments}{HTML}{308495}

\tikzset{elliptic state/.style={draw,ellipse}}

\tikzstyle{loc}=[draw=black,circle,rounded corners=5pt,inner sep=3pt]
\tikzstyle{rloc}=[draw=black,rectangle,rounded corners=5pt,inner sep=3pt]
\tikzstyle{req}=[ rectangle split,
       rectangle split parts=1,
       rectangle split part fill={myorange!20,myorange!20},
       rounded corners=1pt,
       draw=black, 
       dotted,
       minimum height=1em,
       inner sep=2pt,
       text centered]
\tikzstyle{eloc}=[ rectangle split,
       rectangle split parts=3,
       fill=myorange!20,
       rounded corners=1pt,
       draw=black, 
       dotted,
       minimum height=1em,
       inner sep=2pt,
       text centered]       
\tikzstyle{myinit}=[loc,double,inner sep=4pt] 
\tikzstyle{lb}=[pos=1,inner sep=0]
\tikzstyle{mynode}=[fill=red!20,circle,inner sep=0pt,draw]
\tikzstyle{myQ}=[fill=yellow!20, rectangle split, rectangle split parts=3, 
                draw, rectangle split horizontal, 
                rectangle split part align={center, top, bottom}]
\tikzstyle{myarrow}=[fill=red!50, single arrow, draw]
\tikzstyle{myCBQ}=[fill=green!10, draw, minimum height=1.5em, minimum width=3.5em, 
                  double copy shadow={shadow xshift=4pt, shadow yshift=4pt, 
                  fill=green!10, draw}]
\tikzstyle{bx}=[draw=black,inner sep=2.5mm,thick,minimum height=10mm
                ,rounded corners=3pt,font=\footnotesize]
\tikzstyle{pr}=[circle,draw=black,very thick,fill=white,inner sep=2pt]
\tikzstyle{lbl}=[rotate=0,font=\scriptsize]

\tikzstyle{grad}=[draw=dblue,fill=white, postaction={path fading=north, fading angle=45, fill=lblue,draw=dblue}]
\tikzstyle{myhubtkz}=[grad,thick,font={\footnotesize\rm\color{dblue}}]
\tikzstyle{myhublt}=[myhubtkz,minimum height=5mm,inner sep=0.5pt]

\definecolor{myblue}{HTML}{617be3}
\definecolor{myorange}{HTML}{e39c61}
\tikzstyle{var}=[circle,draw=black,font=\bf\ttfamily\footnotesize
                ,minimum height=6mm,fill=myorange!50]
\tikzstyle{fun}=[rectangle,draw=black,inner sep=1.5mm,rounded corners=0.5mm
                ,minimum height=1,font=\bf\ttfamily\footnotesize,fill=myblue!50]
\tikzstyle{source}=[var,regular polygon, regular polygon sides=3,inner sep=2pt]
\tikzstyle{sourcef}=[var,regular polygon, regular polygon sides=3,inner sep=0pt,minimum width=23pt]
\tikzstyle{sourceonce}=[sourcef,ultra thick]
\tikzstyle{sink}=[var,regular polygon, rectangle]
\tikzstyle{arr}=[->,rounded corners=5pt,>=stealth']

\newcommand{\ie}{I.e.\xspace}
\newcommand{\eg}{e.g.,\xspace}
\newcommand{\fancy}[1]{\ensuremath{\mathcal{#1}}\xspace}
\newcommand{\proj}{{\ensuremath{\downharpoonright}\xspace}}
\newcommand{\ab}{\allowbreak}
\newcommand{\mi}[1]{\ensuremath{\mathit{#1}}\xspace}
\newcommand{\mc}[1]{\ensuremath{\mathcal{#1}}\xspace}
\newcommand{\ms}[1]{\ensuremath{\mathsf{#1}}\xspace}

\newcommand{\trFe}[1]{\mi{[\textcolor{green!45!black}{#1}]}}

\newcommand{\tr}[1]{\xrightarrow{\xspace{#1}\xspace}}

\newcommand{\set}[1]{\{#1\}}
\newcommand{\tpl}[1]{\ensuremath{\langle #1 \rangle}\xspace}

\newcommand{\true}{\ensuremath{\mathit{true}}\xspace}
\newcommand{\false}{\ensuremath{\mathit{false}}\xspace}
\newcommand{\Nat}{\ensuremath{\mathbb{N}}\xspace}

\newcommand{\M}{\ensuremath{\mathcal{M}}\xspace}

\newcommand{\kw}[1]{{\textsf{\upshape #1}}}
\newcommand{\bkw}[1]{\textbf{\kw{#1}}}

\usepackage[normalem]{ulem}
\usepackage[textwidth=28.0mm,textsize=scriptsize]{todonotes}
\newcommand{\bgtext}[1]{%
  \bgroup\markoverwith {\textcolor{#1}{\rule[-0.5ex]{2pt}{11pt}}}\ULon}
\newcommand{\mytodon}[2]{\noindent\todo[color=#1,tickmarkheight=1mm]{#2}}

\newcommand{\rolftodo}[2][]{\bgtext{green!25}{#1}\mytodon{green!50}{R: #2}}
\usepackage[commandnameprefix=always,authormarkup=none,final]{changes}
\definechangesauthor[name={}, color=red]{A}

\newcommand{\del}[1]{\chdeleted[id=A]{#1}}

\makeatletter
\@ifpackagewith{changes}{final}{%
  \newcommand{\subcite}[2]{~\cite{#1}}%
}{%
  \newcommand{\subcite}[2]{\textcolor{red}{~\cite{#1}\del{\mbox{\cite{#2}}}}}%
}
\makeatother

\newcommand{\shl}[2][gray]{{\setlength\fboxsep{2pt}\colorbox{#1!20}{{#2}}\setlength\fboxsep{2pt}}}
\newcommand{\mshl}[2][gray]{\shl[#1]{\ensuremath{#2}}\xspace}

\newcommand{\fm}{\mi{fm}}

\newcommand{\verum}{\ensuremath{\mathord{\bm{\top}}}\xspace}
\newcommand{\falsum}{\ensuremath{\mathord{\bm{\perp}}}\xspace}

\newcommand{\FETA}{f\ensuremath{\mkern1mu}ETA\xspace}
\newcommand{\FTS}{f\ensuremath{\mkern1mu}TS\xspace}
\newcommand{\FSTS}{f\ensuremath{\mkern1mu}STS\xspace}
\newcommand{\FSys}{f\ensuremath{\mkern1mu}Sys\xspace}
\newcommand{\FCA}{f\ensuremath{\mkern1mu}CA\xspace}

\newcommand{\mFETA}{\ensuremath{\mathit{fETA}}\xspace}

\newcommand{\mFSTS}{\ensuremath{\mathit{fSTS}}\xspace}

\newcommand{\mFCA}{\ensuremath{\mathit{fCA}}\xspace}
\newcommand{\ETA}{ETA\xspace}

\newcommand{\STS}{STS\xspace}

\newcommand{\CA}{CA\xspace}

\newcommand{\REQs}{Reqs\xspace}

\newcommand{\Aut}{\ensuremath{\mathcal{A}}\xspace}
\newcommand{\LTS}{\ensuremath{\mathcal{L}}\xspace}

\newcommand{\lock}[1][\textcolor{black!50}]{\text{#1{\smaller\faLock}}\xspace}
\newcommand{\unlock}[1][\textcolor{black!50}]{\text{#1{\smaller\faUnlock}}\xspace}
\newcommand\xqed[1]{%
  \leavevmode\unskip\penalty9999 \hbox{}\nobreak\hfill
  \quad\hbox{#1}}
\newcommand\qedx{\xqed{$\vartriangleright$}} 

\newcommand{\K}{\fancy{K}\xspace}
\newcommand{\N}{\fancy{N}\xspace}

\renewcommand{\S}{\ensuremath{\mathcal{S}}\xspace}
\renewcommand{\N}{\ensuremath{\mathcal{N}}\xspace}

\newcommand{\stype}{\bkw{st}\xspace}
\newcommand{\etas}[2]{\bkw{eta}(#2,#1)\xspace}
\newcommand{\ta}[2]{\bkw{ta}(#2,#1)\xspace}
\newcommand{\fta}[2]{\bkw{fta}(#2,#1)\xspace}
\newcommand{\iact}{\Lambda(\syssig,\stype)}
\newcommand{\Labels}[1][\S]{\ensuremath{\Lambda(#1)}\xspace}
\newcommand{\Edges}[1][\S]{\ensuremath{E(#1)}\xspace}
\newcommand{\Labelst}[1][\S,\stype]
{\ensuremath{\Lambda(#1)}\xspace}
\newcommand{\Edgest}[1][\S,\stype]{\ensuremath{E(#1)}\xspace}
\newcommand{\fstype}{\bkw{fst}\xspace}
\newcommand{\lts}{\bkw{lts}\xspace}
\newcommand{\fts}{\bkw{fts}\xspace}

\newcommand{\rcp}{\bkw{rcp}\xspace}
\newcommand{\rsp}{\bkw{rsp}\xspace}
\newcommand{\n}{\mi{n}}
\newcommand{\ki}{\mi{k}}
\newcommand{\In}{\pts{in}}
\newcommand{\Out}{\pts{out}}
\newcommand{\inn}{\mi{I}}
\newcommand{\out}{\mi{O}}
\newcommand{\comm}{\ensuremath{\Sigma}\xspace}
\newcommand{\syssig}{\ensuremath{\Theta}\xspace}

\newcommand{\diam}[1]{\left\langle#1\right\rangle}

\tikzstyle{nitem}=[circle,draw=green!60!black,thick,fill=green!20,inner sep=1pt]

\usepackage{isomath}
\renewcommand{\emptyset}{\varnothing}

\newcommand{\cod}[2][]{\!\!\ensuremath{\text{
\lstinline[morekeywords={#1}]"#2"\xspace
}}}

\makeatletter
\providecommand*{\Dashv}{%
  \mathrel{%
    \mathpalette\@Dashv\vDash
  }%
}
\newcommand*{\@Dashv}[2]{%
  \reflectbox{$\m@th#1#2$}%
}
\makeatother

\usepackage[misc]{ifsym}

\newcommand{\pts}[1]{\mi{{\textcolor{ptcolor}{\mathsf{#1}}}}}
\newcommand{\pt}[1]{\mi{{\textcolor{ptcolor}{#1}}}}
\newcommand{\msg}[1]{\mi{{\textcolor{mscolor}{\mathsf{#1}}}}}

\usepackage{xstring}
\DeclareRobustCommand{\cod}[1]{\mi{
  \IfSubStr{#1}{->}{
    \pt{\StrBefore{#1}{-}}
    \textcolor{opcolor}{\shortrightarrow}
    \pt{\StrBetween{#1}{>}{:}}
    {:}
    \msg{\StrBehind{#1}{:}}
  }{
    \IfSubStr{#1}{?}{
      \pt{\StrBefore{#1}{?}}
      \textcolor{opcolor}{?}
      \msg{\StrBehind{#1}{?}}
    }{
      \IfSubStr{#1}{!}{
        \pt{\StrBefore{#1}{!}}
        \textcolor{opcolor}{!}
        \msg{\StrBehind{#1}{!}}
      }{
        undefined
      }
    }
  }
}}

\newcommand{\goOnline}[2]{\href{http://lmf.di.uminho.pt/ceta/?#2}{#1\,\ensuremath{^{\text{\raisebox{-1pt}{\faExternalLink}}}}}}

\newcommand{\ignore}[1]{}

\tikzstyle{teams}=[
  ->,scale=.5,>=stealth',shorten >=1pt,auto, node distance=1.5cm,
  semithick, every node/.style={scale=0.7},initial text={}]
\tikzstyle{teamState}=[fill=white,draw=black,text=black]
\tikzstyle{lclr}=[rectangle, rounded corners=8pt, inner sep=3pt,fill=localcol]
\tikzstyle{lclp}=[rectangle, rounded corners=8pt, inner sep=3pt,fill=globalcol]
\tikzstyle{select}=[fill=metLightBrown]

\newcommand{\nocontentsline}[3]{}
\let\origcontentsline\addcontentsline
\newcommand\stoptoc{\let\addcontentsline\nocontentsline}
\newcommand\resumetoc{\let\addcontentsline\origcontentsline}

\newcommand{\ai}        {\textit{ai}}             

\theoremstyle{plain} 

\def\eg{{\em e.g.}}
\def\cf{{\em cf.}}
\def\ie{{\em i.e.}}

\def\etal{{\em et al}}

\begin{document}

\title[Overview and Roadmap of Team Automata]{Overview and Roadmap of Team Automata}

\specialissue{Selected Papers of the 44th International Conference on Formal Techniques for Distributed Objects, Components and Systems and the 26th International Conference on Coordination Models and Languages (FORTE and COORDINATION 2024)}

\author[M.H.~ter~Beek]{Maurice H. ter Beek\lmcsorcid{0000-0002-2930-6367}}[a]
\address{CNR--ISTI, Pisa, Italy}
\email{maurice.terbeek@isti.cnr.it}

\author[R.~Hennicker]{Rolf Hennicker}[b]
\address{LMU Munich, Germany}
\email{hennicker@ifi.lmu.de}

\author[J.~Proen\c{c}a]{Jos\'{e} Proen\c{c}a\lmcsorcid{0000-0003-0971-8919}}[c]
\address{CISTER and INESC TEC and University of Porto, Portugal}
\email{jose.proenca@fc.up.pt}

\begin{abstract}\noindent
Team Automata is a formalism for interacting component-based systems proposed in 1997, whereby multiple sending and receiving actions from concurrent automata can synchronise.
During the past 25{\tt +} years, team automata have been studied and applied in many different contexts, involving 25{\tt +} researchers and resulting in 25{\tt +} publications.
In this paper, we first revisit the specific notion of synchronisation and composition of team automata, relating it to other relevant \emph{coordination models}, such as Reo, BIP, Contract Automata, Choreography Automata, and Multi-Party Session Types.
We then identify several aspects that have recently been investigated for team automata and related models. 
These include \emph{communication properties} (which are the properties of interest?), \emph{realisability} (how to decompose a global model into local components?), \emph{tool support} (what has been automatised or implemented?), and \emph{variability} (can a family of concrete product (automata) models be captured concisely?).
Our presentation of these aspects provides a snapshot of the most recent trends in research on team automata, and delineates a roadmap for future research, both for team automata and for related formalisms.
\end{abstract}

\maketitle

\section{Introduction}
\label{sec:introduction}

Team automata (TA) were first proposed by Skip Ellis at the 1997 ACM SIGGROUP Conference on Supporting Group Work~\cite{Ell97} for modelling components of groupware systems and their interconnections. They were inspired by Input/Output (I/O) automata~\cite{LT89,LT26} and in particular inherit their distinction between internal (private, \ie, not observable by other I/O automata) actions and external (\ie, input and output) actions used for communication with the environment (\ie, other I/O automata). Technically, team automata are an extension of I/O automata, since a number of the restrictions of I/O automata were dropped for more flexible modelling of several kinds of interactions in groupware systems. The underlying philosophy is that automata cooperate and collaborate by jointly executing (synchronising) transitions with the same action label (but possibly of different nature, \ie, input or output) as agreed upon upfront. Team automata can be composed using a synchronous product construction that defines a unique composite automaton, the transitions of which are exactly those combinations of component transitions that represent a synchronisation on a common action by all the components that share that action. The effect of a synchronously executed action on the state of the composed automaton is described in terms of the local state changes of the automata that take part in the synchronisation. The automata not involved remain idle and their current states are unaffected.

Team automata were formally defined in Computer Supported Cooperative Work (CSCW)---The Journal of Collaborative Computing~\cite{BEKR03}, in terms of component automata that synchronise on certain executions of actions. Unlike I/O automata, team automata impose hardly any restrictions on the role of actions in components and their composition is not limited to the synchronous product. 
Composing a team automaton requires defining its transitions by providing the actions and synchronisations that can take place from the combined states of the components. Each team automaton is thus a composite automaton defined over component automata. However, a given fixed set of component automata does not define a single unique team automaton, but rather a range of team automata, one for each choice of the team's transitions (individual or synchronising transitions from the component automata). This is in contrast with the usual synchronous product construction. The distinguishing feature of team automata is this very loose nature of synchronisation according to which specific synchronisation policies can be determined, defining how many component automata can participate in
the synchronised execution of a shared external action, either as a sender (\ie, via an output action) or as a receiver (\ie, via an input action). This flexibility makes team automata capable of capturing in a precise manner a variety of notions related to coordination in distributed systems (of systems).

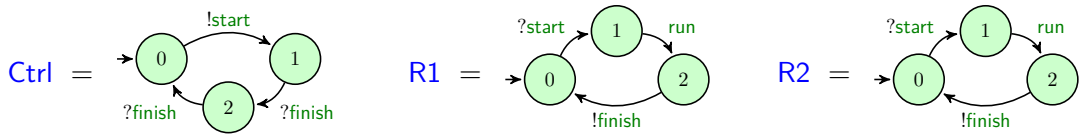
\begin{figure}[b]
  \centering
  $\pts{Ctrl} =\!\!$
   \wrap{

\begin{tikzpicture}[teams]
  \tikzstyle{every state}=[teamState,fill=localcol]
  \node[initial,state]        (0){$0$};
  \phantom{\node[state]       (d)[right=0.5 of 0]{$d$};}
  \node[state]                (2)[below of=d,below=-1.0cm]{$2$}; 
  \node[state]                (1)[right=0.5 of d]{$1$};

  \draw (0) edge[bend left=20,above]            node{{!\msg{start\/}}} (1)
        (1) edge[bend left=20,below right] node{{?\msg{finish\/}}}    (2)
        (2) edge[bend left=20,below left]  node{?\msg{finish\/}}      (0);
\end{tikzpicture}}
  \qquad
  $\pts{R1} =\!\!$
  \wrap{\begin{tikzpicture}[teams]
  \tikzstyle{every state}=[teamState,fill=localcol]
  \node[initial,state] (0){$0$};
  \phantom{\node[state]       (d)[right=0.5 of 0]{$d$};}
  \node[state]                (1)[above of=d,above=-1.0cm]{$1$}; 
  \node[state]                (2)[right=0.5 of d]{$2$};

  \draw (0) edge[bend left=20,above left]   node{{?\msg{start\/}}} (1)
        (1) edge[bend left=20,above right]  node{\msg{run\/}} (2)
        (2) edge[bend left=20,below]   node{!\emph{\msg{finish\/}}} (0);
  \node [fit=(0)(2),draw=none,inner sep=15pt] {};
\end{tikzpicture}}
  \qquad
  $\pts{R2} =\!\!$
  \wrap{\begin{tikzpicture}[teams]
  \tikzstyle{every state}=[teamState,fill=localcol]
  \node[initial,state] (0){$0$};
  \phantom{\node[state]       (d)[right=0.5 of 0]{$d$};}
  \node[state]                (1)[above of=d,above=-1.0cm]{$1$}; 
  \node[state]                (2)[right=0.5 of d]{$2$};

  \draw (0) edge[bend left=20,above left]   node{{?\msg{start\/}}} (1)
        (1) edge[bend left=20,above right]  node{\msg{run\/}} (2)
        (2) edge[bend left=20,below]   node{!\emph{\msg{finish\/}}} (0);
  \node [fit=(0)(2),draw=none,inner sep=15pt] {};
\end{tikzpicture}}
  \caption{Race example: a controller \pts{Ctrl} and two runners \pts{R1} and \pts{R2}}
  \label{fig:race-local}
\end{figure}

To illustrate this, consider the Race example in \cref{fig:race-local}, borrowed from~\cite{BCHK17,BHP24}, which models a controller \pts{Ctrl} that wants to \emph{simultaneously} send to runners \pts{R1} and \pts{R2} a \msg{start} message, after which it is able to receive from each runner \emph{separately} a \msg{finish} message once that runner \emph{individually} has \msg{run}. Here and in all subsequent examples and figures, components have a single initial state, indicated by an incoming arrow head, typically denoted by~$0$, and external actions may be prefixed by \lq\lq !\rq\rq\ (for output) or \lq\lq ?\rq\rq\ (for input). 
Further, more complex examples can also be found in the literature, increasing the comprehension of (meta) access control strategies (\eg, with or without deep revocation) visualised through a virtual spaces metaphor~\cite{BEKR01b}, guaranteeing robustness against security properties (\eg, integrity and secrecy) in dedicated communication protocols, or capturing communication safety in scenarios with multiple clients connecting and interacting with a server~\cite{BCHK17,BHK20b,BCHP21}, possibly with a proxy for buffering~\cite{BK09}.

It is important to note that the synchronous product (as used in I/O automata and many other formalisms) of these three automata has a deadlock: after synchronisation of the three \msg{start} transitions, \pts{Ctrl} is blocked in state~$1$ until both \pts{R1} and \pts{R2} have executed their \msg{run} action; at that point, full synchronisation of the \msg{finish} transitions leads to a deadlock, with \pts{Ctrl} in state~$2$ and \pts{R1} and \pts{R2} in their initial state. Team automata allow to exclude the latter synchronisation, yet at the same time enforcing the full synchronisation of \msg{start}.

\myparagraph{Contribution} 
This paper extends~\cite{BHP24} in the following 
ways, as mentioned in detail below. In particular, we have added examples and details plus two new sections dealing with composition of systems and with variability.
To help readers navigate through this paper, we include a table of contents.

\vspace*{-\baselineskip}
\tableofcontents
\vspace*{-\baselineskip}

\begin{figure}[h]
  \centering
  \resizebox{\textwidth}{!}{\input{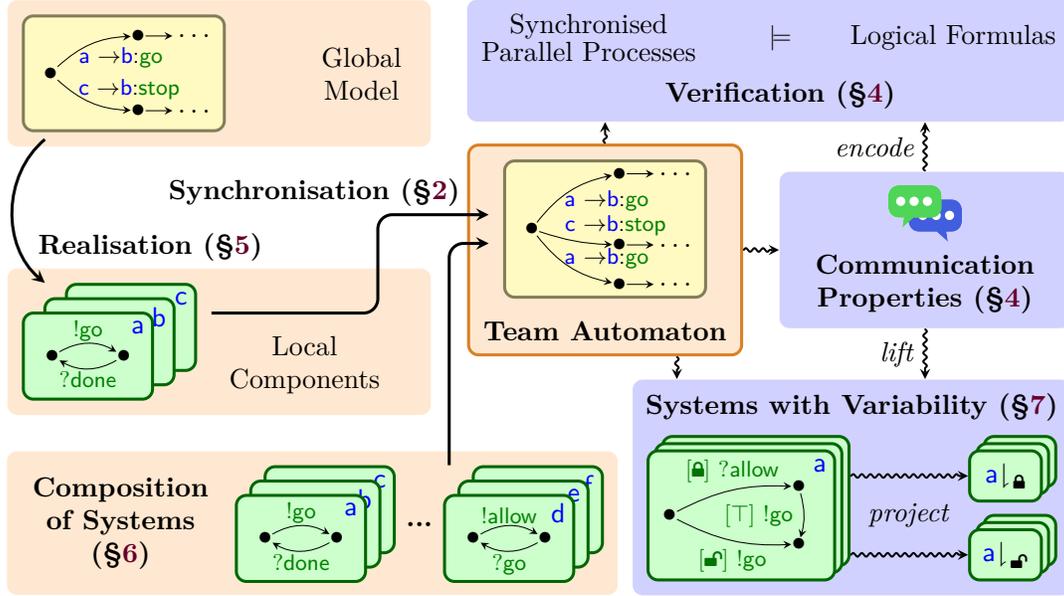}
  }
  \caption{Aspects of team automata addressed in this paper}
  \label{fig:contributions}
\end{figure}

We first revisit the specific notion of synchronisation to build
team automata (Section~\ref{sec:TA}), after which we briefly relate team automata to a selection of coordination models (Section~\ref{sec:formalisms}). 
We then focus on four aspects of team automata that we investigated during the last five years:\linebreak \emph{verification of communication properties} (Section~\ref{sec:comm}), \emph{realisation} (Section~\ref{sec:realisability}), \emph{composition of systems}
(Section~\ref{sec:composition}, not included in~\cite{BHP24}), and \emph{variability} (Section~\ref{sec:var}, not included in~\cite{BHP24}).
\begin{description}
    \item[\textsection\ref{sec:comm}] We report results on verification of compliance with communication requirements in terms of receptiveness (no message loss) and responsiveness (no indefinite waiting), give a thorough comparison with other compatibility notions, incl. deadlock-freedom,
    and give a roadmap for future work on verification of communication properties.
    \item[\textsection\ref{sec:realisability}] We provide results on the decomposition (realisation) of a global interaction model%
    \footnote{This global model is explicitly specified upfront, 
    while the team automaton results from the composition of component automata (\cf\ \cref{sec:realisability})}
    in terms of a (possibly distributed) system of component automata 
    coordinated according to a given synchronisation type specification.
    In particular, we provide a revised and extended comparison of our approach with that of Castellani \etal.~\cite{CMT99} and a roadmap for future work on realisability.
    \item[\textsection\ref{sec:composition}] We discuss results on the composition of \emph{systems} of component automata and the composition of synchronisation type specifications, as well as on the preservation of communication properties from team automata of subsystems to global team automata. We provide a roadmap for future work on composition of systems.
    \item[\textsection\ref{sec:var}] We present results on team automata to support variability in the development and analysis of teams, \ie, featured team automata that concisely describe a family of concrete product (automaton) models for specific configurations determined by feature selection, show how to lift the notion of receptiveness to the level of family models,
    and mention a roadmap for future work on variability.
\end{description}
Finally, we mention further aspects of team automata and of some of the related coordination models and conclude the paper (Section~\ref{sec:conclusion}).  
Figure~\ref{fig:contributions} summarises this paper's contribution.
\cref{app:pub} lists selected team automata publications.

\section{Team Automata in a Nutshell}\label{sec:nutshell}

Team automata were originally introduced by Ellis~\cite{Ell97} and formally defined in~\cite{BEKR03}.
They form an automaton model for systems of reactive components that differentiate input (passive), output (active), and internal (privately active) actions. 
In this section, we recall the basic notions of (extended) team automata. 

\label{sec:ca}
A \emph{labelled transition system} (LTS) is a quadruple $\LTS = (Q,q_0,\Sigma,E)$
such that
$Q$ is a finite set of states,
$q_0 \in Q$ is the initial state,
$\Sigma$ is a finite set of labels, and  
$E\subseteq Q\times \Sigma \times Q$ is a transition relation.
Given an LTS \LTS, we write $q\tr{a}_\LTS q'$, or shortly $q\tr a q'$, to denote $(q,a,q') \in E$. 
Similarly, we write $q\tr{a}_\LTS$ to denote that $a$ is \emph{enabled} in $\LTS$ at state~$q$, \ie, there exists $q'\in Q$ such that $q \tr{a} q'$.
For $\Gamma\subseteq\Sigma$, we write $q\tr{\Gamma}\!\!^{*}\,q'$ if there exist $q\tr{a_1}q_1\tr{a_2}\cdots\tr{a_n}q'$ for some $n\geq0$ and $a_1,\dots,a_n\in \Gamma$.
For $\Gamma\subseteq\Sigma$ and $a \in \Sigma$, we write $q\tr{\Gamma^{*};\,a}\!\,q''$ if there exists $q'$ such that $q\tr{\Gamma}\!\!^{*}\,q'$ and $q'\tr{a} q''$.
A state $q\in Q$ is \emph{reachable by} $\Gamma$ if $q_0\tr{\Gamma}\!\!^*\,q$, it is \emph{reachable} if $q_0\tr{\Sigma}\!\!^*\,q$.
The set of reachable states of \LTS is denoted by~${\mathcal R}(\LTS)$.

\myparagraph{Component Automata.} 
A \emph{component automaton} (CA) is an LTS $\Aut = (Q,q_0,\Sigma,E)$ such that $\Sigma = \Sigma^{?} \cup \Sigma^{!} \cup \Sigma^{\tau}$ is a set of \emph{component actions} (or simply \emph{actions}) with disjoint sets $\Sigma^{?}$ of \emph{input actions}, $\Sigma^{!}$ of \emph{output actions}, and $\Sigma^{\tau}$ of \emph{internal actions}.
{\em Cf.}\ \cref{fig:race-local} for examples of CA. 

\ignore{
\begin{exa}\label{ex:race-ca}
Examples of component automata are shown in~\cref{fig:race} of~\cref{sec:introduction}.
For $i = 1,2$, the action labels of $\Aut_\ms{Ri}$ are
$\Sigma_\ms{Ri} = \Sigma^{?}_\ms{Ri} \uplus \Sigma^{!}_\ms{Ri} \uplus \Sigma^{\tau}_\ms{Ri}$, where
$\Sigma^?_\ms{Ri} = \set{\mi{start}}$, $\Sigma^!_\ms{Ri} = \set{\mi{finish}}$, $\Sigma^{\tau}_\ms{Ri} = \set{\mi{run}}$.
The action labels of $\Aut_\ms{Ctrl}$ are 
$\Sigma_\ms{Ctrl} = \Sigma^{?}_\ms{Ctrl} \uplus \Sigma^{!}_\ms{Ctrl} \uplus \Sigma^{\tau}_\ms{Ctrl}$ where
$\Sigma^?_\ms{Ctrl} = \set{\mi{finish}}$, $\Sigma^!_\ms{Ctrl} = \set{\mi{start}}$, $\Sigma^{\tau}_\ms{Ctrl} = \emptyset$.
\qedx
\end{exa}
}

\myparagraph{Systems.} 
\label{sec:system}
A \emph{system} is a pair $\S = (\N,(\Aut_\n)_{\n\in \N})$, with 
$\N$ a finite, nonempty set of 
names and 
$(\Aut_\n)_{\n\in \N}$ an $\N$-indexed family of CA $\Aut_\n = (Q_\n,q_{0,\n},\Sigma_\n,E_\n)$.
Any system $\S$ 
induces an LTS defined by
$\lts(\S) = (Q, q_0,\Labels,\ab\Edges)$, where
$Q=\prod_{\n\in\N} Q_\n$ is the set of \emph{system states}, 
$q_0 = (q_{0,\n})_{\n \in \N}$ is the \emph{initial system state},
$\Labels$ is the set of \emph{system labels}, and
$\Edges$ is the set of \emph{system transitions}.  
Each system state $q\in Q$ is an $\N$-indexed family $(q_\n)_{\n \in \N}$ of local CA 
states $q_\n \in Q_\n$.
The definitions of $\Labels$ and $\Edges$ follow 
that 
of \emph{system action}.

\myparagraph{System Actions.} 
The set of \emph{system actions} $\Sigma = \bigcup_{\n\in\N} \Sigma_\n$ determines actions that will be part of system labels. 
Within $\Sigma$ we identify $\Sigma^{\bullet} = \bigcup_{\n\in\N} \Sigma^?_\n \cap \bigcup_{\n\in\N} \Sigma^!_\n$ as the set of \emph{communicating actions}.
Hence, an action $a \in \Sigma$ is communicating if it occurs in (at least) one set $\Sigma_k$ of component actions as an input action and in (at least) one set $\Sigma_{\ell}$ of action labels as an output action. 
System actions that are neither communicating nor internal to a component are called \emph{open}.
Hence, the set of open system actions is given by
$\Sigma^{\circ} = \Sigma \setminus (\Sigma^{\bullet} \cup\bigcup_{\n\in\N}\Sigma^{\tau}_\n).$
The system is \emph{closed} if $\Sigma^{\circ} = \emptyset$, \ie, 
all non-communicating actions are internal component actions.

\ignore{
\begin{exa}\label{ex:race-sys}
The Race system of~\cref{sec:introduction} is $\ms{Race} = (\N_\ms{Race}, (\Aut_n)_{n\in\N_\ms{Race}})$, with
$\N_\ms{Race} = \{\ms{R1}, \ms{R2},\ms{Ctrl}\}$
and the CA
$\Aut_\ms{R1},\Aut_\ms{R2}$, and $\Aut_\ms{Ctrl}$
from~\cref{fig:race-loca}. 
The system actions of the Race system are 
$\Sigma_\ms{Race}\,{=}\,\mi{\set{start,finish,run}}$
and its communicating actions are
$\Sigma_\ms{Race}^{\bullet} = \mi{\set{start,finish}}$.
\qedx
\end{exa}
}

\myparagraph{System Labels.} 
We use \emph{system labels} to indicate which components participate (simultaneously) in the execution of a system action.
There are two kinds of system labels.
In a system label of the form $(\Out,a,\In)$,
\Out represents the set of senders of \emph{outputs} and 
\In the set of receivers of \emph{inputs} that synchronise on the action $a \in \Sigma^{\bullet}$.
Either \Out or \In can be empty, but not both.
Obviously, if $a \in \Sigma^{\circ}$ is an open system
action, then either \Out or \In must be empty. 
A system label of the form $(\n,a)$ indicates that component \n executes an internal action $a \in \Sigma^{\tau}_\n$. 
Formally, the set $\Labels$ of system labels of $\S$ is defined as follows: 
\begin{align*}
\Lambda(\S) & = \{\,(\Out,a,\In) \mid \emptyset\neq (\Out\cup \In) \subseteq \N,\, \allowbreak
\forall_{\n\in \Out}\cdot a\in \Sigma^!_\n,\ 
\forall_{\n\in \In}\cdot a\in \Sigma^?_\n\,\} \\ 
& {}\,\cup\{\,(\n,a) \mid \n\in \N,\, a\in \Sigma^{\tau}_\n\,\}
\end{align*}
Note that \Labels depends only on $\N$ and 
the sets $\Sigma_\n$ of component actions for each $\n\in\N$.
If $\Out = \{\n\}$ is a singleton, we write $(\n,a,\In)$ instead of $(\{\n\},a,\In)$; similarly for singleton sets $\In$.\linebreak

In all figures and examples, interactions $(\Out,a,\In)$
are presented by the notation $\syn[a]\Out\In$ and internal labels $(n,a)$ by 
$\alone[a]{n}$.
\ignore{
\begin{exa}\label{ex:race-labels}
The set of system labels of the race system is given by
\begin{align*}
\Lambda(\ms{Race}) =&~
\{(\Out,\mi{start},\In) \mid \emptyset\neq (\Out\cup \In),\, \Out \subseteq \{\ms{Ctrl}\},
\In \subseteq \{\ms{R1,R2}\}\},
\\
\cup&~ 
\{(\Out,\mi{finish},\In) \mid  \emptyset\neq (\Out\cup \In),\, \Out \subseteq \{\ms{R1,R2}\},
\In \subseteq \{\ms{Ctrl}\}\},
\\
\cup&~ 
\{(\ms{R1},\mi{run}), (\ms{R2},\mi{run})\}. \tag*{\qedx}
\end{align*}
\end{exa}
}
System labels provide an appropriate means to describe which components in a system execute, possibly together, a computation step, \ie, a system transition. 

\myparagraph{System Transitions.} 
A \emph{system transition} $t\!\in\!\Edges$ has the form
   $(q_\n)_{\n\in\N}  \tr{\lambda}_{\lts(\S)}  (q'_\n)_{\n\in\N}$
    such that $\lambda \in \Labels$ and
    \begin{itemize}
      \item either
      $\lambda = (\Out,a,\In)$ and:
        \begin{itemize}[label=\hbox{\tiny$\bullet$}]
          \item[] $q_\n \tr{a}_{\Aut_\n} q'_\n$, for all $\n \in \Out\cup \In$, and
          $q_m = q'_m$, for all $m\in\N\backslash(\Out\cup \In)$;
        \end{itemize}
      \item or $\lambda = (n,a)$,
      $a\in \Sigma^{\tau}_\n$ is an internal action of some component $\n \in \N$, and:
      \begin{itemize}[label=\hbox{\tiny$\bullet$}]
        \item[] $q_\n \tr{a}_{\Aut_\n} q_\n'$ and
        $q_m = q_m'$, for all $m\in\N\backslash{\{\n\}}$.
      \end{itemize}
    \end{itemize}
We write $\Lambda$ and $E$ instead of \Labels and \Edges, respectively, if \S is clear from the context.
Surely, at most the components that are in a local state where action~$a$\linebreak is locally enabled can participate in a system transition for $a$.
Since, by definition of system labels, $(\Out\cup \In)\neq\emptyset$, at least one component participates in any system transition.
Given a system transition $t = q\tr{\lambda}_{\lts(\S)} q'$, we write $t.\lambda$ for~$\lambda$.

\begin{exa}\label{ex:sys-transitions}
The Race system in \cref{fig:race-local} is a closed system. 
It has both desired system transitions such as
$(0,0,0) \tr{(\pts{Ctrl},\msg{start},\{\pts{R1},\pts{R2}\})} (1,1,1)$ and
$(1,2,2) \tr{(\pts{R1},\msg{finish},\pts{Ctrl})} (2,0,2)$,
and undesired ones like
$(0,0,0) \tr{(\pts{Ctrl},\msg{start},\emptyset)} (1,0,0)$
and
$(1,2,2) \tr{(\{\pts{R1},\pts{R2}\},\msg{finish},\pts{Ctrl})} (2,0,0)$.
The LTS of the Race system, denoted by $\lts(\ms{Race})$, contains all possible system transitions. 
As mentioned in Section~\ref{sec:introduction}, 
the latter two are undesired since the controller is supposed to start both runners simultaneously, whereas they  
should finish individually. 
These and other system transitions will be discarded based on synchronisation restrictions for teams considered next. 

\noindent\begin{minipage}{0.74\textwidth}
\qquad As an open system, consider a variant of the Race example from \cref{fig:race-local} in which runner \pts{R2} is replaced by $\pts{R2}'$ depicted on the right.\footnotemark\linebreak The difference is that this runner may decide not to wait for the \msg{start} signal of the controller and start by performing an input action \msg{go}. This is an external action but not a communicating action, which may be called by the system's environment (\eg, a false start signal coming from the outside). Thus, this Race system has an open input.
\end{minipage}\ 
\begin{minipage}{0.2\textwidth}
  \centering
  $$\pts{R2}' =$$
  
  \medskip
  \wrap{\begin{tikzpicture}[teams]
  \tikzstyle{every state}=[teamState,fill=localcol]
  \node[initial,state] (0){$0$};
  \phantom{\node[state]       (d)[right=0.5 of 0]{$d$};}
  \node[state]                (1)[above of=d,above=-.5cm]{$1$}; 
  \node[state]                (2)[right=0.5 of d]{$2$};

  \path (0) edge[bend left,above left]   node{{?\msg{start\/}}} (1)
        (0) edge[bend right,below right]   node{{?\msg{go\/}}} (1)
        (1) edge[bend left,above right]  node{\msg{run\/}} (2)
        (2) edge[bend left=45,below]   node{!\emph{\msg{finish\/}}} (0);
  \node [fit=(0)(2),draw=none,inner sep=15pt] {};
\end{tikzpicture}}
\end{minipage}
\footnotetext{Another variant of this Race system as an open system will be described in Section~\ref{sec:composition} (\cf\ Example~\ref{ex:composition}).}

\noindent As a result, the undesired state $(0,0,1)$ becomes reachable with the external \msg{go} action.
\qedx
\end{exa}

\myparagraph{Team Automata.} 
\label{sec:TA}
Synchronisation types specify which synchronisations of components are admissible in a specific system~\S.
A \emph{synchronisation type} $(\out,\inn)\in\kw{Intv}{\times}\kw{Intv}$ is a pair of intervals \out and \inn which determine the number of outputs and inputs that can participate in a communication.
Each interval has the form $[\mi{min},\mi{max}]$
with $\mi{min}\in\Nat$ and $\mi{max}\in\Nat\cup\{*\}$ where $*$ denotes $0$ or more participants.
We write $x\in[\mi{min},\mi{max}]$ if $\mi{min}\,{\leq}\, x\,{\leq}\,\mi{max}$ and $x\in[\mi{min},*]$ if $x\,{\geq}\,\mi{min}$.

A \emph{synchronisation type specification} (STS) over \S is a function $\stype:\Sigma^{\bullet} \to \kw{Intv}{\times}\kw{Intv}$ that assigns to any communicating action $a$ an individual synchronisation type $\stype(a)$.
A system label $\lambda = (\Out,a,\In)$ \emph{satisfies} $\stype(a) = (\out,\inn)$, denoted by $\lambda \models \stype(a)$, if $|\Out|\in \out \land |\In|\in \inn$. 

Each STS 
$\stype$ generates the following subsets
\Labelst of system labels
and \Edgest of corresponding system transitions.
\begin{align*}
  \Labelst & = \set{\,\lambda \in \Lambda \mid
    \lambda = (\Out,a,\In)
    \land\, a \in \Sigma^{\bullet} \Rightarrow
    \lambda \models \stype(a)\,} \\
  \Edgest & = \set{\,t \in E \mid
    t.\lambda \in \Labelst\,}
\end{align*}
Thus, for communicating actions, the set of system transitions is restricted to those transitions whose labels respect the synchronisation type of their communicating action. For internal and open actions no restriction is applied.
The reason is that
(1) an internal action of a component can always be executed when it is locally enabled
and
(2) an open action is meant to be restricted only when composing systems (\cf\ Section~\ref{sec:composition}). 

Components interacting in accordance with an STS $\stype$ over a system \S are seen as a team whose behaviour is represented by the (extended) \emph{team automaton} (TA) $\ta{\stype}{\S}$ generated over \S by $\stype$ and defined by the LTS
$$\ta{\stype}{\S} = (Q,q_0,\Labelst,\Edgest).\footnote{\label{foot:CI}Starting with~\cite{BHK20b}, we have been using the system labels $(\Out, a, \In)$ in $\Labelst$ as the actions in team transitions of what we coined extended team automata (\ETA). This is the main difference with the \lq classical\rq\ team automata from~\cite{Ell97,BEKR03} and subsequent papers, where actions $a\in\Sigma$ have been used in team transitions. However, to study com\-mu\-ni\-ca\-tion properties~\cite{BCHP23}, compositionality~\cite{BHK20b} and realisability~\cite{BHP23}, explicit rendering of the CA that actually participate in a transition of the team turned out useful.}$$

We write \Labelst[\stype] and \Edgest[\stype] instead of \Labelst and \Edgest, respectively, if \S is clear from the context, and assume that $\Labelst[\stype] \neq \emptyset$. 
Labels in \Labelst[\stype] are called \emph{team labels} and transitions in \Edgest[\stype] are called \emph{team~transitions}.

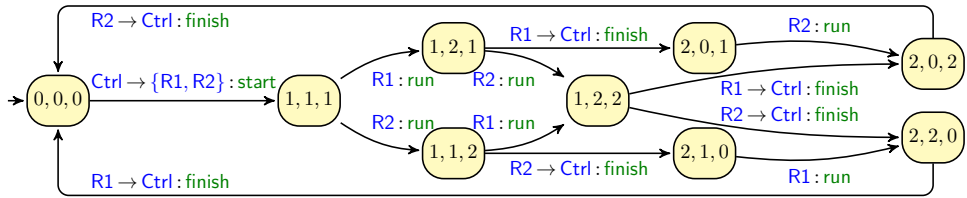
\begin{figure}[hb!]

\begin{tikzpicture}[teams]
  \tikzstyle{every state}=[teamState,fill=globalcol,rectangle,rounded corners=8pt]
  \node[initial,state]            (0){$0,0,0$};
  \node[state,right=2.5 of 0]       (1){$1,1,1$};
  \coordinate[right=1.5 of 1]       (d1);
  \node[state,right=1.5 of d1]      (2){$1,2,2$}; 
  \coordinate[right=1.0 of 2]       (d2);
  \node[state,yshift= 20mm]at(d1) (1u){$1,2,1$};
  \node[state,yshift=-20mm]at(d1) (1d){$1,1,2$};
  \node[state,yshift= 20mm]at(d2) (2u){$2,0,1$};
  \node[state,yshift=-20mm]at(d2) (2d){$2,1,0$};
  \node[state,right=2.5 of 2u]    (3u){$2,0,2$};
  \node[state,right=2.5 of 2d]    (3d){$2,2,0$};

  \newcommand{\lbl}[1]{\mi{\mathsf{#1}}}
  \path (0) edge[bend left=0,above]
              node(st){\lbl{\syn[start]{Ctrl}{\{R1,R2\}}}} (1)
        (1) edge[bend left=15,below right]
              node[pos=0.3,yshift= 2mm]{\lbl{\alone[run]{R1}}} (1u)
            edge[bend right=15,above right]
              node[pos=0.3,yshift=-2mm]{\lbl{\alone[run]{R2}}} (1d)
        (1u) edge[bend left=0,above]
              node{\lbl{\syn[finish]{R1}{Ctrl}}} (2u)
             edge[bend left=15,below left]
              node[pos=0.7,yshift= 2mm]{\lbl{\alone[run]{R2}}} (2)
        (1d) edge[bend left=0,below]
              node{\lbl{\syn[finish]{R2}{Ctrl}}} (2d)
             edge[bend right=15,above left]
              node[pos=0.7,yshift=-2mm]{\lbl{\alone[run]{R1}}} (2)
        (2) edge[bend right=6,below]
              node[pos=0.9,yshift=-4mm]{\lbl{\syn[finish]{R1}{Ctrl}}} (3u)
            edge[bend left=6,above]
              node[pos=0.9,yshift= 4mm]{\lbl{\syn[finish]{R2}{Ctrl}}} (3d)
        (2u) edge[above]
              node{\lbl{\alone[run]{R2}}} (3u)
        (2d) edge[below]
              node{\lbl{\alone[run]{R1}}} (3d)
        ;
  \coordinate[yshift= 12mm](top)at(1u);
  \coordinate[yshift=-12mm](bot)at(1d);
  \draw[rounded corners] (3u) |- (top) -| (0);
  \draw[rounded corners] (3d) |- (bot) -| (0);
  \node[below right] at (st.west|-top)  {\lbl{\syn[finish]{R2}{Ctrl}}};
  \node[above right] at (st.west|-bot)  {\lbl{\syn[finish]{R1}{Ctrl}}};
\end{tikzpicture}
  \medskip
  \caption{Team automaton of the Race system example in \cref{fig:race-local}} 
  \label{fig:ta-race}
\end{figure}

\begin{exa}
\label{ex:race-ta}
  For the Race system $\S_{\ms{Race}}$ in \cref{fig:race-local},
  we define 
  the runners to \mi{start} simultaneously and \mi{finish} individually by 
  the STS $\stype_{\ms{Race}}=\set{\msg{start}\mapsto ([1,1],[2,2]),\allowbreak \msg{finish}\mapsto ([1,1],[1,1])}$.
  The resulting TA $\ta{\S_{\ms{Race}}}{\stype_{\ms{Race}}}$ is shown in \cref{fig:ta-race}, with
  interactions $(n,a,m)$ written as $\syn[a]nm$ and internal labels $(n,a)$ as $\alone[a]{n}$.
  \qedx
\end{exa}

\noindent
Example~\ref{ex:race-ta} uses only STS with intervals of size one, as in all the examples in this paper. Further, more complex examples of STS that exploit larger intervals can be found in the literature (\cf, \eg, \cite{BCHK17,BHK20b}). We now recall some familiar ones from~\cite{BCHK17}.\footnote{The abbreviation \lq \ai\rq\ was introduced in~\cite{BEKR03} and stands for \emph{action indispensable\/}, meaning that all CA that have the action in their set $\Sigma$ of actions have to participate in the synchronisation.}

\begin{description}  
\item[\text{([1,1],\,[0,$\mathbf{*}$])}] multicast communication, meaning that a communicating action can be executed only as a synchronisation involving exactly one component for which it is an output action and any number of the components in which it is an input action.
This is called weak synchronisation in BIP.
\item[\text{([1,1],\,\ai)}] (strong) broadcast communication, meaning that whenever a communicating action is executed it occurs exactly once in its output role in that transition with all input components involved. This is called strong synchronisation in BIP.
\item[\text{(\ai,\,\ai)}] transitions on communicating actions are always \lq full\rq\ synchronisations, meaning that all components that share a communicating action are involved in all transitions on that action. In case all external actions are communicating (\ie, in a closed system), this means that we are dealing with the classical synchronous product of automata (\cf, \eg, \cite{Arn94,CK13,BCK16}). 
\item[\text{([1,$\mathbf{*}$],\,[0,$\mathbf{*}$])}] transitions on communicating actions always involve at least one component where that action is an output action. This is the idea of \lq master-slave\rq\ communication (\cf~\cite{BEKR03}), according to which a master (output) can always be executed and slaves (input) never proceed on their own.
\item[\text{([1,$\mathbf{*}$],\,[1,$\mathbf{*}$])}] as directly above, but now at least one slave has to \lq obey\rq\ (the master).
This is called \lq strong master-slave\rq\ communication (\cf~\cite{BEKR03}), by which a master (output) can always be executed and slaves (input) must be involved.
\end{description}

\ignore{
\begin{exa}\label{ex:race-team}
Recall the race system and its system labels and transitions. 
We require both runners to \mi{start} simultaneously and to \mi{finish} individually by using the STS $\stype_{\mathsf{Race}}$
defined by $\mi{start} \mapsto \mathrm{([1,1],[2,2])}$ 
and $\mi{finish} \mapsto \mathrm{([1,1],[1,1])}$.
Then the team labels of the \ETA $\etas{\stype_{\ms{Race}}}{\ms{Race}}$ are given by 
  $\Labelst[\stype_{\ms{Race}}] =
  \{\,(\ms{Ctrl},\mi{start},\{\ms{R1,R2}\}),\,
  (\ms{R1},\mi{finish},\ms{Ctrl}),\, 
  (\ms{R2},\mi{finish},\ms{Ctrl}),
  (\ms{R1},\mi{run}),\,
  (\ms{R2},\mi{run})\,\}$.
Example transitions are
$$(0,0,0) \tr{(\ms{Ctrl},\mi{start},\{\ms{R1,R2}\})} (1,1,1) \tr{(\ms{R1},\mi{run})} (2,1,1) \tr{(\ms{R1},\mi{finish},\ms{Ctrl})} (0,1,2).$$
\rolftodo{You could drop the notation explanation which is already given above.}
\qedx
\end{exa}
}

\section{Related Coordination Formalisms}
\label{sec:formalisms}

In this section, we briefly relate team automata to a selection of formal coordination models and languages. This selection is based on a more detailed comparison in~\cite{BHP24}, which provides for each formalism,
(1)~the definition of the variant considered,
(2)~the definition of composition (via synchronisation),
(3)~a possible model of our Race example in the formalism,
(4)~a brief relation with team automata, and
(5)~existing tool support. 

\subsection{Reo via Port Automata} 
\label{sec:reo}

Reo~\cite{Arb04,JA12} is a coordination language to specify and compose \emph{connectors}, \ie, patterns of valid synchronous interactions of ports of components or other connectors. 
Constraint automata~\cite{BSAR06} is a reference model for Reo's semantics~\cite{JA12}. 
Many variants of constraint automata exist~\cite[Sect.~3.2.2]{JA12}, some distinguishing inputs from outputs as in team automata. 
A variant called \emph{port automata}~\cite{KC09} abstracts away from \emph{data} constraints, focusing on \textit{synchronisation} (of actions) and \textit{composition} (of automata).
Synchronisation types in team automata restrict the number of inputs and outputs of ports with shared names; synchronisation in port automata force how this is done.  
No variant uses a similar notion to team automata's synchronisation types, although they can be expressed using intermediate ports.

There are tools to \emph{analyse}, \emph{edit}, \emph{visualise}, and~\emph{execute} Reo connectors.
Analyses include model checking, using either the dedicated model checker Vereofy~\cite{KKSB11} or encoding Reo into mCRL2~\cite{KKV10,PM19}, and simulation of extensions with parameters~\cite{PC17} and with reactive programming notions~\cite{PC20}, many accessible online at \url{http://arcatools.org/reo}.
Editors and visualisation engines include an Eclipse-based implementation~\cite{FKMMP08} and editors based on JavaScript that run in a browser~\cite{S18,CP18}.
Execution engines for Reo include a Java-based implementation~\cite{DA18} and a distributed engine using actors in Scala~\cite{P11,PCVA12}.%

\subsection{BIP (without Priorities)}
\label{sec:bip}

BIP~\cite{BBS06,BS08} is formal language to specify architectures for interacting components. A program describes the \textbf{\underline{B}}ehaviour of each component, the valid \textbf{\underline{I}}nter\-actions between their ports, and the \textbf{\underline{P}}riority among interactions.
Multiple formal models for specifying interactions exist, such as an algebra of connectors~\cite{BS08}.
We have previously compared team automata with BIP in~\cite{BCHK17}, describing how some explicit patterns of interaction of BIP, such as broadcasts, are modelled in team automata (\cf\ Section~\ref{sec:nutshell}), and with BIP without priorities~\cite{KKWVBS16} in~\cite{BHP24}, presenting a formalisation of BIP without priorities in a style similar to team automata that facilitated the comparison.
There are some technical core differences between these formalisations. 

BIP's formalisation does not include explicit internal actions, although they do exist at 
implementation level (\eg, in JavaBIP~\cite{JavaBIP-spe}). BIP's \emph{synchronisation} mechanism 
ignores
inputs and outputs.
However, the flow of \emph{data} at each interaction is sometimes described orthogonally~\cite{JavaBIP-spe}, where ports can either be \emph{enforceable} by the environment (similar to input actions) or \emph{spontaneous} by the components (similar to output actions). A more thorough comparison of the expressiveness of different BIP formalisations is given by Baranov and Bliudze~\cite{BarBliu20-express}. 
These differences reflect a different 
focus: less emphasis on \emph{communication properties} (Section~\ref{sec:comm}), internal behaviour of local components, and \emph{realisability} notions (Section~\ref{sec:realisability}); yet more focus on the exploration of different formalisms for the \emph{composition} of interactions and programs, supported by tools to connect to running systems~\cite{JavaBIP-spe}. 
Similar to team automata's synchronisation types, BIP has a formalisation parameterised on the number of components~\cite{MBBS16}. Ports can have multiple instances, and are enriched with a bound on the number of allowed agents they can synchronise with, and a bound on the number of interactions they can be involved in. 

Several tools exist for BIP, including verification tools that traverse the state space of BIP programs~\cite{BCJMRSW15} or use the VerCors model checker~\cite{BBHRS23}, and a toolset LALT-BIP for verifying freedom from global and local deadlocks~\cite{ABBJSZ18}. BIP also has a C{\tt ++} reference engine~\cite{BBBCJNS11} and a Java engine called JavaBIP~\cite{JavaBIP-spe}.

\subsection{Contract Automata} 
\label{sec:cat}

Contract automata~\cite{BDFT14,BDF16} are a finite state automata dialect proposed to model multi-party composition of contracts that can perform \emph{request\/} 
or \emph{offer\/} 
actions, which need to match 
to achieve \emph{agreement} among a composition of contracts. Contract automata have been equipped with variability in~\cite{BGGDF17,BBDLFGD20} by \emph{modalities} to specify when an action \emph{must} be matched (necessary) and when it \emph{may} be withdrawn (optional) in a composition, and with \emph{real-time constraints} in~\cite{BBL20}. 
\emph{Composition} of contract automata is 
a variant of the synchronous (automata) product, which 
interleaves or matches the transitions of the component (contract) automata such that whenever two components are enabled to execute their respective request/offer action, then the match (sychronisation) must happen. 
A composition is in \emph{agreement} if each request is matched with an offer.
Contract automata with committed states 
as introduced in~\cite{BB24} 
can
mimick multi-party \emph{synchronisation} 
(by forcing two concatenated transitions with an intermediate committed state to be executed atomically).
Internal actions can be modelled as offer actions (which do not interfere with agreement) or as silent~($\tau$) actions as introduced in~\cite{BB24}.

In~\cite{BDFT16}, contract automata are compared with communicating machines~\cite{BZ83}.
To guarantee that a composition of contract automata 
corresponds to a well-behaving (\ie, \emph{realisable}, \cf\ Section~\ref{sec:realisability}) choreography,\footnote{A choreography can be seen as a global viewpoint specifying the interactions among distributed participants in terms of a \emph{contract} defining their expected
communication behaviour in terms of message exchanges. Such a choreography is well-behaving if it is realisable by local behaviour conforming to the global viewpoint.} a
\emph{branching condition} is used.
This condition requires contract automata to perform their offers independently of the other component automata in the composition. As noted in~\cite{BBDLFGD20}, 
this condition 
is related to the phenomenon of 
\emph{state sharing} in team automata~\cite{EG02}, meaning that 
system components 
influence potential synchronisations through their local (component) states 
even if 
not involved in the actual global (system) transition. 
While a synchronous product of (I/O) automata can directly 
be seen as a Petri net, 
for team automata this only holds for 
non-state-sharing vector team automata~\cite{CK04b,BK12}.
The relation between the branching condition of contract automata and (non-)\allowbreak\hspace{0pt}state-sharing in (vector) team automata needs further study. 

Contract automata are supported by a software API called Contract Automata Library (CATLib)~\cite{BB22},
which a developer can exploit to specify contract automata and perform operations like composition and synthesis. 
The synthesis operation uses 
supervisory control theory~\cite{RW87}, 
properly revisited in~\cite{BBP20} for synthesising orchestrations and choreographies of contract automata.
An application developed with CATLib is thus formally validated \emph{by-construction} against well-behaving properties from the theory of contract automata~\cite{BDF16,BBDLFGD20,BBP20}. 
CATLib was designed to be easily extendable to support related 
formalisms; it currently supports synchronous communicating machines~\cite{LTY15}. 

\subsection{Choreography Automata}\label{sec:chor-aut}

Choreography automata~\cite{BLT20} are automata with labels that describe interactions (sender, receiver, and message name), very similar to contract automata. 
Internal actions are not captured by choreography automata (as in Reo and BIP's formalisations) and only binary \emph{synchronisation} (of actions) is supported (as in contract automata): each such interaction has a single sender (the agent that \emph{offers}) and a single receiver (the agent that \emph{requests}). Desirable properties of choreography automata include deadlock-freedom, among others, focused on the language accepted by these automata~\cite{BLT22}. Consequently, properties that rely on observational equivalence notions such as bisimilarity are not covered by these analyses. 

Corinne~\cite{OPBLT21} can be used to visualise choreography automata and to automatise operations like composition (to combine local automata into a single global automaton representing combined behaviour, \eg, through synchronisation of actions), projection (to derive local automata from a global composite automaton), and checking for well-formedness (conditions to ensure communication properties like deadlock-freedom, \cf\ Section~\ref{sec:comm}).\footnote{As admitted by the authors themselves in~\cite{BLT23}, the well-formedness condition for choreography automata reported in~\cite[Def.~4.13]{BLT20} was flawed, invalidating \cite[Thm~4.15]{BLT20} on communication properties (liveness, lock freedom, and deadlock freedom) of choreography automata. In~\cite{BLT23}, the authors show that choreography automata can be conceived as specifications of formal choreographic languages, which can be used to prove the communication properties of a given choreography automaton.}

\subsection{Multi-Party Session Types}
\label{sec:mpst}

Multi-party session types~\cite{HYC08,BY09,SY19} are a well-established family of typing disciplines  defined on top of session calculi to describe communication protocols involving multiple session participa{nts} (\emph{multi-party}).
Some variants of multi-party session types are asynchronous~\cite{HYC08}, in which  the communication channel{s} on which the parties interact  a{re} buffer{ed}, allowing participants to  read or write messages (which may carry \emph{data}).
Other variants are synchronous~\cite{BY09,SY19}, more similar to team automata, in which communication is entirely  synchron{ous} without the use of buffered channel{s}.
Some of the early asynchronous variants (like the ones used by Honda \etal.~\cite{HYC08})  and some of the early synchronous variants
(like~\cite{BY09}) support some form of multicasting to multiple (explicit) receivers.

Most approaches to multi-party session types
distinguish (1)~a \emph{global type}, often a starting point, describing the whole communication protocol of the \emph{composed} system; and (2)~the \emph{local types},
often derived from the global type, describing the contributions of the individual participants to the protocol.
Multi-party session types typically require \emph{projectability}, \ie, a global type can be projected onto consistent local types for each participant, a property closely related to realisability (\cf\ Section~\ref{sec:realisability}).
Synchronisation types in team automata, on the other hand, are essentially \emph{capacity types for communicating actions, which specify a range for the numbers of senders and a range for the number of receivers that can participate in such actions.}
A special case is unicast communication.
 In contrast to team automata, neither
global nor local types support internal actions, since their focus is on inter-process communication within a session, thus deliberately ignoring the internal behaviour of processes.
Team automata, however, do not support communication of data,
which is an important ingredient of the typing discipline of session types.

Synchronous multi-party session types support a limited set 
of 
communication patterns when compared to team automata (\cf\ Section~\ref{sec:nutshell}). 
Multi-party session types allow useful behavioural properties to be statically verified, like the absence of communication errors and, in case of simple session types without session interleaving, also deadlock-freedom
(\eg, in the synchronous multi-party session types by Dezani-Ciancaglini \etal.~\cite{DGJPY15,SD19}).
To ensure deadlock-freedom in general multi-party session types, additional type systems (like the interaction typing proposed in~\cite{BCADDY08,CDYP16}) are required to keep track of causality across sessions and forbid causality cycles in the overall behaviour. 

There exists a rich variety of tools on variants of multi-party session types, as reported by Yoshida in a recent survey~\cite{Y24} of 
integrations of multi-party session types into different programming languages. This survey reflects a core focus in session types theory
on
increasing trustworthiness of implementations of distributed systems by statically guaranteeing properties like deadlock-freedom.

\subsection{Other Coordination Formalisms} 

Many other coordination formalisms involve similar notions of composition and synchronisation of agent behaviour. These include the specification languages provided by model checkers such as Uppaal~\cite{BDL04,LLN18} and mCRL2~\cite{GM14,AG23}, as well as more generic formalisms like message sequence charts~\cite{HT03,ITU11}, event structures~\cite{NPW81,Win88}, Petri nets~\cite{Rei13,BD24}, and I/O automata~\cite{LT89,KLSV10,LT26}. 
Without pretending completeness, we discuss some of these in this section.

\myparagraph{Uppaal.} Uppaal accepts systems modelled by (stochastic, timed) automata, with matching input-output actions that must synchronise (either $1$-to-$1$ or $1$-to-many). The latter requires a sender to synchronise with all, possibly zero, available receivers at that time, which differs from the synchronisation policies of team automata. Contract automata's committed states stem from Uppaal's concept of committed states. Uppaal also provides partial support for priorities over actions, but not over interactions as in BIP.

\myparagraph{mCRL2.} mCRL2 accepts systems modelled as a parallel composition of algebraic processes, with special operators to allow the synchronous execution of groups of actions, and the restriction of given actions. This is powerful enough to enumerate all valid synchronisations between concurrent processes (components), yet quite verbose, which we exploited to verify communication properties of team automata~\cite{BCHP23}.

  \myparagraph{Message Sequence Charts.} Message Sequence Charts are visual diagrams commonly used to describe scenarios with interacting agents which have historically been used to describe telecommunication protocols. They are not always precise, and can be enriched with constructs to denote loops, choices, and parallel threads.
  Katoen and Lambert~\cite{KL98} have used \emph{pomsets} to formalise message sequence charts, where each possible trace is described as a (multi-)\allowbreak\hspace{0pt}set of actions with a partial order.
  Guanciale and Tuosto used a pomset semantics for choreographies to reason over realisability~\cite{GT19}, and Edixhoven \etal.\ 
  extended pomsets with a hierarchical structure~\cite{EJPI24}, reasoned over realisability, and compared it to event 
structures~\cite{NPW81}
(often used to give semantics to Petri nets).

\myparagraph{Petri Nets.} Petri nets come in many flavours, and provide a compact representation of a global model of interaction that avoids the explosion of states caused by the interleaving of independent actions.
In~\cite{BK12}, a subclass of team automata---non-state-sharing vector team automata---has been encoded into \emph{Individual Token Net Controllers}---a model of vector-labeled Petri nets---covering team automata in which the synchronisation of a set of agents cannot be influenced by the remaining agents (\cf\ Section~\ref{sec:cat}).
\emph{Zero-safe nets} are another extension of Petri nets with a transactional mechanism that distinguishes observable from hidden states, used to formalise Reo~\cite{C07}. This extension could also be used to model the synchronisation mechanism of team automata, although the analysis of these nets is non-trivial.

\myparagraph{I/O Automata.} I/O automata and related 
formalisms like I/O systems~\cite{Jon87}, interface automa\-ta~\cite{DH01}, reactive transition systems~\cite{CC02}, interacting state machines~\cite{Ohe02}, and com\-po\-nent-interaction (CI) automata~\cite{BCVZ06} all distinguish input, output, and internal actions. However, I/O automata are, by definition, \emph{input-enabled}: in every state of the automaton every input action of the automaton is enabled (\ie, executable). In fact, team automata are a generalisation of I/O automata~\cite{BK05}. All formalisms define composition as the synchronous product of automata except for interface automata~\cite{DH01}, which restrict product states to compatible states,
and  CI automata, which were specifically designed to have this distinguishing feature of team automata. CI automata however restrict communication to binary synchronisation between a pair of input and output actions. Contrary to \lq classical\rq\ team automata, CI automata use system labels to preserve the information about their communication, a feature which inspired the introduction of extended team automata in~\cite{BHK20b} (\cf\ \cref{foot:CI}). 

\section{Communication Properties}
\label{sec:comm}

Compatibility of components is an important issue for systems to guarantee successful (\emph{safe}) communication~\cite{BSBM04,CGP09,DOS12,CK13,BCZ15,BCK16,BCHK17,BHK20a}, \ie, free from message loss (output actions not accepted as input by some other component) and indefinite waiting (for input to be received in the form of an appropriate output action provided by another component). 

In~\cite{BCHK17}, we identified 
representative \emph{synchronisation types} to classify synchronisation policies that are realisable in team automata (\eg, binary, multicast and broadcast communication, synchronous product) in terms of 
ranges for the number of sending and receiving components participating in synchronisations. Moreover, we provided a generic procedure to derive, for each synchronisation type, requirements for \emph{receptiveness} and for \emph{responsiveness} of team automata that prevent outputs not being accepted and inputs not being provided, respectively, \ie, guaranteeing safe communication. 
This allowed us to define a notion of \emph{compatibility} for team automata in terms of their compliance with communication requirements, \ie, receptiveness and responsiveness. A team automaton was said to be \emph{compliant} with a given set of communication requirements if in each of its reachable states, 
the desired communications can immediately occur; 
in~\cite{BCHK17} it was said to be \emph{weakly compliant} if the communication can eventually occur after some internal actions have been performed (akin to weak compatibility~\cite{BMSH10,HB18} or agreement of lazy request actions~\cite{BBDLFGD20}).
This idea was generalised in~\cite{BHK20a,BHK20b}
where \emph{arbitrary} actions,
not only internal ones, are allowed before the required communication happens.
In this paper, we use this more general notion of weak compliance.
Since communication requirements are derived from synchronisation types, we get a family of compatibility notions indexed by synchronisation types.

We revisited the definition 
of \emph{safe communication} 
in terms of receptive- and responsiveness requirements in~\cite{BHK20a,BHK20b}.

\subsection{Communication Requirements}
\label{sec:comreqs}
The assignment of a single synchronisation type to a team automaton was deemed too restrictive, so we decided to fine tune the number of synchronising sending and receiving components \emph{per action}. For this purpose we introduced, in~\cite{BHK20b}, \emph{synchronisation type specifications} which assign a synchronisation type individually to each communicating action. As we have seen in Section~\ref{sec:TA}, such specifications uniquely determine a team automaton. Any synchronisation type specification generates communication requirements to be satisfied by the team. 

\myparagraph{Receptiveness.}
If, in a reachable state $q$ of $\ta{\stype}{\S}$, a group $\{\,\Aut_n \mid n \in \Out\,\}$ of CA 
with $\emptyset \neq \Out \subseteq \N$ is (locally) enabled to perform a communicating output action~$a\in \Sigma^\bullet$, \ie, for all $n\in \Out$ holds 
$a \in \Sigma^!_n$ and $q_n\tr{a}_{\Aut_n}$, and if
moreover both
(1)~the number of CA in $\Out$ fits that 
of the senders allowed by the synchronisation type $\stype(a) = (\out,\inn)$, \ie, $|\Out|\in\out$, and  
(2)~the CA need at least one receiver to join the communication, \ie, $0\notin\inn$, then we get a \emph{receptiveness requirement}, denoted by $\rcp(\Out,a)@q$.
If $\Out = \{\n\}$, we write $\rcp(\n,a)@q$ for $\rcp(\{\n\},a)@q$.

\myparagraph{Responsiveness.}
If, in a reachable state $q$ of $\ta{\stype}{\S}$, a group $\{\,\Aut_n \mid n \in \In\,\}$ of CA 
with $\emptyset \neq \In \subseteq \N$ is (locally) enabled to perform a communicating input action~$a \in \Sigma^\bullet$, \ie, for all $n\in \In$ holds 
$a \in \Sigma^?_n$ and $q_n\tr{a}_{\Aut_n}$, and if
moreover both
(1)~the number of CA in $\In$ fits that 
of the senders allowed by the synchronisation type $\stype(a) = (\out,\inn)$, \ie, $|\In|\in\inn$, and  
(2)~the CA need at least one sender to join the communication, \ie, $0\notin\out$, then we get a \emph{responsiveness requirement}, denoted by $\rsp(\In,a)@q$.
If $\In = \{\n\}$, we write $\rsp(\n,a)@q$ for $\rsp(\{\n\},a)@q$.

\subsection{Compliance with Communication Requirements}
\label{sec:compliance}

The TA $\ta{\stype}{\S}$ is compliant with a receptiveness requirement $\rcp(\Out,a)@q$ 
if the group of components (with names in~$\Out$) can find partners in the team which synchronise with the group by taking (receiving) $a$ as input.
If reception is immediate, then we speak of receptiveness;
if the other components may still perform \emph{arbitrary intermediate actions} (\ie, not limited to internal ones
like in~\cite{BCHK17})
before accepting $a$, then we speak of weak receptiveness~\cite{BHK20b}.
We now formally define (weak) compliance, (weak) receptiveness and (weak) responsiveness.

The TA $\ta{\stype}{\S}$ is \emph{compliant} with $\rcp(\Out,a)@q$ if
$\exists_{\In}\,.\,q\tr{(\Out,a,\In)}_{\ta{\stype}{\S}}$
while it is \emph{weakly compliant} with $\rcp(\Out,a)@q$ if 
$\exists_{\In}\,.\,q \tr{(\Lambda(\stype)_{\setminus \Out})^{*}\,;\, (\Out,a,\In)}_{\ta{\stype}{\S}}$, 
where $\Lambda(\stype)_{\setminus \Out}$ denotes the set of team labels in which no component of $\Out$ participates.
Formally, $\Labelst[\stype]_{\backslash\Out} = \{\,(\Out',a,\In)\in\Labelst[\stype] \mid (\Out'\cup\In)\cap\Out=\emptyset\,\}
\cup\{\,(n,a)\in\Labelst[\stype] \mid n\notin\Out\,\}$.

The TA $\ta{\stype}{\S}$ is \emph{compliant} with
$\rsp(\In,a)@q$ if
$\exists_{\Out}\,.\,q\tr{(\Out,a,\In)}_{\ta{\stype}{\S}}$,
while it is \emph{weakly compliant} with $\rsp(\In,a)@q$ if
$\exists_{\Out}\,.\,q \tr{(\Lambda(\stype)_{\setminus \In})^{*}\,;\, (\Out,a,\In)}_{\ta{\stype}{\S}}$, 
where
$\stype(\Lambda)_{\backslash\In}$ denotes the set of team labels in which no component of $\In$ participates. Formally,
$\stype(\Lambda)_{\backslash\In} =
    \{\,(\Out,a,\In') \in \stype(\Lambda) \mid (\Out\cup\In')\cap\In = \emptyset\,\} \cup
    \{\,(n,a) \in \stype(\Lambda) \mid n \notin \In\,\}$.

\myparagraph{TA: (Weak) Receptiveness.}
The TA $\ta{\stype}{\S}$ is \emph{(weakly) receptive} if for all reachable states $q\in \mathcal R(\ta{\stype}{\S})$,
the TA $\ta{\stype}{\S}$ is (weakly) compliant with \emph{all} receptiveness requirements $\rcp(\Out,a)@q$ established for~$q$.  

\begin{exa}[Receptiveness and Compliance]\label{ex:race-receptiveness}
In the initial state $(0,0,0)$
of the Race team (\cf\ \cref{fig:ta-race}), 
there is a receptiveness requirement of the controller who wants to start the competition, 
expressed by $\rcp(\pts{Ctrl},\msg{start})@(0,0,0)$.
The TA $\ta{\stype_{\ms{Race}}}{\ms{Race}}$ is compliant with this requirement.
When the first runner is in state~$2$, the desire to send $\msg{finish}$
leads to three receptiveness requirements: 
$\rcp(\pts{R1},\msg{finish})@(1,2,1)$, 
$\rcp(\pts{R1},\msg{finish})@(1,2,2)$, and 
$\rcp(\pts{R1},\msg{finish})@(2,2,0)$. 
If the second runner is in state~$2$, we get three more receptiveness requirements: 
$\rcp(\pts{R2},\msg{finish})@(1,1,2)$, 
$\rcp(\pts{R2},\msg{finish})@(1,2,2)$, and
$\rcp(\pts{R2},\msg{finish})@(2,0,2)$. 
The TA $\ta{\stype_{\ms{Race}}}{\ms{Race}}$ is compliant also with these. 
\qedx 
\end{exa}

\noindent
Unlike output actions, the selection of an input action of a component is not controlled by the component but by the environment, \ie, there is an external choice. 
If, for a choice of enabled inputs $\{a_1,\ldots,a_n\}$, 
\emph{only one of them} can be supplied with a corresponding output of the environment this suffices to guarantee progress of each component waiting for input.

\myparagraph{TA: (Weak) Responsiveness.}
The TA $\ta{\stype}{\S}$ is \emph{(weakly) responsive} if for all reachable states $q\in \mathcal R(\ta{\stype}{\S})$
and for all $n \in \N$ the following holds:
if there is a responsiveness requirement $\rsp(\In,a)@q$ established for~$q$ with $n \in \In$,
then the TA $\ta{\stype}{\S}$ is (weakly) compliant with at least one of these requirements.%
\footnote{This version of (weak) responsiveness is slightly stronger than in our previous work,
driven by the comparison with local deadlock-freedom below.}

\begin{exa}[Responsiveness and Compliance]\label{ex:race-responsiveness}
In the initial state $(0,0,0)$ of the Race team, 
there is a responsiveness requirement of the two runners who want the competition to start, expressed by $\rsp(\{\pts{R1},\pts{R2}\},\msg{start})@(0,0,0)$.
The TA $\ta{\stype_{\ms{Race}}}{\ms{Race}}$ is compliant with this requirement.
When the controller is in state~$1$, there are responsiveness requirements
$\rsp(\pts{Ctrl},\msg{finish})@(1,q_1,q_2)$ for any $q_1, q_2\in\{1,2\}$.
In state $(1,1,1)$, at least one $\msg{run}$ must happen before a $\msg{finish}$ is sent; in all other cases, this requirement is immediately fulfilled.
Hence, the TA $\ta{\stype_{\ms{Race}}}{\ms{Race}}$ is weakly compliant with these responsiveness requirements.
There are four more responsiveness requirements when
the controller is in state~$2$, two of which are only weakly fulfilled.
\qedx 
\end{exa}

\noindent
As far as we know, such powerful compliance notions for I/O-based, synchronous component systems were not studied before, which we will explain better in \cref{sub:related_work_comm}. For one, I/O automata are, by definition, input-enabled. This means that, differently from component automata and team automata, in any state any input action will be accepted. Hence any I/O automaton is receptive.

In case of open systems the arbitrary immediate actions 
before a desired communication happens may be output or input actions open to the environment. Then local communication properties could be violated upon composition with other team automata. 
This led us to consider \emph{composition} of open team automata and to investigate conditions ensuring \emph{compositionality} of communication properties~\cite{BHK20b,BHK20c,BCHP21x}.

\begin{exa}[Compliance of Open Systems]\label{ex:open-receptiveness}
Consider the open system variant $\text{Race}'$ of the Race example from Example~\ref{ex:sys-transitions}. Due to the open input action, state $(0,0,1)$ is reachable and has a receptiveness requirement of the controller who wants to start the competition, 
expressed by $\rcp(\pts{Ctrl},\msg{start})@(0,0,1)$. Clearly, the TA $\ta{\stype_{\ms{Race}}}{\ms{Race}'}$ is \emph{not} (weakly) compliant with this receptiveness requirement at $(0,0,1)$, since the second runner would also be needed to \msg{start}. So we must choose a different synchronisation policy. The solution is simple. We just omit transition $(0,0,0) \tr{(\emptyset,\msg{go},\pts{R2}')} (0,0,1)$. Then $(0,0,1)$ is no longer reachable and the new team is compliant with the receptiveness requirements. Removal of the \lq bad\rq\ open input transition matches the approach of interface automata in~\cite{DH01}, where components are considered to be \emph{compatible} if they can work properly together in a \lq helpful\rq\ environment.
\qedx 
\end{exa}

\subsection{Tool Support}
Automatically verifying communication properties is non-trivial, as it may involve traversing networks of interacting automata with large state spaces. 
We pursued a different approach by providing a \emph{logical} characterisation of receptiveness and responsiveness in terms of formulas of a (test-free) propositional dynamic logic~\cite{HKT00} using (complex) interactions as modalities (\cf~\cite{BCHP22,BCHP23}).
Verification of communication properties
then relies on model checking receptive- and responsiveness formulas against a system of component automata taking into account a given synchronisation type specification.

We developed an open-source prototypical tool~\cite{BCHP23-tool}
to support our theory for closed systems. It implements a transformation of
CA, systems, and TA into mCRL2~\cite{AG23} processes and of the characterising dynamic logic formulas into $\mu$-calculus formulas. The latter is straightforward, whereas the former uses mCRL2's \mi{\textsf{allow}} operator to suitably restrict the number of multi-action synchronisations such that the semantics of systems of CA is preserved. Then we can automatically check communication properties with the model-checking facilities offered by mCRL2, which outputs the result of the formula as well as a witness or counterexample.

\subsection{Related Work}\label{sub:related_work_comm}
The genericity of our approach with respect to synchronisation policies allows us to capture compatibility notions for various multi-component coordination strategies.
In the literature, compatibility notions are mostly considered for systems relying on peer-to-peer communication, \ie, all synchronisation types are ([1,1],[1,1]), meaning that a communicating action can be executed only as a synchronisation involving exactly one component for which it is an output action and exactly one for which it is an input action (\cf\ our discussion in~\cite{BCHK17} and in Section~\ref{sec:TA}).
Our notion of receptiveness is inspired by the compatibility notion of \emph{interface automata}~\cite{DH01}
and indeed both notions coincide for closed systems and $1$-to-$1$ communication.
It also coincides with receptiveness in~\cite{CC02}.

Weak receptiveness is inspired by the notion
of weak compatibility in~\cite{BMSH10} and also corresponds
to unspecified reception in the context of $n$-protocol compatibility in~\cite{DOS12}
and lazy request in \emph{contract automata}~\cite{BBDLFGD20}.
We are not aware of 
compatibility notions concerning
responsiveness. In~\cite{CC02}, it is captured by deadlock-freedom and 
in~\cite{DOS12} it is expressed by part of the definition of bidirectional complementary compatibility
which, however, does not support choice of inputs as we do.

The relationship between deadlock-freedom, used 
in different variations in the literature, and our communication properties inspired by de Alfaro and Henzinger~\cite{DH05} is 
subtle.
Note that the distinction between our input and output actions is not relevant for the more traditional notions of deadlock 
and that two 
types of deadlocks are often distinguished: 
global and local 
(\cf, \eg, \cite{ABBJSZ18}).
In short, global deadlock is a property of the whole system/composition (\ie, no further interaction is enabled), whereas local deadlock is a liveness failure of a component or subsystem within that system/composition (\ie, some component(s) can no longer progress while the system/composition as a whole can). Clearly, global deadlock implies local deadlock but not vice versa, which illustrates the importance of studying local deadlocks: proving only global deadlock-freedom may ignore situations in which part of the system is permanently deadlocked, while the rest can continue to execute.

Weak receptiveness and weak responsiveness can be used to guarantee (global) deadlock-freedom, in the sense of BIP~\cite{GS05}, choreography automata~\cite{BLT20}, or stuck-freeness in multi-party session types~\cite{SY19}. This requires some assumptions (\eg, on the synchronisation types) and we refer to~\cite{BHK20a} for a more detailed discussion.
Global deadlock-freedom does, however, not imply (weak) receptiveness.
Consider, for instance, a system state where Alice can send either a message $a$ or a message $b$ to Bob, but Bob can only receive $a$. This state is deadlock-free but it is not receptive, since Alice can never send $b$.
Similarly, global deadlock-freedom does not imply (weak) responsiveness.
Consider, for instance, a system state where Alice can receive $a$ forever, Bob can receive $b$ forever, and Carol can send $a$ forever. If no other participants exist, then this state is deadlock-free but it is not responsive, since Bob can never receive $b$.

These examples 
point out the crucial difference between
global and local deadlocks, covered by the notions of ``individual deadlock'' in BIP~\cite{GS05},
``lock'' in choreography automata~\cite{BLT20}, and ``strong lock''
in multi-party session types~\cite{SY19}. The difference between the latter two is that~\cite{BLT20} assumes fair runs. The notion of individual deadlock in~\cite{GS05} is defined differently,
but seems equivalent to 
a ``lock'' in~\cite{BLT20} 
for
$1$-to-$1$ communication. 
Weak receptiveness together with weak responsiveness
is equivalent to individual deadlock-freedom in BIP
if the interaction model fits to the synchronisation types.
Hence, this is also equivalent to lock-freedom in~\cite{BLT20}. Strong lock-freedom in multi-party session types~\cite{SY19} is indeed
a stronger requirement.

The compatibility notions 
above formalise general requirements for safe communication. Some approaches prefer to formulate individual compatibility requirements tailored to
particular applications by formulas in 
a
logic for dynamic systems using model-checking tools for 
verification (\cf, \eg, \cite{AKM08,KKSB11,KKV10,PM19}).%

\subsection{Roadmap}
\label{sec:roadmap:comm}

In~\cite{BHK20a}, we argued that it may be the case that (local) enabledness of an action indicates only readiness for communication and not so much that communication is required. To make this distinction between possible and required communication explicit, we proposed to introduce designated states in CA, where execution can stop but may also continue, next to states where progress is required.\footnote{In~\cite{BHK20a}, we called these final states, but we now prefer to call them \emph{safe states}. Roughly speaking, these are states where automata can break a communication requirement without raising a communication error. A typical example would be when a server component is ready to receive an input which never arrives.} 
The introduction of such \emph{safe} states to component automata has significant consequences for the derivation of communication requirements and for our compliance notions, which would have to be adjusted accordingly.

\vspace*{-.25\baselineskip}
\section{Realisation}
\label{sec:realisability}

In this section, we consider a top-down method by which first a global model $\M$ for the intended interaction behaviour of a
 system is provided
(on the basis of a system signature and synchronisation type specification $\stype$).
Then we look for a system $\S$ of component automata, to be used as a \emph{realisation of}
$\M$, whose generated team 
automaton $\ta{\stype}{\S}$
is bisimilar to
the global model $\M$. 
Our approach is restricted to the case
in which components do not contain internal actions. Hence bisimilarity
is strong.

For our realisation method 
we instantiate the generic approach investigated in~\cite{BHP23}
by applying the localisation style with so-called ``poor'' local actions of the form $!a$ for outputs and
$?a$ for inputs; localisations with ``rich'' local actions,
which mention in the local context of a component $n$ the receiver $m$ of a message (\eg, by $nm!a$) or the sender $m$ of a message (\eg, by $mn?a$), are treated in~\cite{BHP23} as well, but they are not relevant for team automata.

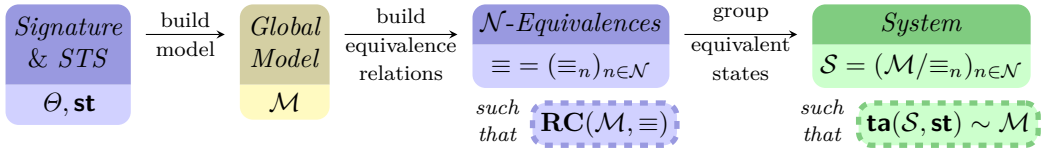
\begin{figure}[htb!]
  \centering\noindent
\begin{tikzpicture}[x=1.5cm]

\newcommand{\mydef}[1]{\textcolor{gray}{#1}}

\tikzstyle{spl}=[rectangle split,rectangle split parts=2]

\tikzstyle{objBase} = [inner sep=3pt, outer sep=2mm,font=\small,
    align=center,rounded corners=5pt] 

\tikzstyle{obj} = [spl,objBase,
    rectangle split part fill={dbluecol, bluecol}]
\tikzstyle{objm} = [obj,
    rectangle split part fill={dglobalcol, globalcol}]
\tikzstyle{objs} = [obj,
    rectangle split part fill={dlocalcol, localcol}]
\tikzstyle{arr} = [inner sep=0pt, -stealth, rounded corners=5pt]
\tikzstyle{lbl} = [inner sep=2pt, align=center,font=\scriptsize]
\tikzstyle{bold} = [font=\scriptsize\bf]
\tikzstyle{slbl}=[below,inner sep=8pt,font=\sf\smaller]
\tikzstyle{sqgl}=[arr,decoration={snake,amplitude=0.7pt,segment length=1.2mm,
                  post=lineto,post length=3pt},decorate] 

\tikzstyle{objPO} = [objBase, line width=2pt, dashed, minimum width=0mm]
\tikzstyle{objmPO} = [objPO, draw=dlocalcol, fill=localcol]
\tikzstyle{objsPO} = [objPO, draw=dbluecol, fill={bluecol}]

\node[objm,minimum width=2mm]
  (LTS) {\wrap{\mi{Global}\\\mi{Model}}
        \nodepart{second}\mi{\M}};

\node[obj,left=0.7 of LTS.north west, below left]
  (sign) {\wrap{\mi{Signature}\\\mi{\&~STS}}
         \nodepart{second}\mi{\Theta, \stype}};

\node[obj,right=1 of LTS.north east,below right] 
  (equiv) {
          \mi{\mathcal{N}\text{-}Equivalences}
          \nodepart{second}\mi{{\equiv} = (\equiv_n)_{n\in \mathcal{N}}}};

\node[objsPO,anchor=north east,yshift=2mm,xshift=1mm] at (equiv.south east)
  (RC) {\mi{\mathbf{RC(\M,\equiv)}}};
\node[xshift=1mm,yshift=0.5mm,align=center,anchor=west]
  at (RC.west-|equiv.west)
  (suchthat) {\scriptsize\mi{such}\\[-3pt]\scriptsize\mi{that}};

\node[objs,right=1 of equiv.north east,below right]
  (Sys) {\mi{System}
         \nodepart{second}$\S=(\M/{\equiv_n})_{n\in \mathcal{N}}$};
\node[objmPO,anchor=north east,yshift=2mm,xshift=1mm] at (Sys.south east)
  (bis) {\mi{\ta{\stype}{\S} \sim \M}};
\node[xshift=-1mm,yshift=0.5mm,align=center,anchor=west]
  at (bis.west-|Sys.west)
  (suchthat2) {\scriptsize\mi{such}\\[-3pt]\scriptsize\mi{that}};

\coordinate[yshift=-6mm](arrLvl) at (LTS.north) ;

\draw[arr] (LTS.east|-arrLvl)
    to node [lbl,yshift=-1.3mm] {build\\[0.3mm]equivalence\\[0.3mm]relations} (arrLvl-|equiv.west);
\draw[arr] (arrLvl-|sign.east)
    to node [lbl] {build\\[0.3mm]model} (LTS.west|-arrLvl);
\draw[arr] (equiv.east|-arrLvl)
    to node [lbl,yshift=-1.3mm] {group\\[1mm]equivalent\\[0.3mm]states} (arrLvl-|Sys.west)
     ;
\end{tikzpicture}
  \caption{Realisation method}
  \label{fig:realisation-approach}
\end{figure}

Recall from Section~\ref{sec:nutshell} the notions of component automaton (CA), system $\S$, synchronisation type specification \stype, and generated team automaton (TA) $\ta{\stype}{\S}$.
Our realisation method is summarised in~\cref{fig:realisation-approach} and will be explained in more detail in the next sections.

\vspace*{-.5\baselineskip}
\subsection{Global Models of Interaction}
\label{sec:reqspec}

Our method starts with a \emph{system signature}
$\syssig = (\N,(\Sigma_\n)_{\n\in \N})$, where 
$\N$ is a finite, nonempty set of component names and 
$(\Sigma_\n)_{\n\in \N}$ is an $\N$-indexed family of action sets $\Sigma_n = \Sigma_n^{?} \cup \Sigma_n^{!}$ 
split into disjoint sets
$\Sigma_n^{?}$ of \emph{input actions} and $\Sigma_n^{!}$ of \emph{output actions}.
As in Section~\ref{sec:system}, let $\Sigma^{\bullet} = \bigcup_{\n\in\N} \Sigma^?_\n \cap \bigcup_{\n\in\N} \Sigma^!_\n$ 
be the set of \emph{communicating actions}.
We do not consider internal actions here and we assume that all system actions
$a \in \bigcup_{\n\in\N} \Sigma_\n$ are communicating.

Together with the system signature
a synchronisation type specification \stype
must be provided
assigning to each $a \in \Sigma^{\bullet}$ a pair
of intervals $\stype(a) = (\out,\inn)$,
as explained in Section~\ref{sec:TA}.
The system signature $\syssig$ together with the STS $\stype$ determine the following set $\iact$ of \emph{(multi-)\allowbreak\hspace{0pt}interactions} respecting the constraints of the given synchronisation types:
\begin{align*}
    \iact &=
      \{\,(\Out,a,\In) \mid \emptyset\neq (\Out\cup \In) \subseteq \N,\,\allowbreak
\forall_{\n\in \Out}\cdot a\in \Sigma^!_\n,\,
\forall_{\n\in \In}\cdot a\in \Sigma^?_\n,\,\\
    &\qquad
      \stype(a) = (\out,\inn) \Rightarrow |\Out|\in \out \land |\In|\in \inn\,\}
  \end{align*}
We model the global interaction behaviour of the intended system by an LTS whose labels are (multi-)interactions in $\iact$.
Hence, a \emph{global interaction model}
over $\syssig$ and $\stype$ is an LTS of the form
$\M\,{=}\,(Q,q_0,\iact,E)$.

\begin{exa}\label{ex:race-global-model}
To develop the Race system we would start
with system signature $\syssig_{\mathsf{Race}}$ with component names \pts{Ctrl}, \pts{R1}, \pts{R2}
and action sets
$\Sigma_{\pts{Ctrl}}^{?} = \{\msg{finish}\}
= \Sigma_{\pts{R1}}^{!} = \Sigma_{\pts{R2}}^{!}$ and
$\Sigma_{\pts{Ctrl}}^{!} = \{\msg{start}\}
= \Sigma_{\pts{R1}}^{?} = \Sigma_{\pts{R2}}^{?}$.
We do not consider the internal $\msg{run}$ action.  
As in Example~\ref{ex:race-ta}, we use the STS
$\stype_{\mathsf{Race}}$ with
$\stype_{\mathsf{Race}}(\msg{start}) = \mathrm{([1,1],[2,2])}$ and
$\stype_{\mathsf{Race}}(\msg{finish}) = \mathrm{([1,1],[1,1])}$.

\cref{fig:race-global} shows on the left the induced interaction set
$\Lambda(\syssig_{\mathsf{Race}},\stype_{\mathsf{Race}})$
(abbreviated by $\Lambda_{\mathsf{Race}}$)
and on the right a global interaction model $\M_\ms{Race}$.
It imposes the system to start in a (global) state,
where the controller $\msg{start}$s both runners at once.
Each runner separately sends a $\msg{finish}$ signal
to the controller 
(in arbitrary order).
After that a new run can $\msg{start}$.
\qedx

\tikzstyle{teams}=[
  ->,scale=.5,>=stealth',shorten >=1pt,auto, node distance=1.5cm,
  semithick, every node/.style={scale=0.7},initial text={}]
\tikzstyle{teamState}=[fill=white,draw=black,text=black]
\tikzstyle{lclr}=[rectangle, rounded corners=8pt, inner sep=3pt,fill=localcol]
\tikzstyle{lclp}=[rectangle, rounded corners=8pt, inner sep=3pt,fill=globalcol]
\tikzstyle{select}=[fill=metLightBrown]

\begin{figure}[tb]
\medskip\centering
\wrap{
  $\Lambda_{\mathsf{Race}} = \left\{\begin{array}{@{}l@{}l}
    \syn[start]{Ctrl}{\{\pts{R1},\pts{R2}\}}\,,\\[.25em]
    \syn[finish]{R1}{Ctrl}\,,\\[.25em]
    \syn[finish]{R2}{Ctrl}
  \end{array}\right\}$%
  }
\qquad\qquad
\wrap{
  \begin{tikzpicture}[teams]
    \tikzstyle{every state}=[teamState,fill=globalcol]
    \node[initial,state]        (0){$0$};
    \phantom{\node[state]                 (d)[right=1 of 0]{$d$};}
    \node[state]                 (2)[above of=d,above=-.5cm]{$2$}; 
    \node[state]                 (3)[below of=d,below=-.75cm]{$3$}; 
    \node[state]               (1)[right=1 of d]{$1$};

    \path (0) edge[above]  node[inner sep=1pt]{\syn[start]{Ctrl}{\{R1,R2\}}\/} (1)
          (1) edge[bend right=15,above right] node{\syn[finish]{R1}{Ctrl}}  (2)
              edge[bend left=15,below right] node{\syn[finish]{R2}{Ctrl}}  (3)
          (2) edge[bend right=15,above left]  node{\syn[finish]{R2}{Ctrl}}    (0)
          (3) edge[bend left=15,below left] node{\syn[finish]{R1}{Ctrl}}  (0)
          ;
  \end{tikzpicture}
}
\caption{Interaction set $\Lambda_{\mathsf{Race}}$
and global interaction model $\M_\ms{Race}$}
\label{fig:race-global}
\end{figure}
\end{exa}

\begin{rem}
Our development methodology can be extended by providing
first an abstract specification of desired and forbidden interaction properties from a global perspective.
In~\cite{BHP23}, we proposed a propositional dynamic logic for this purpose where interactions are used as
atomic actions in modalities such that we can express 
safety and (a kind of) liveness properties.
For instance, we could express the following requirements
for the Race system by dynamic logic formulas, using the usual box modalities $[\cdot]$ and $\tpl{\cdot}$, sequential composition~(;), choice~(\texttt{+}), and iteration~($\cdot^{*}$):%
\begin{enumerate}
\item
\emph{``For any started runner, it should be possible to finish her/his run.''}

\smallskip
$\Big[some^{*} ; \syn[start]{Ctrl}{\{\pts{R1},\pts{R2}\}}\Big]
\left(\mwrap{
  \diam{some^{*} ; \syn[finish]{R1}{Ctrl}} \true ~ \land \,\\
  \diam{some^{*} ; \syn[finish]{R2}{Ctrl}} \true~~~\,\ \ }\right)$

\medskip
    \item
    \emph{``No runner should finish before she/he was started
            by the controller.''}

\smallskip
$
\left[\Big(-(\syn[start]{Ctrl}{\{\pts{R1},\pts{R2}\}})\Big)^{*} ~;~
\left(\mwrap{
  \syn[finish]{R1}{Ctrl} ~\texttt{+}\\
  \syn[finish]{R2}{Ctrl}~~~\,}\right)\right]\, \false
$
\end{enumerate}

\noindent
To check that a global interaction model satisfies a specification,
we propose to use the mCRL2 toolset~\cite{GM14,AG23}.
For this purpose, as explained in~\cite{BCHP23},
we can use the translation from LTS models into process expressions and
the translation from our interaction-based
dynamic logic into the syntax used by mCRL2. 
\qedx
\end{rem}

\subsection{Realisation of Global Models of Interaction}
\label{sec:realisations}

Our central task concerns the realisation (decomposition) of a global interaction model $\M$ in terms of a (possibly distributed) system
$\S$ of component automata which are coordinated according to the given synchronisation type specification.

In order to formulate our realisation notion, we briefly recall the standard notion of bisimulation. 
Let $\LTS_n\,{=}\,(Q_n,q_{n,0},\Sigma,E_n)$ be two LTS
(for $n = 1,2$) over the same action set~$\Sigma$.
A \emph{bisimulation relation} between $\LTS_1$ and $\LTS_2$ is a relation $B \subseteq Q_1 \times Q_2$ such that for all
$(q_1,q_2) \in B$ and for all $a \in \Sigma$ the following holds:
\begin{enumerate}
    \item 
    if $q_1\tr{a}_{\LTS_1} q_1'$, then there exist
    $q_2' \in Q_2$ and $q_2\tr{a}_{\LTS_2} q_2'$ such that
    $(q_1',q_2') \in B$;
    \item 
    if $q_2\tr{a}_{\LTS_2} q_2'$, then there exist
    $q_1' \in Q_1$ and $q_1\tr{a}_{\LTS_1} q_1'$ such that
    $(q_1',q_2') \in B$.
\end{enumerate}

\noindent
$\LTS_1$ and $\LTS_2$ are \emph{bisimilar},
denoted by $\LTS_1\sim\LTS_2$,
if there exists a bisimulation relation $B$ between
$\LTS_1$ and $\LTS_2$ such that $(q_{1,0},q_{2,0}) \in B$.

Now, we assume given
a system signature $\syssig = (\N,(\Sigma_\n)_{\n\in \N})$, an STS \stype and a global interaction model
$\M$ with labels $\iact$.
A system
$\S = (\N,(\Aut_\n)_{\n\in \N})$ of component automata $\Aut_\n$ with actions $\Sigma_\n$ is a \emph{realisation} of $\M$ with respect to \stype if
 the team automaton $\ta{\stype}{\S}$
generated over \S by $\stype$ (as defined in Section~\ref{sec:TA}) is bisimilar to $\M$, \ie, $\ta{\stype}{\S} \sim \M$.
Note that the team labels $\Labelst$ of $\ta{\stype}{\S}$ are exactly the interactions in $\iact$, \ie, the actions of the LTS $\M$.
The global model $\M$ is called \emph{realisable}
if such a system $\S$ exists.

\begin{rem}
Technically, the definition of realisability in~\cite{BHP23} uses a synchronous $\Gamma$-composition $\otimes_\Gamma (\Aut_n)_{n \in \N}$~\cite[Def.~7]{BHP23} of the component automata. Transferred to the context of synchronisation types, $\Gamma$ would be $\iact$. Moreover, $\otimes_\Gamma (\Aut_n)_{n \in \N}$
contains only reachable states, which need not to be the case for the TA $\ta{\stype}{\S}$. However, we can restrict $\ta{\stype}{\S}$ to its reachable sub-LTS
which coincides with $\otimes_\Gamma (\Aut_n)_{n \in \N}$. Note also that any LTS is bisimilar to its reachable sub-LTS, such that this restriction does not harm.
\qedx
\end{rem}
 
\noindent
Since our realisability notion relies on bisimulation,
we are able to deal with non-deterministic behaviour.
Note that, according to the invariance of propositional dynamic logic under bisimulation (\cf~\cite{BES94}), we obtain that global models and their realisations satisfy the same propositional dynamic logic formulas when (multi-)interactions are used as atomic actions 
as proposed in~\cite{BCHP23,BHP23}.

Next, we consider the following two important questions:
\begin{enumerate}
    \item How can we check whether a given global model $\M$ is realisable?
    \item If it is, how can we build/synthesise a concrete realisation?
\end{enumerate}

\noindent
To tackle the first question, we propose to find a family $\equiv\ = (\equiv_n)_{n \in \N}$
of equivalence relations on the \emph{global} state space $Q$ of $\M$
such that, for each component name $n \in \N$ and states $q, q'\in Q$, $q \equiv_n q'$ expresses that  
$q$ and $q'$ are indistinguishable from the viewpoint of $n$.
It is required that at least 
any two global states $q, q' \in Q$
which are related by a global transition 
$q  \tr{(\Out,a,\In)}_\M  q'$
should be indistinguishable for any
$n \in \N$ which does not participate in the interaction,
\ie, $q \equiv_n q'$ for all $n \notin \Out\cup\In$.
A family $\equiv\ = (\equiv_n)_{n \in \N}$
of equivalence relations $\equiv_n\, \subseteq Q \times Q$
with this property is called an \emph{$\N$-equivalence}.

We can now formulate our sufficient realisability condition for the global model
$\M\,{=}\,(Q,q_0,\iact,E)$.
Our goal is to find an $\N$-equivalence $\equiv\ = (\equiv_n)_{n \in \N}$ over $\M$ such that
the following \emph{realisability condition}
$\mi{\mathbf{RC(\M,\equiv)}}$ holds.

    \myparagraph{$\mi{\mathbf{RC(\M,\equiv)}}$:} For each interaction $(\Out,a,\In) \in \iact$,
whenever there is (1)~a map $\ell: \Out\cup\In \to Q$ assigning a global state $q_n = \ell(n)$ to each $n \in \Out\cup\In$
together with (2)~a global \emph{``glue''} state $g$, \ie,
$q_n \equiv_n g$ for each $n \in \Out\cup\In$,
then we expect:
for all $n \in \Out\cup\In$ and
all global transitions $q_n \tr{(\Out_n,a,\In_n)}_\M  q'_n$
such that
$n \in (\Out_n\cap\Out) \cup (\In_n\cap\In)$,
there is a global transition $g \tr{(\Out,a,\In)}_\M  g'$ such that $q'_n \equiv_n g'$ for each $n \in \Out\cup\In$.

The intuition for this requirement is that if component
$n$ can participate in executing an 
output (input, resp.) action $\msg{a}$ in state $q_n$, then $n$ should also be able to participate
in executing
output (input, resp.) action $\msg{a}$ in state~$g$, since $n$ cannot distinguish $q_n$ and $g$. As this should hold for all $n \in \Out\cup\In$,
the interaction $(\Out,a,\In)$ should be enabled
in $g$ and preserve indistinguishability of states
for all $n \in \Out\cup\In$.

As a consequence of our results in~\cite{BHP23}, in particular Thm.~3, we obtain that if there is an
$\N$-equivalence $\equiv\ = (\equiv_n)_{n \in \N}$ such that the realisability condition $\mi{\mathbf{RC(\M,\equiv)}}$ holds, then the global model
$\M\,{=}\,(Q,q_0,\iact,E)$ can indeed be realised by the system $\S_{\equiv} = (\M/{\equiv_n})_{n \in \N}$ of component automata 
constructed as local quotients of $\M$, \ie, $\ta{\stype}{\S_{\equiv}} \sim \M$.
More precisely, the \emph{local n-quotient} of $\M$
is the component automaton 
$\M/{\equiv_n} = (Q/{\equiv_n},[q_0]_{\equiv_n},\Sigma_n,(E/{\equiv_n})$, where
\begin{itemize}
    \item $Q/{\equiv}_n = \{\,[q]_{\equiv_n} \mid q \in Q\,\}$ and
\vspace*{-.75\baselineskip}
  \item $E/{\equiv_n}$ is the least set of transitions generated by 
  rule
    $\dfrac{
 q\tr{(\Out,a,\In)}_\M q' \qquad n \in \Out\cup\In
}
{
[q]_{\equiv_n}\tr{\msg {a}}_{(\M/{\equiv_n})} [q']_{\equiv_n}
}$ 
\end{itemize}
\enlargethispage*{\baselineskip}
\begin{exa}\label{ex:race-poor-rc}
Take the global LTS $\M_\ms{Race}$
in~\cref{fig:race-global}.
We use the family of equivalences
${\equiv} = ({\equiv}_n)_{n\in\{\pts{Ctrl,R1,R2}\}}$
that obeys $\mi{\mathbf{RC(\M_\ms{Race},\equiv)}}$ (see below)
and partitions the state space $Q$ in
$
Q/{\equiv}_\pts{Ctrl} = \{\{0\},\{1\},\{2,3\}\}$, 
$Q/{\equiv}_\pts{R1} = \{\{0,2\},\{1,3\}\}$, and 
$Q/{\equiv}_\pts{R2} = \{\{0,3\},\{1,2\}\}$.
Using these equivalences, the local quotients 
are:

{\centering 
  $(\M/{\equiv_{\pts{Ctrl}}}) =~~$
  \raisebox{0mm}{\wrap
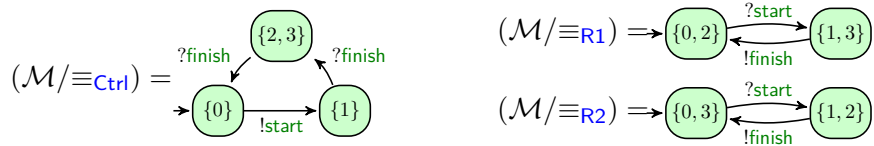
{
  \begin{tikzpicture}[teams]
    \tikzstyle{every state}=[teamState]
    \node[initial,state,lclr]        (0){$\{0\}$};
    \node[state,lclr]               (1)[right=1.0 of 0]{$\{1\}$};
    \coordinate(d) at ($(0)!0.5!(1)$);
    \node[state,lclr]                 (2)[above of=d,above=-4mm]{$\{2,3\}$}; 

    \path (0) edge[below]            node{{!\msg{start\/}}} (1)
          (1) edge[bend right=15,above right] node{{?\msg{finish\/}}}    (2)
          (2) edge[bend right=15,above left]  node{?\msg{finish\/}}      (0);
    \pgfresetboundingbox
    \node [fit=(0)(1)(2),draw=none,inner sep=15pt] {};
  \end{tikzpicture}
  }}
  \qquad\qquad
  \wrap{
  $(\M/{\equiv_{\pts{R1}}}) =~~$
  \raisebox{-1mm}{\wrap{
  \begin{tikzpicture}[teams]
    \tikzstyle{every state}=[teamState,inner sep=2pt]
    \node[initial,state,lclr] (0){$\{0,2\}$};
    \node[state,lclr]                (2)[right=1.1 of 0]{$\{1,3\}$};

    \path (0) edge[bend left=10pt,above]   node{{?\msg{start\/}}}(2)
          (2) edge[bend left=10pt,below]   node{{!\msg{finish\/}}} (0);
    \node [fit=(0)(2),draw=none,inner sep=15pt] {};
    \pgfresetboundingbox
    \node [fit=(0)(2),draw=none,inner sep=15pt] {};
  \end{tikzpicture}
  }}
  \\[-3mm]
  $(\M/{\equiv_{\pts{R2}}}) =~~$
  \raisebox{-1mm}{\wrap{
  \begin{tikzpicture}[teams]
    \tikzstyle{every state}=[teamState,inner sep=2pt]
    \node[initial,state,lclr] (0){$\{0,3\}$};
    \node[state,lclr]                (2)[right=1.1 of 0]{$\{1,2\}$};

    \path (0) edge[bend left=10pt,above]   node{{?\msg{start\/}}} (2)
          (2) edge[bend left=10pt,below]   node{{!\msg{finish\/}}} (0);
    \node [fit=(0)(2),draw=none,inner sep=15pt] {};
    \pgfresetboundingbox
    \node [fit=(0)(2),draw=none,inner sep=15pt] {};
  \end{tikzpicture}
  }}
  }
  \\
  }

\noindent
 So we obtained a system that is
  a realisation of $\M_\ms{Race}$.
This means the team automaton 
$\ta{\stype_{\mathsf{Race}}}{\S_\equiv}$ generated by
$\S_\equiv$ and $\stype_{\mathsf{Race}}$ is bisimilar to
$\M_\ms{Race}$. Indeed, both are the same LTS
up to renaming of states: state
$(\{0\},\{0,2\},\{0,3\})$ in $\ta{\stype_{\mathsf{Race}}}{\S_\equiv}$
instead of state $0$ in $\M_\ms{Race}$,
$(\{1\},\{1,3\},\{1,2\})$ instead of $1$,
$(\{2,3\},\{0,2\},\{1,2\})$ instead of $2$, and
$(\{2,3\},\{1,3\},\{0,3\})$ instead of $3$.
  
  We show how to check
  $\mi{\mathbf{RC(\M_\ms{Race},\equiv)}}$
using interaction 
$\syn[finish]{R1}{Ctrl}$ as example.
We have $1\tr{\syn[finish]{R1}{Ctrl}}_{\M_\ms{Race}}2$ and 
$1\tr{\syn[finish]{R2}{Ctrl}}_{\M_\ms{Race}}3$.
Taking $1$ as (trivial) glue state, we thus have, as required, 
$1\tr{\syn[finish]{R1}{Ctrl}}_{\M_\ms{Race}}2$,
\emph{but also} $2 \equiv_\pts{Ctrl} 3$
must hold, which is the case.
Note that we would not have succeeded here if we had
taken the identity for $\equiv_\pts{Ctrl}$.
\qedx
\end{exa}

\vspace*{-.75\baselineskip}
\subsection{Tool Support}
We implemented a prototypical tool, called Ceta, to perform re\-ali\-sa\-bility checks and system synthesis (\cf~\cite{BHP23,BHP23z}).
It is open source, available at\linebreak \url{https://github.com/arcalab/choreo/tree/ceta}, and executable by navigating to \url{https://lmf.di.uminho.pt/ceta}.
It provides 
a web browser where one can
input a global protocol described in a choreographic language, resembling regular expressions of interactions.
It offers automatic visualisation of the protocol as a state machine
representing a global model
and it includes examples with descriptions. 

Ceta implements a \emph{constructive} approach to the \emph{declarative} description of the realisability conditions. It builds a family of equivalence relations, starting with one that groups states connected by transitions in which the associate participant is not involved.
This family of minimal equivalence relations is checked
for satisfaction of the realisability condition
with respect to the global model. If it fails, a new attempt
is started after extending the equivalence relations appropriately.
If no failure occurs, then 
the realisability condition is satisfied and
the resulting equivalence classes are used to join equivalent states in the global model, yielding local quotients
which can again be visualised.
Thus a realisation of the global model is constructed.
There are several widgets that provide further insights, such as the intermediate equivalence classes, the synchronous composition of local components, and bisimulations between global models and team automata. Readable error messages are given when a realisability condition does not hold.

\subsection{Related Work}
\label{sec:realisability-related-work}

Our approach is driven by specified sets of multi-interactions supporting any kind of synchronous communication between multiple senders and multiple receivers.
To the best of our knowledge, realisations of global models with arbitrary multi-interactions have not yet been studied in the literature. There are, however, specialised approaches that deal with realisations of global models or decomposition of transition systems. In this section, we first provide a revised and extended comparison of our approach with that of Castellani \etal.~\cite{CMT99}, followed by a brief comparison with other approaches.

\myparagraph{Relationship to~\cite{CMT99}.}
Our realisability condition $\mi{\mathbf{RC(\M,\equiv)}}$, 
based on the notion of $\N$-equivalence $\equiv$, is strongly related to
a condition for implementability in~\cite[Theorem~3.1]{CMT99}.
The main differences are:
\begin{enumerate}
\item In~\cite{CMT99}, there is no distinction between input and output actions.
\item In~\cite{CMT99}, interactions are always full synchronisations on an action $a$,
while we deal with individual multi-interactions $(\Out,a,\In)$ specified by an STS. Of course, we can also
use an STS $\stype_{\mathit{full}}$ for full synchronisation. Then we define, for each action $a$,
$\stype_{\mathit{full}}(a) = ([\#\mathit{out}(a),\#\mathit{out}(a)],[\#\mathit{in}(a),\#\mathit{in}(a)])$,
where $\#\mathit{out}(a) = |\{n\in \N \mid a \in \Sigma_n^{!}\}|$ is the number of components having $a$ as an output action and $\#\mathit{in}(a) = |\{n\in \N \mid a \in \Sigma_n^{?}\}|$ is the number of components having $a$ as an input action.
\item In~\cite{CMT99}, they provide a characterisation of implementability up to isomorphism, while we provide a criterion for realisability modulo bisimulation, thus supporting non-determinism. 
To achieve this, we 
basically omitted condition~(ii) of~\cite[Theorem~3.1]{CMT99} which requires
that whenever two global states $q$ and $q'$
are $n$-equivalent for all $n \in \N$, then $q = q'$.
In Example~\ref{ex:not-iso} below
we provide a simple example for a global interaction model which satisfies our realisability condition but for which there is no realisation up to isomorphism.
In fact, Example~\ref{ex:not-iso} shows that bisimulation
allows the merging of of states which would be distinct under isomporphism.
Note that~\cite[Theorem~6.2]{CMT99} provides a proposal to deal with a characterisation of implementability modulo bisimulation  under the assumption of deterministic
product transition systems. The authors also report on
a result to characterise implementability for non-deterministic product systems which uses infinite execution trees and is thus, unfortunately, not effective.

\begin{exa}
\label{ex:not-iso}
There is a basic example for a situation, depicted in \cref{tab:bisim-vs-iso}, in which a global LTS \M satisfies our realisability condition but there is no
realisation that is isomorphic to \M (only bisimilar).
In this example, let $\Sigma = (I,M)$, with $I = \{\pts p,\pts q\}$ and $M = \{\msg a\}$, and let
$\Gamma =
\{\syn[a]{p}{\emptyset},\syn[a]{q}{\emptyset}\}$
be the interaction set.
The global LTS $\M$ is depicted on the left side of \cref{tab:bisim-vs-iso} with five states and four transitions.
As equivalences for $\pts{p}$ and $\pts{q}$  we take the reflexive, symmetric, and transitive closures of
$1 \equiv_\pts{p} 3$, $0 \equiv_\pts{p} 2$, and $1 \equiv_\pts{p} 4$,
and of $0 \equiv_\pts{q} 1$, $2 \equiv_\pts{q} 4$, and $2 \equiv_\pts{q} 3$. Note that $1 \equiv_\pts{p} 4$ and $2 \equiv_\pts{q} 3$ are enforced to satisfy $\mathit{RC}(\M,\equiv)^\ms r$. By symmetry and transitivity, we also get
$3 \equiv_\pts{p} 4$ and $3 \equiv_\pts{q} 4$.
Since the realisability condition holds, $\M$ is realisable
with rich local actions up to bisimilarity. The local components $\M_\pts{p}$ and $\M_\pts{q}$ are depicted in the middle of \cref{tab:bisim-vs-iso}.
However,
$\M$ is not realisable up to isomorphism. Otherwise, states $3$ and $4$ should be the same (\cf~\cite[Theorem~3.1]{CMT99}).
\qedx
\end{exa}

\begin{table}[t!]
  \centering
  \caption{A \goOnline{global model}
  {\%28p-\%3E:a;\%20q-\%3E:a\%29\%20+\%20\%28q-\%3E:a;\%20p-\%3E:a;1\%29}
  (left)~\cite[Example~8]{BHP23z} that is realisable for bisimilarity but not for isomorphism; the realisation (middle); and the re-composed TA with
  $\S = (\{p,q\}, \{\M_p,\M_q\})$ and $\stype(\msg{a}) = \mathrm{([1,1],[0,0])}$}
  \label{tab:bisim-vs-iso}
  \begin{tabular}{c@{\hspace*{8mm}}c@{\hspace*{6mm}}c@{\hspace*{5mm}}c}
    \toprule
      Global $\M$ 
    & Local $\M_\pts{p}$
    & Local $\M_\pts{q}$
    & $\ta{\stype}{\S}$
    \\
    \midrule
    \wrap{
      \begin{tikzpicture}[teams]
        \tikzstyle{every state}=[teamState,lclp]
        \node[initial,state]        (0){$0$};
        \coordinate[right=1 of 0] (d);
        \node[state]                (1)[above of=d,above=-.95cm]{$1$}; 
        \node[state]                (2)[below of=d,below=-.95cm]{$2$}; 
        \node[state]                (3)[right=0.6 of 1]{$3$};
        \node[state]                (4)at(2-|3){$4$};

        \path (0) edge[bend left=20,above left,pos=0.8]
                    node{\syn[a]{\{p\}}{\emptyset}} (1)
                  edge[bend right=20,below left,pos=0.8]
                    node{\syn[a]{\{q\}}{\emptyset}} (2)
              (1) edge[above]
                    node[yshift= 15pt]{\syn[a]{\{q\}}{\emptyset}} (3)
              (2) edge[below]
                    node[yshift=-15pt]{\syn[a]{\{p\}}{\emptyset}} (4);
      \end{tikzpicture}
    }
    &
    \wrap{\begin{tikzpicture}[teams]
      \tikzstyle{every state}=[teamState]
      \node[initial above,state,lclr]        (0){$\{0,2\}$};
      \node[state,below=0.6 of 0,lclr]   (1){$\{1,3,4\}$}; 

      \path (0) edge[right]   node[yshift=3pt]{\snd[a]{}{}} (1);
      \pgfresetboundingbox
      \draw [draw=none] (0.north west) rectangle (1.south east);
    \end{tikzpicture}}
    &
    \wrap{\begin{tikzpicture}[teams]
    \tikzstyle{every state}=[teamState]
      \node[initial above,state,lclr]        (0){$\{0,1\}$};
      \node[state,below=0.6 of 0,lclr]   (1){$\{2,3,4\}$}; 

      \path (0) edge[right]   node[yshift=3pt]{\snd[a]{}{}} (1);
      \pgfresetboundingbox
      \draw [draw=none] (0.north west) rectangle (1.south east);
    \end{tikzpicture}}
    &
    \wrap{\begin{tikzpicture}[teams]
        \tikzstyle{every state}=[teamState,lclp,align=center,inner sep=5pt]
        \node[initial,state]        (0){$\{0,2\}$\\[1mm]$\{0,1\}$};
        \coordinate[right=1 of 0] (d);
        \node[state] (1)[above of=d,above=-.95cm]{$\{1,3,4\}$\\[1mm]$\{0,1\}$}; 
        \node[state] (2)[below of=d,below=-.95cm]{$\{0,2\}$\\[1mm]$\{2,3,4\}$}; 
        \node[state] (3)[right=1 of d]{$\{1,3,4\}$\\[1mm]$\{2,3,4\}$};

        \path (0) edge[bend left=20,above left,pos=0.8]
                    node{\syn[a]{\{p\}}{\emptyset}} (1)
                  edge[bend right=20,below left,pos=0.8]
                    node{\syn[a]{\{q\}}{\emptyset}} (2)
              (1) edge[bend left=20,above right,pos=0.2]
                    node{\syn[a]{\{q\}}{\emptyset}} (3)
              (2) edge[bend right=20,below right,pos=0.2]
                    node{\syn[a]{\{p\}}{\emptyset}} (3);
    \end{tikzpicture}}
    \\
    \bottomrule
  \end{tabular}
\end{table}
\end{enumerate}
\myparagraph{Relationship to Other Approaches.}
\begin{enumerate}

    \item Our correctness notion for realisation of global models by systems of communicating local components is based on bisimulation, beyond language-based approaches like~\cite{BLT22,DBLP:journals/corr/abs-1203-0780} with realisability expressed by language equivalence.
    
    \item For realisable global models, we construct realisations in terms of systems of local quotients.
    This technique differs from projections used, \eg, for multi-party session types, 
    where projections are partial operations depending on syntactic conditions (\cf, \eg, \cite{BY09}). Our approach assumes no restrictions on the form of global models. 
    However, the syntactic restrictions used for global types guarantee some kind of communication properties of a resulting system which we consider separately (\cf\ Section~\ref{sec:comm}).

    \item There are other papers in the literature in the context of different formalisms dealing with decomposition of port automata~\cite{KC09}, Petri nets, or algebraic processes into (indecomposable) components~\cite{MM93,Lut16} used for the efficient verification and parallelisation of concurrent systems \cite{CGM98,GM92} or to obtain better (optimised) implementations~\cite{TCV21,TCV22}. Moreover, Saeedloei and Klu{\'{z}}niak recently tackled the problem of realisability of sets of (sequential) timed scenarios as a timed scenario expression that realises the set~\cite{SK26}.\footnote{A timed scenario is a sequence of events with a set of constraints on the times at which the events occur.}
\end{enumerate}

\subsection{Roadmap}
\label{sec:roadmap:realisabilty}
\begin{enumerate}
\setlength\itemsep{1.2mm}
\item 
Our current realisability approach does not deal with internal actions, which are however also an ingredient
of the team automata framework and represented by system labels of the form $(n,a)$ (\cf\ Section~\ref{sec:system}).
They naturally appear when we build a team of CA which have internal actions. We believe, however, that internal actions should not be part of a global interaction model whose purpose is to present the \emph{observable} interaction behaviour of an intended system.
To bridge the gap, the idea is to relax the realisation notion by requiring only a weak bisimulation relation between a global model and the team automaton of a system realisation with internal actions.

\item
Another aspect concerns the fact that,
in general, it may happen that a global interaction model does not satisfy the realisability condition but is nevertheless realisable. Therefore, we want to look for a weaker version of our realisability condition making it necessary and sufficient for realisations based on bisimulation.
The following example shows that our current realisability condition is only sufficient.

\begin{exa}
Consider the system signature $\syssig$ with component names
$\N = \{\ms{p}, \ms{q}, \ms{r}\}$ and with the action sets
$\Sigma_{n}^{!} = \{\msg{a}\}, \Sigma_{n}^{?} = \emptyset$ for $n \in \N$ and
we use the STS with $\stype(\msg{a}) = \mathrm{([2,2],[0,0])}$.
Then the interaction set is
$\iact =
\big\{\syn[a]{\{p,q\}}{\emptyset},\syn[a]{\{q,r\}}{\emptyset},\syn[a]{\{p,r\}}{\emptyset}\big\}.$
The global model $\M$ 
in~\cref{tab:example-realisation} (left)
is realisable by the system $\S = \{\M_\pts{p},\M_\pts{q},\M_\pts{r}\}$, whose CA are shown in~\cref{tab:example-realisation} (middle). 

\begin{table}[!hbt]
  \centering
  \caption{
      Global $\M$ violates 
      $\mi{\mathbf{RC(\M,\equiv)}}$, but 
    $\S\!=\!\{\M_\pts{p},\M_\pts{q},\M_\pts{r}\}$ realises~$\M$}
  \label{tab:example-realisation}
  \begin{tabular}{c@{\hspace*{15mm}}c@{\hspace*{3mm}}c@{\hspace*{3mm}}c@{\hspace*{12mm}}c}
    \toprule
      \qquad Global $\M$ 
    & Local $\M_\pts{p}$
    & Local $\M_\pts{q}$
    & Local $\M_\pts{r}$
    & $\ta{\stype}{\S}$
    \\
    \midrule
    \wrap{
    \begin{tikzpicture}[teams]
    \tikzstyle{every state}=[teamState,fill=globalcol]
    \node[initial,state]  (0){$0$};
    \node[state] (2)[below=1.2 of 0]  {$1$};
    \path (0)
      edge[bend left=65 ,below,inner sep=2pt] node[fill=white,xshift=2pt,yshift=-5pt]{\syn[a]{\{p,r\}}{\emptyset}} (2)
      edge[bend left=0, anchor=center, inner sep=2pt] node[fill=white]{\syn[a]{\,\{q,r\}}{\emptyset}} (2)
      edge[bend right=65,above,inner sep=2pt] node[fill=white,xshift=2pt,yshift=5pt]{\syn[a]{\{p,q\}}{\emptyset}} (2);
      \pgfresetboundingbox
      \draw [draw=none] (0.north west) rectangle (2.south east);
    \end{tikzpicture}
    }
    &
    \wrap{\begin{tikzpicture}[teams]
    \tikzstyle{every state}=[teamState]
    \node[initial,state,inner sep=1pt,lclr]  (0){$0$};
    \coordinate[right=1 of 0] (2);
    \node[state,lclr] (3)[below=1.2 of 0]{$1$}; 
    \path (0)
        edge[bend right=0,anchor=center,inner sep=2pt]
          node[fill=white,yshift=0pt]{\snd[a]{}{}} (3);
      \pgfresetboundingbox
      \draw [draw=none] (0.north west) rectangle (3.south east);
    \end{tikzpicture}}
    &
    \wrap{\begin{tikzpicture}[teams]
    \tikzstyle{every state}=[teamState]
    \node[initial,state,inner sep=1pt,lclr]  (0){$0$};
    \coordinate[right=1 of 0] (2);
    \node[state,lclr] (3)[below=1.2 of 0]{$1$}; 
    \path (0)
      edge[bend right=0,anchor=center,inner sep=2pt]
        node[fill=white,yshift=0pt]{\snd[a]{}{}} (3);
      \pgfresetboundingbox
      \draw [draw=none] (0.north west) rectangle (3.south east);
    \end{tikzpicture}}
    &
    \wrap{\begin{tikzpicture}[teams]
    \tikzstyle{every state}=[teamState]
    \node[initial,state,inner sep=1pt,lclr]  (0){$0$};
    \coordinate[right=1 of 0] (2);
    \node[state,lclr] (3)[below=1.2 of 0]{$1$}; 
    \path (0)
      edge[bend right=0,anchor=center,inner sep=2pt]
        node[fill=white,yshift=0pt]{\snd[a]{}{}} (3);      
      \pgfresetboundingbox
      \draw [draw=none] (0.north west) rectangle (3.south east);
    \end{tikzpicture}}
    &
    \wrap{\begin{tikzpicture}[teams]
    \tikzstyle{every state}=[teamState]
    \node[initial,state,lclp]  (0){$0,0,0$};
    \node[state,lclp] (2)[below=1.2 of 0]  {$0,1,1$};
    \node[state,lclp] (1)[left=0.2 of 2]{$1,1,0$}; 
    \node[state,lclp] (3)[right=0.2 of 2]{$1,0,1$}; 
    \path (0)
      edge[bend right=25 ,above,inner sep=2pt] node[fill=white,yshift=3pt]{\syn[a]{\{p,q\}}{\emptyset}} (1)
      edge[bend left=0  ,anchor=center,inner sep=2pt] node[fill=white]{\syn[a]{\{q,r\}}{\emptyset}} (2)
      edge[bend left=25,below,inner sep=2pt] node[fill=white,yshift=-8pt]{\syn[a]{\{p,r\}}{\emptyset}} (3);
      \pgfresetboundingbox
      \draw [draw=none] (0.north west) rectangle (0.south east);
      \draw [draw=none] (1.north west) rectangle (3.south east);
    \end{tikzpicture}}
    \\
    \bottomrule
    \\
  \end{tabular}
\end{table}

\noindent
To see this, we compute the TA $\ta{\stype}{\S}$ shown in~\cref{tab:example-realisation} (right).
Obviously, $\M$ and $\ta{\stype}{\S}$ are bisimilar and hence
$\M$ is realisable.
However, there is no $\N$-equivalence $\equiv$ such that
the realisability condition holds. 
We now prove this by contradiction.

Assume that $\equiv\ = \{\equiv_\pts{p},\equiv_\pts{q},\equiv_\pts{r}\}$ is an  $\N$-equivalence such that $\mi{\mathbf{RC(\M,\equiv)}}$ holds.
Now consider the interaction $\syn[a]{\{p,q\}}{\emptyset}$,
 the global state $0$ of $\M$ and the transition
$0 \tr{\syn[a]{\{p,q\}}{\emptyset}}_{\M} 1$.
Obviously, $0 \equiv_\pts{p} 1$ and $1 \equiv_\pts{q} 0$ must hold
because of the transition $0 \tr{\syn[a]{\{q,r\}}{\emptyset}}_{\M} 1$ in which $\pts{p}$
does not participate and the transition $0 \tr{\syn[a]{\{p,r\}}{\emptyset}}_{\M} 1$ in which $\pts{q}$ does not participate.
So we can take $1$ as a glue state
for both $\pts{p}$ and $\pts{q}$ because state $1$ is
indistinguishable from state $0$ for both components $\pts{p}$ and $\pts{q}$. 
Then we consider the interaction $\syn[a]{\{p,q\}}{\emptyset}$ one time for
component $\pts{p}$ and starting in state 
$q_1 = 0$ and one time for
component $\pts{q}$ and starting in state
$q_2 = 0$.
Since we assumed $\mi{\mathbf{RC(\M,\equiv)}}$,
there must be a transition  
$1 \tr{\syn[a]{\{p,q\}}{\emptyset}}_{\M}$ leaving the glue state,
which is not the case. Contradiction!
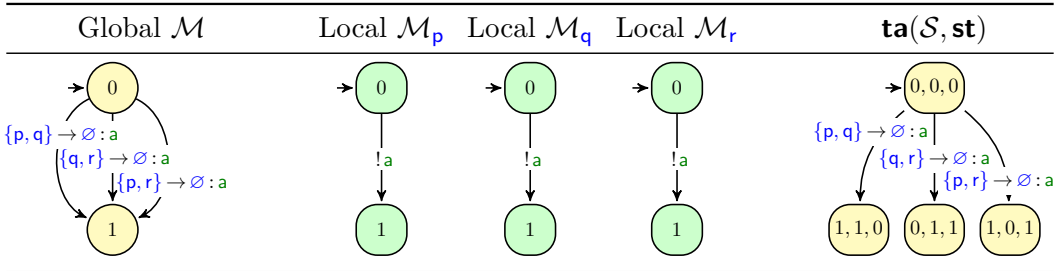
\qedx
\end{exa}

\item
We are interested in a compositional approach to construct larger realisations from smaller ones. Then compositionality of realisability is important. 

\item
We want to study under which conditions on global models
and synchronisation types our communication properties 
can be guaranteed for realisations.
\end{enumerate}

\section{Composition of Systems}
\label{sec:composition}

In this section, we consider how to compose systems of component automata and also how to compose synchronisation type specifications.
We provide results concerning the preservation of communication properties.
For this purpose we instantiate ideas presented in~\cite{BHK20b,BHP23}
\textbf{}by looking more closely to synchronisation type specifications and their composition instead of considering arbitrary synchronisation policies for team automata. 
Moreover, besides open system actions, we also allow internal actions here.

\subsection{System Composition}
We assume given a finite set $\K$ of indices and a family $(\S_\ki)_{\ki \in \K}$ of systems $\S_\ki = (\N_\ki,(\Aut_{\n})_{\n \in \N_\ki})$
with component automata
$\Aut_{\n} = (Q_{\n},q_{0,\n},\Sigma_{\n},E_{\n})$ and component actions
$\Sigma_{\n} = \Sigma^{?}_{\n} \cup \Sigma^{!}_{\n} \cup \Sigma^{\tau}_{\n}$ for each $\n \in \N_\ki$.
In particular, for each $\ki \in \K$,
$Q_\ki=\prod_{\n\in\N_\ki} Q_{\n}$
is the set of \emph{system states} of $\S_\ki$,
$\Sigma_\ki = \bigcup_{\n\in\N_\ki} \Sigma_{\n}$
its set of \emph{system actions},
$\Sigma_\ki^{\bullet} = \bigcup_{\n\in\N_\ki} \Sigma^?_{\n} \cap \bigcup_{\n\in\N_\ki} \Sigma^!_{\n}$
its set of \emph{communicating actions},
and
$\Sigma_\ki^{\circ} = \Sigma_\ki \setminus (\Sigma_\ki^{\bullet} \cup\bigcup_{\n\in\N_\ki}\Sigma^{\tau}_\n)$
its set of \emph{open actions}.

To compose the system family $(\S_\ki)_{\ki \in \K}$
we assume that component names are unique and that
system actions which are 
communicating
in one system cannot 
occur
in another system. Formally,
the family $(\S_\ki)_{\ki \in \K}$ is \emph{composable}
if for all distinct $\ki,\ell \in \K$,
$\N_\ki \cap \N_\ell = \emptyset$
and $\Sigma_\ki^{\bullet} \cap \Sigma_\ell = \emptyset$.
In this case, we write $\N_\K$
for the disjoint union $\bigcup_{\ki\in\K}\N_\ki.$ %
The \emph{composition} of a composable family
$(\S_\ki)_{\ki \in \K}$ of systems is defined as the system
$$\bigotimes_{\ki\in\K} \S_\ki =
(\N_\K,
(\Aut_{\n})_{\n\in\N_\K}).$$
The state space of $\bigotimes_{\ki\in\K} \S_\ki$ is
$Q_\K=\prod_{\n\in\N_\K} Q_{\n}$,
the set of system actions 
of $\bigotimes_{\ki\in\K} \S_\ki$ is 
$\Sigma_\K = \bigcup_{\n\in\N_\K}\Sigma_{\n}$, 
the set of communicating actions of $\bigotimes_{\ki\in\K} \S_\ki$ is
$$\Sigma_\K^{\bullet} = \bigcup_{\n\in\N_\K}
\Sigma^?_{\n} \cap \bigcup_{\n\in\N_\K}\Sigma^!_{\n}$$
and the set of open actions of $\bigotimes_{\ki\in\K} \S_\ki$ is $\Sigma_\K^{\circ} = \Sigma_\K \setminus (\Sigma_\K^{\bullet} \cup\bigcup_{\n\in\N_\K}\Sigma^{\tau}_\n).$
\cref{tab:K-notation} lists the notions and notations of single systems and their lifted versions used for composed systems.

Obviously $\Sigma^{\bullet}_\ki \subseteq \Sigma_\K^{\bullet}$ for all $\ki \in \K$, \ie,
the communicating actions of the composed system contain
the communicating actions of each of its subsystems.
It is, however, important to notice that the composed system
may have communicating actions which do not belong to the communicating actions of some subsystem, \ie, the inclusion $\Sigma^{\bullet}_\ki \subseteq \Sigma_\K^{\bullet}$ can be strict for at least one $\ki \in \K$. 
This happens if a system action occurs as an open input action in (at least) one component of (at least) one subsystem and as open output action in (at least) one component of (at least) one \emph{other} subsystem. Due to the composability conditions these actions cannot be communicating actions of a subsystem. They are called
\emph{interface actions}
and formally defined by
$$\Sigma_{\mathit{inf}} = \Sigma_\K^{\bullet} \setminus \bigcup_{\ki \in \K}\Sigma^{\bullet}_\ki.$$
Of course, there might also remain system actions 
in $\Sigma_\K$ which are not in $\Sigma_\K^{\bullet}$
and also not internal to a component. Those actions still remain open for further system compositions. They are given by the set of open actions $\Sigma_\K^{\circ}$ defined above.

\begin{table}[tb]
  \caption{Lifting from systems to composed systems}
  \label{tab:K-notation}
  \begin{tabular}{rr@{~}lr@{~}l}
    \toprule
    & \multicolumn{2}{c}{\textbf{Systems}} & \multicolumn{2}{c}{\textbf{Composition of systems}}
    \\\midrule
    Automata names  & $\N_\ki$ && $\N_\K$ &= $\bigcup_{n\in\K}\N_k$
    \\
    System states & 
      $Q_k$&= $\prod_{n\in\N_k}Q_n$ & $Q_\K$&= $\prod_{n\in\N_\K}Q_n$
    \\
    System actions &
      $\Sigma_k$&= $\bigcup_{n\in\N_k}\Sigma_n$ & $\Sigma_\K$&= $\bigcup_{n\in\N_\K}\Sigma_n$
    \\
    Communicating actions &
      $\Sigma_\ki^{\bullet}$ &=
      $\bigcup_{\n\in\N_\ki} \Sigma^?_{\n} \cap \bigcup_{\n\in\N_\ki} \Sigma^!_{\n}$ &
      $\Sigma_\K^{\bullet}$ &=
      $\bigcup_{\n\in\N_\K} \Sigma^?_{\n} \cap \bigcup_{\n\in\N_\K} \Sigma^!_{\n}$
    \\
    Open actions &
      $\Sigma_\ki^{\circ}$ &=
      $\Sigma_\ki \setminus (\Sigma_\ki^{\bullet} \cup\bigcup_{\n\in\N_\ki}\Sigma^{\tau}_\n)$ &
      $\Sigma_\K^{\circ}$ &=
      $\Sigma_\K \setminus (\Sigma_\K^{\bullet} \cup\bigcup_{\n\in\N_\K}\Sigma^{\tau}_\n)$
    \\\bottomrule
  \end{tabular}
  \medskip
\end{table}

\subsection{Composition of STS}
The considerations so far did not consider synchronisation type specifications which are our crucial instrument to express which synchronisations are
permitted in a system.
Assume that
each subsystem $\S_\ki$ of a composable family $(\S_\ki)_{\ki \in \K}$ comes with a synchronisation type specification
$\stype_\ki:\Sigma^{\bullet}_\ki \to \kw{Intv}\times\kw{Intv}$ (\cf\ Section~\ref{sec:nutshell}).
Due to the composability conditions, each single
$\stype_\ki$ can be uniquely extended to define a synchronisation
type for each communicating action in
$\bigcup_{\ki \in \K}\Sigma^{\bullet}_\ki.$
Still missing are synchronisation types for the interface actions which must be provided by the system architect who composes the family of systems. 
Formally, this is done by providing a function
$\stype_{\mathit{inf}}:\Sigma_{\mathit{inf}} \to \kw{Intv}\times\kw{Intv}$.
This gives rise to an STS
$\bigotimes_{\ki\in\K}^{\stype_{\mathit{inf}}}\stype_\ki: 
\Sigma_\K^{\bullet} \to \kw{Intv}\times\kw{Intv}$
for the composed system
defined by:
$$
\bigotimes_{\ki\in\K}^{\stype_{\mathit{inf}}}\stype_\ki(a) = \left\{
  \begin{array}{ll}
  \stype_\ki(a) \,\,&\mathrm{if}\,\,a \in \Sigma^{\bullet}_\ki \,\,\mathrm{for~some}\,\,\ki\in\K
  \\
  \stype_{\mathit{inf}}(a) \,\,&\mathrm{if}\,\,a \in \Sigma_{\mathit{inf}}
  \end{array}\right.
$$

\noindent
Thus we can construct the team automaton
$\ta{\bigotimes_{\ki\in\K}^{\stype_{\mathit{inf}}}\stype_\ki}{\bigotimes_{\ki\in\K} \S_\ki}.$
By construction, any transition of
$\ta{\bigotimes_{\ki\in\K}^{\stype_{\mathit{inf}}}\stype_\ki}{\bigotimes_{\ki\in\K} \S_\ki}$
is an extension of some transition in a team automaton
$\ta{\stype_\ki}{\S_\ki}$ generated over some single system $\S_\ki$ by the STS $\stype_\ki$.
In particular, all synchronisation transitions using
interface actions $a \in \Sigma_{\mathit{inf}}$ may extend the number of participants in accordance with the synchronisation type $\stype_{\mathit{inf}}(a)$.\footnote{Transitions with open actions in subsystems
that became interface actions but do not fit 
$\stype_{\mathit{inf}}$ are omitted.}
All other transitions\footnote{These other transitions are transitions labelled with internal actions, communicating actions in a subsystem, and open actions in a subsystem which remain open in the composed system.} 
are just lifted from subsystems to the larger state space of $\bigotimes_{\ki\in\K}\S_\ki.$

As a consequence, projections of globally reachable states of $\ta{\bigotimes_{\ki\in\K}^{\stype_{\mathit{inf}}}\stype_\ki}{\bigotimes_{\ki\in\K} \S_\ki}$ are reachable states
in any team automaton $\ta{\stype_\ki}{\S_\ki}$.
Therefore, we get that receptiveness and responsiveness
requirements satisfied by the team automata of subsystems are also satisfied by the global team automaton. Moreover, if we have checked that the communication requirements for all interface actions $a \in \Sigma_{\mathit{inf}}$ with respect to $\stype_{\mathit{inf}}(a)$ are satisfied by the global team automaton, it is ensured that the global team automaton is receptive and responsive. 
More details, including also the propagation of weak receptiveness and weak responsiveness, can be found in\subcite{BHK20b}{BHP23z}.
In particular, for the preservation of weak communication properties one must be careful to avoid the introduction of cyclic waits.

\begin{exa}\label{ex:composition}
We consider two systems, $\S_\pts{RaceV}$ and $\S_\pts{Arb}$, depicted in \cref{fig:racev}.
The system
$\S_\pts{RaceV}$ is a variant
of the Race system in Example~\ref{ex:sys-transitions}.
There are three component names \pts{R1}, \pts{R2},
and \pts{CtrlV}. \pts{R1} and \pts{R2} are the names
of the two runner components known from~\cref{fig:race-local}.
The behaviour of the controller \pts{CtrlV} is shown in the upper part of system $\S_{\pts{RaceV}}$ in~\cref{fig:racev}. It is a variant of the controller in~\cref{fig:race-local}.
The idea is that before the controller can \msg{start} a race it must \msg{ask} for a \msg{grant}.
The system $\S_\pts{RaceV}$ has two communicating actions,
\msg{start} and \msg{finish}, and three open actions
\msg{ask}, \msg{grant}, and \msg{reject}.
The system $\S_\pts{Arb}$, instead, has only one component, named \pts{Arbiter}, whose behaviour is shown by the CA on the right side of \cref{fig:racev}. It is the task of the arbiter to
\msg{grant} or \msg{reject} a race after being asked.
The system $\S_\pts{Arb}$ has only open actions:
\msg{ask}, \msg{grant}, and \msg{reject}.
We now compose the two systems resulting
in the system $\S_\pts{RaceV} \otimes \S_\pts{Arb}.$
The three actions \msg{ask}, \msg{grant}, and \msg{reject} are the interface actions of the composed system (which is closed). All five actions \msg{start}, \msg{finish}, \msg{ask}, \msg{grant}, and \msg{reject} are communicating actions of
$\S_\pts{RaceV} \otimes \S_\pts{Arb}$.
Synchronisation types for the communicating actions \msg{start} and \msg{finish} of the subsystem
$\S_\pts{RaceV}$
are those provided in Example~\ref{ex:race-ta}.

For the composition with $\S_\pts{Arb}$, we still have
to provide synchronisation types for the interface actions \msg{ask}, \msg{grant}, and \msg{reject}.
We use binary communication for
all of them, expressed by the synchronisation type ([1,1],[1,1]).
Thus we have obtained an STS, which we abbreviate by $\stype_\otimes$, for the composed system. 
The resulting team automaton for $\S_\pts{RaceV} \otimes \S_\pts{Arb}$, denoted by $\ta{\stype_\otimes}{\S_\pts{RaceV} \otimes \S_\pts{Arb}}$, is a variant of the team automaton
in~\cref{fig:ta-race}. The difference appears
in the first phase when the team automaton of the composed system performs
the interaction between the controller and the arbiter. This leads to two more states and three more transitions compared with~\cref{fig:ta-race}. 

To analyse communication properties we first point out
that, similarly to Examples~\ref{ex:race-receptiveness}
and~\ref{ex:race-responsiveness},
the team automaton of the subsystem $\S_\pts{RaceV}$ is receptive and weakly responsive. By our preservation results from~\cite{BHK20b} (Theorems~2 and~3),
it remains to check that both properties are also
satisfied for the three interface actions.
Those actions can only occur in the initial phase of $\ta{\stype_\otimes}{\S_\pts{RaceV} \otimes \S_\pts{Arb}}$.
As an example, we consider the receptiveness requirement
$\rcp(\pts{Ctrl},\msg{ask})@(0,0,0,0)$, where all components are in their initial states and the controller wants to \msg{ask} the arbiter for approval. Obviously, the TA $\ta{\stype_\otimes}{\S_\pts{RaceV} \otimes \S_\pts{Arb}}$ is compliant with this requirement since, in its initial state, the arbiter is ready to receive the message.
As an example of a responsiveness requirement, we
consider $\rsp(\pts{Arbiter},\msg{ask})@(0,0,0,0)$.
It is clear that the TA $\ta{\stype_\otimes}{\S_\pts{RaceV} \otimes \S_\pts{Arb}}$ is also compliant with this requirement since, in its initial state, the controller is ready to \msg{ask} the arbiter.
Now assume that the team automaton of the composed system is in state $(0,2,0,0)$,
where the controller is in state~2, \ie, it got the grant, and the arbiter is back
to its initial state~0 (indicated by the first component of $(0,2,0,0)$). In such a state,
the next possible action would be a \msg{start} communication between the controller and the two runners. On the other hand, there is a responsiveness requirement of the arbiter expressed by $\rsp(\pts{Arbiter},\msg{ask})@(0,2,0,0)$ saying
that the arbiter expects an \msg{ask} message. But this is not immediately possible, since the controller must first
start and control a race. After that the controller will ask the arbiter again for getting a grant for the next race. Hence, the TA $\ta{\stype_\otimes}{\S_\pts{RaceV} \otimes \S_\pts{Arb}}$
is weakly compliant with this responsiveness requirement.
\qedx 
\end{exa}

\begin{figure}[tb]
\centering
\begin{tikzpicture}[
  ->,scale=.5,>=stealth',shorten >=1pt,auto,node distance=65pt,
  semithick, initial text={},
  align=center,
  every node/.style={scale=.9,font=\footnotesize}
  ]
  \tikzstyle{every state}=[teamState,fill=localcol]

  \node[initial above,state]    (c0){$0$};
  \node[state]                  (c1)[right of=c0,xshift=-2mm]{$1$};
  \node[state]                  (c2)[right of=c1,xshift=2mm]{$2$};
  \node[state]                  (c3)[below of=c2,yshift=8mm]{$3$};
  \node[state]                  (c4)at(c3-|c1) {$4$};  

  \path (c0) edge
                                    node{$!\msg{ask\/}$}     (c1)
        (c1) edge[bend right=60]    node[above](rej){$?\msg{reject\/}$} (c0)
        (c1) edge                   node[xshift=-4pt]{~~$?\msg{grant\/}$}   (c2)
        (c2) edge                   node{$!\msg{start\/}$}   (c3)
        (c3) edge                   node[xshift=4pt]{$?\msg{finish\/}~~~~$}  (c4)
        (c4) edge[bend left=50]     node[xshift=-2pt,yshift=2pt,above right,inner sep=1pt]
                                        {$?\msg{finish\/}$}  (c0);

  \node[initial,state,xshift=-7mm,yshift=-25mm] at (c0|-c4) (0){$0$};
  \phantom{\node[state]       (d)[right=0.1 of 0]{$d$};}
  \node[state]                (1)[above=0mm of d]{$1$}; 
  \node[state]                (2)[right=0.1 of d]{$2$};

  \path (0) edge[bend left,above left]   node{{?\msg{start\/}}} (1)
        (1) edge[bend left,above right]  node{\msg{run\/}} (2)
        (2) edge[bend left,below]   node{!\emph{\msg{finish\/}}} (0);
  \node[initial,state,xshift=18mm] at (2) (0b){$0$};
  \phantom{\node[state]       (d)[right=0.1 of 0b]{$d$};}
  \node[state]                (1b)[above=0mm of d]{$1$}; 
  \node[state]                (2b)[right=0.1 of d]{$2$};

  \path (0b) edge[bend left,above left]   node{{?\msg{start\/}}} (1b)
        (1b) edge[bend left,above right]  node{\msg{run\/}} (2b)
        (2b) edge[bend left,below]   node(fin){!\emph{\msg{finish\/}}} (0b);

  \node[initial above,state,xshift=45mm,yshift=-10mm]   (a0) at(c2) {$0$};
  \node[state]                  (a1)[below of=a0]{$1$};

  \path (a0) edge[bend left=90]     node[yshift=-15pt,xshift=-2pt](ask){$?\msg{ask\/}$}   (a1)
        (a1) edge                   node[right,xshift=-2pt]{$!\msg{reject\/}$}      (a0)
        (a1) edge[bend left=90]   node[yshift=15pt,xshift=2pt](grant){$!\msg{grant\/}$}     (a0);

  \node at ($(c1)!0.5!(c3)$){\pts{CtrlV}};
  \node[below,yshift=-1mm] at (1.south){\pts{R1}};
  \node[below,yshift=-1mm] at (1b.south){\pts{R2}};
  \node[below,yshift=-1mm] at (a1.south){\pts{Arbiter}};

  \coordinate(top1)at(rej.north);
  \coordinate[xshift=-2mm](left1)at(0.west);
  \coordinate(right1)at(2b.east);
  \coordinate(bot1)at(fin.south);
  \coordinate(top2)at(a1|-top1);
  \coordinate(left2)at(grant.west);
  \coordinate(right2)at(ask.east);
  \coordinate(bot2)at(a1|-bot1);
  \begin{scope}[on background layer]
    \node[fit=(top1)(left1)(right1)(bot1),draw=black!50,very thick,
          rounded corners=6pt,fill=black!8,inner sep=15pt](s1) {};
    \node[fit=(top2)(left2)(right2)(bot2),draw=black!50,very thick,
          rounded corners=6pt,fill=black!8,inner sep=15pt](s2) {};
    \node[anchor=north west,draw=black!50,very thick,inner sep=5pt,
          rounded rectangle,
          rounded rectangle west arc=none,fill=white]at(s1.north west)
      {$\S_\pts{RaceV}$};
    \node[anchor=north west,draw=black!50,very thick,inner sep=5pt,
          rounded rectangle,
          rounded rectangle west arc=none,fill=white]at(s2.north west)
      {$\S_\pts{Arb}$};
  \end{scope}

\end{tikzpicture}
\caption{\label{fig:racev}Two systems (adapted from~\cite{BHK20a,BHK20b})
}

\end{figure}

\subsection{Tool Support}
Our current tool (\url{http://arcatools.org/feta}) supports the composition of CA within a single system, with explicit external (input and output), and internal actions. However, there is no support yet to compose such systems.

\subsection{Related Work}

Composition of systems has also been studied for several related coordination models, such as Reo, BIP, and contract automata.
Reo~\cite{Arb04} does not fully support systems, which are simply viewed as connectors.
However,
Reo supports the composition of connectors at its core, and further includes a hiding operator that restricts which ports remain available for external interactions.
Hence the composition of systems is at a different level with respect to the related formalisms.

In the framework of BIP, composition of components has been studied extensively in~\cite{GS05}.
Composition is based on interaction models which must be respected by the interaction behaviour of sets of components.
The spirit of our synchronisation types is similar,
but BIP relies on connectors between ports with disjoint names instead of synchronisation types for shared actions. Differently from our communication properties, BIP studies preservation of different kinds of deadlock freedom.

In~\cite{BDF16}, Basile \etal.\ introduce two different operators for composing contract automata, representing two different orchestration policies. 
The first composition operator 
($\otimes$, called \emph{product} in~\cite{BDF16}) composes 
a 
singleton
contract automaton~$\mc{C}$ 
and a composite contract automaton~$\mc{S}$ 
such that $\mc{C}$ cannot interact with the components of $\mc{S}$ but only with $\mc{S}$ as a whole. On the contrary, the second composition operator ($\boxtimes$, called \emph{a-product} in~\cite{BDF16}) first extracts from its operands the 
singleton
automata 
they are composed of, by projection, and then reassembles these according to the first composition operator. This means that matching request and offer actions in a composite contract automaton~$\mc{S}$ become available for interaction with complementary actions in the components of the contract automaton it is composed with. Hence, for two 
singleton
contract automata $\mc{C}_1$ and $\mc{C}_2$, 
$\mc{C}_1\otimes\mc{C}_2 = \mc{C}_1\boxtimes\mc{C}_2$. Moreover, while $\boxtimes$ is associative, $\otimes$ obviously is not.
Composition of systems is not in the focus of \emph{choreography automata}~\cite{BLT20,BLT23}, whose theory is more related to questions of projection of global behaviour to local behaviour.

Multi-party session types, finally, offer parallel composition of systems expressed by $\S\,\|\,\S$. 
But, to the best of our knowledge, they are not concerned with preservation of communication 
safety by composition since, similarly to choreography automata, their focus is more on the relation between global types and their local projections.

\subsection{Roadmap}
\label{sec:roadmap:composition}

Our composition operator is based on flattening system compositions such that the result is again a set of CA.
From the software engineer's perspective we are also interested in hierarchical designs where sub-teams are first encapsulated into CA by hiding communicating actions. We believe this could simplify the analysis of behavioural properties of larger systems, \eg, by using techniques of minimisation with respect to weak bisimulation.
Furthermore, we intend to extend our tool to support the composition of systems.

Finally, it would be interesting to compare our approach to composition with the notion of so-called interleaving (weaving) composition, which uses a form of synchronisation based on assertions (\ie, orthogonal to actions) and expresses directional (\ie, rely-guarantee style) dependencies~\cite{BOV23}. According to the authors, these interleaving compositions are not characterisable by team automata synchronisations and---vice versa---cannot capture the full range of synchronisations offered by team automata.

\section{Variability}
\label{sec:var}

We recently proposed featured team automata~\cite{BCHP21} to support \emph{variability} in the development and analysis of teams. 
A featured team automaton can be projected into different variations, each from a given configuration determined by a selection of desired features that filter desired transitions; thus concisely
capturing a family of concrete product (automata) models, 
as is common in software product line engineering~\cite{ABKS13}. 

\begin{figure}[ht]
  \centering
  \begin{tikzpicture}[x=2.50cm, y=-1.4cm]
  \tikzstyle{obj} = [inner sep=0pt, outer sep=1mm,font=\small]
  \tikzstyle{arr} = [inner sep=0pt, -stealth]
  \tikzstyle{lbl} = [inner sep=2pt, font=\scriptsize]
  \tikzstyle{slbl}=[below,inner sep=8pt,font=\sf\smaller]
  \tikzstyle{sqgl}=[arr,decoration={snake,amplitude=0.7pt,segment length=1.2mm,
                    post=lineto,post length=3pt},decorate] 

  \node [obj] (FCA) at (0.05,0) {$A\,{:}\,\mFCA$};
  \node [obj] (FS) at (1.1,0) {$\S {=} (A_n)_{n\in\N}{:}\,\textit{\FSys}$ };
  \draw [arr] (FCA) to node [lbl, above=1.0mm] {} (FS);
  
  \node [obj] (CA) at (0.05,1) {$A\proj_p\,{:}\,\mi{\CA}$};
  \node [obj] (S) at (CA-|FS) {$\S\proj_p {=}(A_n\proj_p)_{n\in\N}{:}\, \textit{Sys}$ };
  \draw [arr] (CA) to node [lbl, above=1.0mm] {} (S);
  
  \node [obj] (FETA) at (2.76,0) {$\!\mFETA\times f\ensuremath{\mkern-3mu}Reqs\xspace$};
  \node [obj,xshift=-0pt] (ETA) at (CA-|FETA) {$\,\mi{TA}\,\times\REQs$};

  \node [obj,right=0.2 of FETA,align=center] (frec)
    {\smaller featured\\[-4pt]\smaller (weakly)\\[-4pt]\smaller receptive};
  \node [obj,align=center]at(ETA-|frec) (rec)
    {\smaller (weakly)\\[-4pt]\smaller receptive};
  \draw [stealth-stealth,double] (frec) to (rec);
  \draw [slbl,inner sep=2pt,draw=none]
        (frec.center) to node[yshift=0pt,right](thm){\scriptsize theorem} (rec.center);
  \draw [slbl,inner sep=2pt,draw=none]
        (frec.center) to node[yshift=0pt,left](thm){\scriptsize Core} (rec.center);        
  \coordinate[xshift=10mm](thm2)at(thm.east);

  \draw[arr] (FS) to node[lbl,below,xshift=3pt] (fst) {$\fstype{:}\mFSTS$} (FETA);
  \draw[arr] (S)  to node[lbl,above,xshift=-2pt] (st) {$\stype{:}\textit{\STS}$}  (ETA);

  \draw[arr] (FCA)  to node[lbl,right,xshift=-1pt] {$\proj_p$}  (CA);
  \draw[arr] (FS)   to node[lbl,right,xshift=-1pt] {$\proj_p$}  (S);
  \coordinate (fetasw) at ($(FETA.south west)+(15pt,0)$);
  \draw[arr] (fetasw) to node[lbl,right,xshift=-1pt] {$\proj_p$}  (ETA.north-|fetasw);
  \draw[arr] (fst)  to node[lbl,right,xshift=-1pt] {$\proj_p$}  (st.north-|fst);

  \draw[sqgl] (FETA) to (frec);
  \draw[sqgl] (ETA) to (rec);

  \begin{scope}[on background layer]
    \tikzstyle{bg}=[draw=none]
    \coordinate (case) at ($(CA.south east)-(4pt,0)$);
    \coordinate (fetanw) at ($(FETA.north west)-(4pt,0)$);
    \coordinate (etaL) at ($(ETA.center)-(9pt,0)$);
    \coordinate (etaR) at ($(ETA.center)+(5pt,0)$);
    \coordinate (stL) at ($(st.west)+(1pt,0)$);
    \coordinate (SR) at ($(FS.east)+(2pt,0)$);
    \coordinate (SL) at ($(FS.west)+(-2pt,0)$);
    \coordinate (frecR) at ($(frec.east)+(8pt,0)$);
    \coordinate (recR) at ($(rec.east)+(8pt,0)$);
    \node[fit=(FCA.north west|-FETA.north)(case|-ETA.south),
          fill=yellow!30,draw=yellow!75!gray,bg](comp){};
    \node[fit=(FETA.north-|S.center)(SL)(SR)(ETA.south-|S.center),
          fill=orange!18,draw=orange!75!gray,bg](sys){};
    \node[fit=(stL)(fetanw)(ETA.south west)(etaL),
          fill=red!15,draw=red!75!gray,bg](team){};
    \node[fit=(etaR)(FETA.north east|-FETA.north)(frecR)(rec.south east|-ETA.south)(thm2),
          fill=purple!15,draw=purple!75!gray,bg](req){};
    \node[slbl]at(comp.south){Components}; 
    \node[slbl]at(sys.south){Systems}; 
    \node[slbl]at(team.south){Teams}; 
    \node[slbl]at(req.south){Receptiveness}; 
    \node[fit=(FCA)(FETA)(frecR),inner sep=4pt,draw=black!30,thick,dotted]{};
    \node[fit=(CA.north-|FCA.west)(ETA.south)(recR),inner sep=4pt,draw=black!30,thick,dotted]{};
    \node[lbl,left,align=right](other)at(CA.west){Earlier work\,\\from~\cite{BHK20b}:\,};
    \node[lbl,left,align=right](this)at(other.east|-FCA){Contribution\,\\from \cite{BCHP21}:\,};
  \end{scope}

\end{tikzpicture}
  \caption{Overview of the contributions in our previous work~\cite{BCHP21} with respect to earlier work~\cite{BHK20b}, using a valid product $p$}
  \label{fig:var-overview}
\end{figure}
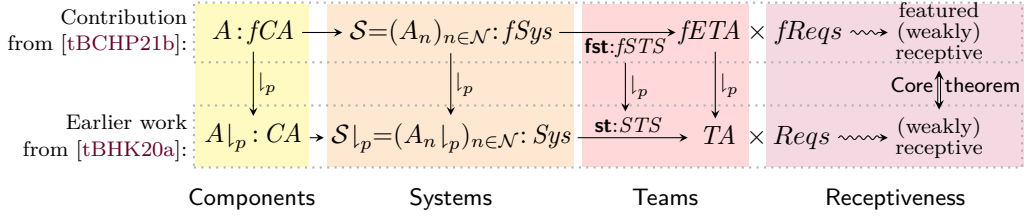

Figure~\ref{fig:var-overview} 
shows 
the contributions of~\cite{BCHP21} in relation to earlier work~\cite{BHK20b}.

\noindent In particular, we 
extended TA (as presented in this paper) 
by enriching them with variability, proposing a new model called \emph{featured (extended) TA} (\FETA) to allow the specification of---and reasoning on---a family of TA parameterised by a set of features. We define~\emph{projections}~%
$\proj_p$ (for any valid product $p$) to relate the featured setting of~\cite{BCHP21}
to that without in~\cite{BHK20b}.
We start by illustrating in the example below the notion of projection of
featured component automata (fCA) to
CA, covering the left part of \cref{fig:var-overview} over \textsf{Components}.

\begin{exa}\label{ex:variability}
Recall the two versions of a controller: the original \pts{Ctrl} (introduced in \cref{fig:race-local}) and its variation \pts{CtrlV} (introduced in \cref{fig:racev}) with a new approval phase before each race. Using a family model we can describe both variants in a single \emph{superimposed} model---an \mFCA---presented on the left side of \cref{tab:fca-ex} (\pts{CtrlF}).
This \mFCA includes all possible transitions from both variants, each guarded by a Boolean formula over features (\eg, $\trFe{\lock[]} \msg{!ask\/}$ is only active when the feature \lock is selected, $\trFe{\unlock[]} \msg{!start\/}$ is only active when \unlock is selected, and $\trFe{\verum} \msg{?finish\/}$ is always active).
The semantics of \pts{CtrlF} is given by the semantics of its \emph{projections} to 
sets of features, each filtering the set of active transitions (\eg, projecting \pts{CtrlF} on the (singleton) set $\set{\unlock}$ yields the \pts{Ctrl} component, while projecting it on the set $\set{\lock}$ yields the \pts{CtrlV} component, both depicted in \cref{tab:fca-ex}).
\qedx
\end{exa}

\begin{table}[H]
\centering
  \caption{Family model of a controller \pts{CtrlF} and its two products obtained by projecting on the feature \unlock (identical to \cref{fig:race-local}) and on the feature \lock (identical to \cref{fig:racev})}
  \label{tab:fca-ex}
  \scalebox{0.9}{%
  \begin{tabular}{c@{\hspace*{1mm}}c@{\hspace*{-2mm}}c}
    \toprule
    \mFCA (\pts{CtrlF})
    & \CA ($\pts{CtrlF}\proj_{\,\unlock} = \pts{Ctrl}$)
    & \CA ($\pts{CtrlF}\proj_{\,\lock} = \pts{CtrlV}$)
    \\
    \midrule
    \begin{tikzpicture}[
  ->,scale=.5,>=stealth',shorten >=1pt,auto,node distance=65pt,
  semithick, initial text={},
  align=center,
  every node/.style={scale=.9,font=\footnotesize}
  ]
  \tikzstyle{every state}=[teamState,fill=localcol]

  \node[initial above,state]    (c0){$0$};
  \node[state]                  (c1)[right of=c0,xshift=-2mm]{$1$};
  \node[state]                  (c2)[right of=c1,xshift=2mm]{$2$};
  \node[state]                  (c3)[below of=c2,yshift=4mm]{$3$};
  \node[state]                  (c4)at(c3-|c1) {$4$};  

  \path (c0) edge
                                    node{$\trFe{\lock[]} \msg{!ask\/}$}     (c1)
        (c0) edge                   node[inner sep=1pt]{$\trFe{\unlock[]} \msg{!start\/}$}   (c3)
        (c1) edge[bend right=60]    node[above]{$\trFe{\lock[]} \msg{?reject\/}$} (c0)
        (c1) edge                   node[xshift=-4pt]{~~$\trFe{\lock[]} \msg{?grant\/}$}   (c2)
        (c2) edge                   node{$\trFe{\lock[]} \msg{!start\/}$}   (c3)
        (c3) edge                   node[xshift=4pt]{$\trFe{\verum} \msg{?finish\/}~~~~$}  (c4)
        (c4) edge[bend left=50]     node[xshift=-2pt,yshift=2pt,above right,inner sep=1pt]
                                        {$\trFe{\verum} \msg{?finish\/}$}  (c0);
\end{tikzpicture}
    &
    \raisebox{1mm}{\begin{tikzpicture}[
  ->,scale=.5,>=stealth',shorten >=1pt,auto,node distance=65pt,
  semithick, initial text={},
  align=center,
  every node/.style={scale=.9,font=\footnotesize}
  ]
  \tikzstyle{every state}=[teamState,fill=localcol]

  \node[initial above,state]    (c0){$0$};
  \phantom{
  \node[state]                  (c1)[right of=c0,xshift=-10mm]{$1$};
  \node[state]                  (c2)[right of=c1,xshift=-3mm]{$2$};}
  \node[state]                  (c3)[below of=c2,yshift=4mm]{$3$};
  \node[state]                  (c4)at(c3-|c1) {$4$};  

  \path (c0) edge                   node{$\msg{!start\/}$}   (c3)
        (c3) edge                   node[xshift=1pt]{$\msg{?finish\/}~~~~$}  (c4)
        (c4) edge[bend left=50]     node[xshift=1pt,above right,inner sep=1pt]
                                        {$\msg{?finish\/}$}  (c0);
\end{tikzpicture}}
    &
    \raisebox{1mm}{\begin{tikzpicture}[
  ->,scale=.5,>=stealth',shorten >=1pt,auto,node distance=65pt,
  semithick, initial text={},
  align=center,
  every node/.style={scale=.9,font=\footnotesize}
  ]
  \tikzstyle{every state}=[teamState,fill=localcol]

  \node[initial above,state]    (c0){$0$};
  \node[state]                  (c1)[right of=c0,xshift=-4mm]{$1$};
  \node[state]                  (c2)[right of=c1,xshift=-3mm]{$2$};
  \node[state]                  (c3)[below of=c2,yshift=4mm]{$3$};
  \node[state]                  (c4)at(c3-|c1) {$4$};  

  \path (c0) edge
                                    node{$\msg{!ask\/}$}     (c1)
        (c1) edge[bend right=60]    node[above]{$\msg{?reject\/}$} (c0)
        (c1) edge                   node[xshift=-4pt]{~~$\msg{?grant\/}$}   (c2)
        (c2) edge                   node{$\msg{!start\/}$}   (c3)
        (c3) edge                   node[xshift=0pt]{$\msg{?finish\/}$}  (c4)
        (c4) edge[bend left=50]     node[xshift=1pt,above right,inner sep=1pt]
                                        {$\msg{?finish\/}$}  (c0);
\end{tikzpicture}}
    \\
    \bottomrule
  \end{tabular}
  }
\end{table}

\noindent
Example~\ref{ex:variability} illustrates how CA are projected in the \textsf{Components} part of the overview diagram in \cref{fig:var-overview}. This is trivially lifted for systems of CA (\cf\ the \textsf{System} part in \cref{fig:var-overview}), and required a notion of \emph{featured} STS (synchronisation type specification) when enriching TA with variability (\cf\ the \textsf{TA} part in \cref{fig:var-overview}). Our core result related notions of receptiveness in \FETA and its projections (\cf\ the \textsf{Receptiveness} part in \cref{fig:var-overview}), supporting a single analysis over families of systems (instead of over each possible projection).

The work on featured TA~\cite{BCHP21} describes a more complex example of a system, in which two users join and leave a server using two possible products: either with or without requiring permission from the server. The resulting superimposed featured system (\textit{\FSys}) counts 142~transitions and 18~states, which are reduced when building the \FETA and projecting onto each TA, counting 10~transitions and 8~states with the simplest product. 

This ability to present many variations of a state machine based on annotations of features over the transitions has been shown to be more efficient for model checking all variations in a single model (in a so-called \emph{family-based} fashion~\cite{TAKSS14}) rather than model checking each of the product variants individually (\cf, \eg, \cite{CHSLR10,CCSHLR13,CDKB18,BDLMP22}).

\subsection{Featured Component Automata and Featured Systems} 
\label{sub:fca}\label{sub:fsystems}

We assume a finite set of \emph{features} $F$, each 
regarded as a Boolean variable that represents a unit of variability. In our running example, $F = \set{\unlock,\lock}$. 

A \emph{product} is a finite combination of (desirable) features $p\subseteq F$.
A \emph{feature expression}~$\psi$ over a set of features $F$ 
is a Boolean expression over features with the usual Boolean connectives and constants \verum and \falsum interpreted by the truth values \kw{true} and \kw{false}.
A product~$p$ satisfies a feature expression $\psi$, denoted by $p\models \psi$, if and only if $\psi$ is evaluated to \verum
when \verum is assigned to every feature in $p$ and \falsum to the features not in $p$. A feature expression $\psi$ is \emph{satisfiable} if there exists a product $p$
such that $p\models \psi$.

A \emph{feature model} $\fm$ is a feature expression that determines 
the set of products 
that satisfy the feature model. 

A \emph{featured transition system} (\FTS) is a septuple $A\,{=}\,(Q,q_0,\Sigma,E,\mshl{F},\mshl{\fm},\mshl{\gamma})$%
\footnote{Throughout this section we use \shl{grey backgrounds} to highlight extensions with features.}
such that $(Q,q_0,\Sigma,E)$ is an LTS, 
\shl{$F$} is a finite set of features,
\shl{$\fm$} is a feature model, 
and
\shl{$\gamma$} is a mapping assigning feature expressions 
to transitions in~$E$. 
A product $p \subseteq F$ is \emph{valid} for the feature model $\fm$
if $p \in {\fm}$.
A transition $t \in E$ is \emph{realisable} for a valid product $p$
if $p \models \gamma(t)$.
An \FTS $A$ can be \emph{projected} to a valid product $p \in \fm$
by using $\gamma$ to filter realisable transitions,
resulting in
the LTS $A\proj_p = (Q,I,\Sigma,E\proj_p)$, where
$E\proj_p = \{\,t\in E \mid$ $p \models \gamma(t)\,\}$.

A \emph{featured component automaton} (\FCA) is an \FTS $A = (Q,q_0,\Sigma,E,\mshl{F},\mshl{\fm},\mshl{\gamma})$
such that $\Sigma=\Sigma^?\cup\Sigma^!$ 
are the disjoint sets of \emph{input} and \emph{output} actions as before.
For simplicity, we do not consider
internal actions here.
Example~\ref{ex:variability} contains an \FCA and its projections.

A \emph{featured system} (\FSys) is a pair $\S = (\N,(A_n)_{n\in \N})$,
where 
$(A_n)_{n\in \N}$ is an $\N$-indexed family of
    \FCA $A_n = (Q_n,q_{0,n},\Sigma_n,E_n,\mshl{F},\mshl{\fm},\mshl{\gamma_n})$
    over a shared set of features~$\mshl{F}$ and feature model~$\mshl{\fm}$.
Any such \FSys 
$\S = (\N,(A_n)_{n\in \N})$ induces an \FTS
$\fts(\S) = (Q,q_0, \Lambda, E, 
\mshl{F},\mshl{\fm},\mshl{\gamma})$,
where
$Q=\prod_{n\in\N} Q_n$ is the set of \emph{system states},
writing $q_n$ to represent the projection of a state $q\in Q$ to the \FCA $A_n$,
$q_0 = \prod_{n\in\N} q_{0,n}$ is the \emph{initial system state},
$\Lambda$ 
is the set of \emph{system labels},
$E$ 
is the set of \emph{system transitions},
and 
$\mshl{\gamma}$
holds the \emph{system transition constraints}.
Formally, 
$\Lambda$
and 
$E$
are defined as in Section~\ref{sec:TA}, since they do not refer to features, and
$\gamma$ is built out of the individual $\gamma_n$ by mapping each transition $t = q\tr{(\Out,a,\In)}q' \in E$
to $\bigwedge_{n\in (\Out\cup \In)} \gamma_n(q_n\tr{a}_{A_n} q_n')$.
The \emph{projection} of
  an \FSys $\S = (\N,(A_n)_{n\in \N})$ on a
  product $p \in {\fm}$
is the system $\S\proj_p = (\N,(A_n\proj_{p})_{n\in \N})$. 

\subsection{Featured Team Automata} 
\label{sub:featured_eta}

Similarly to TA, \FETA restrict the behaviour of (featured) systems with synchronisation types, here enriched with constraints over features and called \emph{featured synchronisation type specification} (\FSTS). An \FSTS
over an \FSys 
\S, is a total function, 
$\fstype\!:\mshl{{\fm}} \times \comm \to \kw{Intv}\!\times\!\kw{Intv}$, mapping each product $p\in{\fm}$ and 
action $a\in\comm$ to a traditional synchronisation type, \ie, to a pair of intervals as before.
Thus, an \FSTS is parameterised by (valid) products and therefore
supports variability of synchronisation conditions.
For any product $p\in {\fm}$,
an \FSTS \fstype can be projected on an \STS $\fstype\proj_p$
such that $\fstype\proj_p(a) = \fstype(p,a)$ for all $a \in \Sigma$. 

Given
  an \FSys $\S = (\N,(A_n)_{n\in \N})$ and
  an \FSTS \fstype over \S,
the \emph{featured team automaton} (\FETA) generated by $\S$ and \fstype is the \FTS $\fta{\fstype}{\S} =
(Q,\ab q_0,\ab \Lambda,\ab E,\ab \mshl{F},\ab \mshl{\fm},\ab \mshl{{\gamma\mkern1mu_\fstype}})$,
such that
$\fts(\S) = (Q,\ab q_0,\ab \Lambda,\ab E,\ab \mshl{F},\ab \mshl{\fm},\ab \mshl{\gamma})$ is the \FTS induced by \S, and $\gamma\mkern1mu_\fstype$ is a variation of $\gamma$ that takes $\fstype$ into account.
Formally, for any $t \in E$, $\gamma\mkern1mu_\fstype(t)$ holds iff $\gamma(t)$ holds and if, for every valid product $p \in {\fm}$, the senders and receivers of $t$ are bounded by $\fstype(p,t.a)$.%

Receptiveness of \FETA has been analysed in our previous work~\cite{BCHP21}, corresponding to the right-most part of \cref{fig:var-overview}. Our core theorem shows that receptiveness of a family of TA can be checked by
verifying, once and for all, receptiveness on the featured level.
For doing so, we introduced a notion of featured receptiveness; \cf~\cite{BCHP21} for more details.

\subsection{Tool Support}
We implemented a prototypical tool to specify and analyse \FETA. It can be used online and downloaded at \url{http://arcatools.org/feta}. It offers
a 
DSL to describe a set of \FCA, a feature model as a constraint, and an \FSTS. The tool can calculate and draw the resulting \FETA and the requirements for each state that must hold to guarantee receptiveness. It is implemented in Scala, using the Play Framework to generate an interactive website with a server that uses a Java SAT solver to reason over valid products.

\subsection{Related Work}

Variability has also been studied for 
related coordination models.
BIP~\cite{KKWVBS16,MBBS16} has been extended with parameters to represent families of systems, one for each actual parameter, including approaches to model check such families.
Contract automata~\cite{BGGDF17,BBDG17,BBDLFGD20} have also been lifted to a featured version, further investigating systems where the selected product can be updated at runtime.
Reo~\cite{PC17} has been extended with numeric parameters describing the number of inputs and outputs of connectors, formalised using category theory.
Petri nets~\cite{MPC11,MPC16} have been extended to featured nets, enriching arcs with conditions over features akin to \FCA.
I/O\linebreak automata~\cite{LNW07,LPT09} have been lifted to product lines and analysed by model checking.

\subsection{Roadmap}
\label{sec:roadmap:var}

In the future, we intend to extend our theory of featured team automata to address receptiveness and responsiveness as well as compositionality, \ie, extend \FETA to composition of systems
and investigate conditions under which communication safety is preserved by \FETA composition.
We would also like to lift to \FETA our approaches to (1)~model check communication requirements~\cite{BCHP23}, and (2)~decompose a realisable global model into a system of CA~\cite{BHP23}.
Moreover, we plan to further develop the tool and analyse the practical impact of \FETA on the basis of larger case studies. This involves a thorough study of the efficiency of featured receptiveness checking compared to product-wise receptiveness checking.
Finally, we would like to implement a family-based analysis algorithm that computes, for a given \FETA, the set of all product configurations that yield communication-safe systems.

\section{Conclusion}
\label{sec:conclusion}

We provided an overview of team automata, a model for capturing a variety of notions related to coordination in distributed systems with decades of history (\cf\ \cref{app:pub}) as witnessed by 25{\tt +} publications by 25{\tt +} researchers,\footnote{\url{http://fmt.isti.cnr.it/~mtbeek/TA.html}} 
and discussed a number of aspects: communication properties, realisation, verification, composition of systems, variability, and tool support. 
\cref{tab:formalism-overview} shows the relations between the main formalisms discussed in Section~\ref{sec:formalisms} (and included in the more detailed comparison in~\cite{BHP24}) and the team automata aspects discussed in Sections~\ref{sec:comm}--\ref{sec:var}, \ie, \emph{verification} of compliance of \emph{communication properties} in terms of receptiveness and responsiveness, \emph{verification} of the possible decomposition (\emph{realisation}) of global models in terms of component models, \emph{composition of systems} of component models, \emph{variability} in the development and analysis of families of component models, and \emph{tool support} for all these aspects).
We did not discuss aspects like \emph{data}, for which Reo and BIP provide native support, since team automata cannot currently deal with data. 

In the past, we also studied in detail the computations and behaviour of team automata in relation to that of their constituting component automata~\cite{BK03,BK09}, identifying several types of team automata that satisfy \emph{compositionality} in terms of (synchronised) shuffles of their computations (\ie, formal languages). Moreover, a process calculus for modelling team automata was proposed~\cite{BGJ06,BGJ08}, extending some classical results on I/O automata as well as enlarging the family of team automata that guarantee a degree of compositionality. Team automata have also been used for the analysis of \emph{security} aspects in communication protocols~\cite{BLP03,BLP05,BLP06,EP06}, in particular for spatial and spatio-temporal access control~\cite{BEKR01b,JVK10}. 

\begin{table}[t]
  \centering
  \caption{Coordination formalisms and aspects analysed in this paper} 
  \label{tab:formalism-overview}
  \small
  \newcommand{\hdone}[1]{\textbf{\rotatebox{70}{\lwrap{\!\!#1}}}}
  \newcommand{\hd}[2]{\textbf{\rotatebox{70}{\lwrap{\!\!#1\\[-2pt]{\scriptsize{~(Section #2)}}}}}}
  \newcommand{\pd}[2]{\textbf{\rotatebox{70}{\lwrap{\!\!#1\\[-2pt]{\scriptsize{~(Papers #2)}}}}}}
  \newcommand{\hdthree}[3]{\textbf{\rotatebox{70}{\lwrap{\!\!#1\\[-2pt]{\scriptsize{~(Sections #2 \& #3)}}}}}}
  \newcommand{\hdfour}[4]{\textbf{\rotatebox{70}{\lwrap{\!\!#1\\[-2pt]{\scriptsize{~(Sections #2, #3 \& #4)}}}}}}
  \newcommand{\hdtwo}[3]{\textbf{\rotatebox{70}{\lwrap{\!\!#1\\[-2pt]{\scriptsize{~(Sections #2~--~#3)}}}}}}
  \newcommand{\chk}{\multicolumn{1}{c}{\checkmark}}
  \newcommand{\y}{& \multicolumn{1}{c}{\checkmark}}
  \renewcommand{\n}{& ~}\newcommand{\?}{& \multicolumn{1}{c}{?}}
  \begin{tabular}{p{61.5mm}p{10mm}p{6.5mm}p{6.5mm}p{10mm}p{6.5mm}p{6.5mm}p{6.5mm}p{6.5mm}}
    \toprule
    \hdthree{Coordination\\[-1pt] Formalism}{\ref{sec:TA}}{\ref{sec:formalisms}}
      & \hd{Communication\\[-2pt] Properties}{\ref{sec:comm}}
      & \hd{Realisation}{\ref{sec:realisability}}
      & \hdthree{Verification}{\ref{sec:comm}}{\ref{sec:realisability}} 
      & \hd{Composition of\\[-2pt] Systems}{\ref{sec:composition}}
      & \hd{Variability}{\ref{sec:var}}
      & \hdtwo{Supporting Tools}{\ref{sec:formalisms}}{\ref{sec:var}} 
      & \hdone{Data} 
   \\\midrule
      Team Automata~\cite{Ell97,BEKR03}
     \y\y\y\y\y\y\n\,
   \\ Reo via Port Automata~\cite{Arb04,KC09}
     \n\n\y\n\y\y\y
   \\ BIP (without priorities)~\cite{BBS06,BS08}
     \n\n\y\y\y\y\y
   \\ Contract Automata~\cite{BDFT14,BDF16}
     \y\y\y\y\y\y\n\,
   \\ Choreography Automata~\cite{BLT20,BLT23}
     \y\y\y\n\n\y\n
   \\ Multi-Party Session Types~\cite{SY19,SD19}
     \n\y\y\y\y\y\y
   \\\bottomrule
  \end{tabular}
  \vspace*{1\baselineskip}
\end{table}

In the future, we want to address the roadmaps identified in Sections~\ref{sec:roadmap:comm}, \ref{sec:roadmap:realisabilty}, \ref{sec:roadmap:composition}, and~\ref{sec:roadmap:var}.
Recently, 
we have 
investigated \emph{asynchronous} team automata~\cite{BBP26}, in which components use either FIFO channels or multisets to act as buﬀers 
for asynchronous communication, like asynchronous multi-party session types~\cite{HYC08}
or systems of communicating finite state machines (CFSMs)~\cite{BZ83}, rather than the synchronous communication based on simultaneous execution of shared actions used so far in TA and throughout the current paper. As a basis, we 
used CFSMs which, however, are based on peer-to-peer communication. This means that it was 
challenging to understand how to generalise the approach by taking into account multi-action communication specified by synchronisation types. 
In such an asynchronous context, we 
obtained different kinds of communication properties and challenging investigations for
their preservation by composition of systems
(following, \eg, ideas in the context of CFSMs from~\cite{BdLH19,BH26}). We also provided tool support for modelling, animation, and automated checks for well-formedness and well-behavedness of asynchronous team automata~\cite{BBP26-artefact}.

\stoptoc
\section*{Acknowledgements}
We would like to thank the four reviewers for their careful reading and detailed comments, which have allowed us to considerably improve the paper.

Maurice ter Beek was funded by MUR PRIN 2020TL3X8X project T-LADIES (Typeful Language Adaptation for Dynamic, Interacting and Evolving Systems) and 
CNR project \lq\lq Formal Methods in Software Engineering~2.0\rq\rq, CUP~B53C24000720005. 

Jos\'{e} Proen\c{c}a
was funded by National
Funds through the FCT (Portuguese Foundation for Science and Technology) within
the project POISE, with reference 2024.15783.PEX (https://doi.org/10.54499/2024.15783.PEX);
by National Funds through FCT/MCTES, within the CISTER Unit (UIDP/UIDB/04234/2020);
by the EU/Next Generation, within the Recovery and Resilience Plan, within project Route~25 (TRB/2022/00061~-- C645463824-00000063); and
by national funds through FCT and European funds through EU Chips JU, within project Shift2SDV (Grant Agreement No. 101194245).

\resumetoc

\clearpage

\appendix

\section{Selected Publications from 25{\tt +} Years of Team Automata}
\label{app:pub}
\vspace*{-0.5\baselineskip}
\renewcommand{\baselinestretch}{0.55}
\renewcommand{\arraystretch}{2.0}
\renewcommand{\tabcolsep}{1pt}
\newcommand{\ttl}[1]{#1}
{\footnotesize
\begin{longtable}{llp{98mm}l}
\toprule
 \multicolumn{2}{l}{\ \ \textbf{Year}} & \ \ \textbf{Title} & \ \ \textbf{Venue} 
 \\
 \cmidrule(lr){1-2} \cmidrule(lr){3-3} \cmidrule(lr){4-4}
 \endhead
 \midrule
 \multicolumn{4}{r}{(continues)}
 \endfoot
 \bottomrule
 \endlastfoot
2026 & \cite{BBP26} & \ttl{Asynchronous Team Automata} & ~FM \\
2024 & \cite{BHP24} & \ttl{Team Automata: Overview and Roadmap} & ~COORDINATION \\
2023 & \cite{BHP23} & \ttl{Realisability of Global Models of Interaction} & ~ICTAC \\
2023 & \cite{P23} & \ttl{Overview on Constrained Multiparty Synchronisation in Team Auto\-mata} & ~FACS \\
2023 & \cite{BCHP23} & \ttl{Can we Communicate? Using Dynamic Logic to Verify Team Auto\-mata} & ~FM \\
2021 & \cite{BCHP21} & \ttl{Featured Team Automata} & ~FM \\
2020 & \cite{BHK20a} & \ttl{Team Automata@Work: On Safe Communication} & ~COORDINATION \\
2020 & \cite{BHK20b} & \ttl{Compositionality of Safe Communication in Systems of Team Automata} & ~ICTAC \\
2017 & \cite{BCHK17} & \ttl{Communication Requirements for Team Automata} & ~COORDINATION \\
2016 & \cite{BCK16} & \ttl{Conditions for Compatibility of Components: The Case of Masters and Slaves} & ~ISoLA \\
2014 & \cite{BK14} & \ttl{{O}n {D}istributed {C}ooperation and {S}ynchronised {C}ollaboration} & ~JALC \\
2013 & \cite{CK13} & \ttl{Compatibility in a multi-component environment} & ~TCS \\
2012 & \cite{BK12} & \ttl{Vector {T}eam {A}utomata} & ~TCS \\
2010 & \cite{JVK10} & \ttl{Team {A}utomata {B}ased {F}ramework for {S}patio-{T}emporal {RBAC} {M}odel} & ~BAIP \\
2009 & \cite{BK09} & \ttl{Associativity of {I}nfinite {S}ynchronized {S}huffles and {T}eam {A}utomata} & ~Fundam. Inform. \\
2008 & \cite{Sha08} & \ttl{Extending {T}eam {A}utomata to {E}valuate {S}oftware {A}rchitectural {D}esign} & ~COMPSAC \\
2008 & \cite{BGJ08} & \ttl{A calculus for team automata} & ~ENTCS \\
2007 & \cite{SSM07} & \ttl{A {R}eview on {S}pecifying {S}oftware {A}rchitectures {U}sing {E}xtended {A}utomata-{B}ased {M}odels} & ~FSEN \\
2006 & \cite{EP06} & \ttl{Modelling a {S}ecure {A}gent with {T}eam {A}utomata} & ~ENTCS \\
2006 & \cite{BLP06} & \ttl{A {T}eam {A}utomaton {S}cenario for the {A}nalysis of {S}ecurity {P}roperties in {C}ommunication {P}rotocols} & ~JALC \\
2006 & \cite{BGJ06} & {A calculus for team automata} & ~SBMF \\
2005 & \cite{BLP05} & \ttl{Team {A}utomata for {S}ecurity -- {A} {S}urvey --} & ~ENTCS \\
2005 & \cite{BK05} & \ttl{Modularity for {T}eams of {I}/{O} {A}utomata} & ~IPL \\
2005 & \cite{Len05} & \mbox{Integration of {A}nalysis {T}echniques in {S}ecurity and {F}ault-{T}olerance} \mbox{(Chapter~6: \emph{Security Analysis with Team Automata})} & ~PhD thesis \\
2005 & \cite{Pet05} & \mbox{{A}spects of {M}odeling and {V}erifying {S}ecure {P}rocedures} \mbox{(Chapter~4: \emph{The Team Automata Chapter})} & ~PhD thesis \\
2004 & \cite{BCM04} & \ttl{Teams of {P}ushdown {A}utomata} & ~IJCM \\
2004 & \cite{CK04b} & \ttl{Interactive {B}ehaviour of {M}ulti-{C}omponent {S}ystems} & ~Workshop ToBaCo \\
2003 & \cite{BCM03} & {Teams of {P}ushdown {A}utomata} & ~PSI \\
2003 & \cite{Bee03} & \ttl{Team {A}utomata: {A} {F}ormal {A}pproach to the {M}odeling of {C}ollaboration {B}etween {S}ystem {C}omponents} & ~PhD thesis \\
2003 & \cite{BK03} & \ttl{Team {A}utomata {S}atisfying {C}ompositionality} & ~FME \\
2003 & \cite{BLP03} & \ttl{Team {A}utomata for {S}ecurity {A}nalysis of {M}ulticast/{B}roadcast {C}ommunication} & ~Workshop WISP \\
2003 & \cite{Kle03} & \ttl{Team {A}utomata for {CSCW} -- {A} {S}urvey --} & ~LNCS \\
2003 & \cite{BEKR03} & \ttl{Synchronizations in {T}eam {A}utomata for {G}roupware {S}ystems} & ~CSCW\\
2002 & \cite{EG02} & \ttl{Towards {T}eam-{A}utomata-{D}riven {O}bject-{O}riented {C}ollabora\-tive {W}ork} & ~LNCS \\
2001 & \cite{BEKR01b} & \ttl{Team {A}utomata for {S}patial {A}ccess {C}ontrol} & ~ECSCW \\
2001 & \cite{BEKR01a} & \ttl{Team {A}utomata for {CSCW}} & ~Workshop \\ 
2000 & \cite{HB00} & \ttl{A {C}onflict-{F}ree {S}trategy for {T}eam-{B}ased {M}odel {D}evelopment} & ~Workshop PDTSD \\
1999 & \cite{BEKR99-12} & \ttl{Synchronizations in {T}eam {A}utomata for {G}roupware {S}ystems} & ~Technical Report \\
1997 & \cite{Ell97} & \ttl{Team {A}utomata for {G}roupware {S}ystems} & ~GROUP\\

\end{longtable}
}

\bibliographystyle{alphaurl}
\bibliography{src/TA}

@inproceedings{Ell97, 
  author = {Clarence (Skip) Ellis}, 
  title = {{Team Automata for Groupware Systems}}, 
  booktitle = {Proceedings of the International ACM SIGGROUP Conference on Supporting Group Work: The Integration Challenge (GROUP 1997)},
  OPTeditor = {Stephen C. Hayne and Wolfgang Prinz and Mark Pendergast and Kjeld Schmidt},
  publisher = {ACM}, 
  year = {1997},
  pages = {415--424},
  OPTurl = {http://www.acm.org/pubs/citations/proceedings/cscw/266838/p415-ellis/},
  doi = {10.1145/266838.267363}
}

@techreport{BEKR99-12,
  author = {ter Beek, Maurice H. and Clarence A. Ellis and Jetty Kleijn and Grzegorz Rozenberg},
  title = {{Synchronizations in Team Automata for Groupware Systems}},
  institution = {Leiden Institute of Advanced Computer Science, Leiden University},
  number = {TR-99-12},
  year = {1999},
  pages = {49},
  OPTurl = {http://fmt.isti.cnr.it/~mtbeek/TR-99-12.ps.gz}
}

@inproceedings{HB00,
  author = {'t Hoen, Pieter Jan and ter Beek, Maurice H.}, 
  title = {{A Conflict-Free Strategy for Team-Based Model Development}},
  booktitle = {Proceeedings of the International Workshop on Process support for Distributed Team-based Software Development (PDTSD 2000)}, 
  OPTbooktitle = {Proceeedings of the International Workshop on Process support for Distributed Team-based Software Development (PDTSD 2000) in Volume~IX: Industrial Systems of the Proceedings of the World MultiConference on Systemics, Cybernetics and Informatics (SCI 2000), Orlando, FL, U.S.A.}, 
  OPTeditor = {B. Sanchez and R. Hammel~II and M. Soriano and P. Tiako},
  publisher = {IIIS}, 
  year = {2000},
  pages = {720--725},
  OPTurl = {http://fmt.isti.cnr.it/~mtbeek/pdtsd2000.ps.gz}
}

@inproceedings{BEKR01a, 
  author = {ter Beek, Maurice H. and Clarence A. Ellis and Jetty Kleijn and Grzegorz Rozenberg}, 
  title = {{Team Automata for {CSCW}}}, 
  booktitle = {Proceedings of the 2nd International Colloquium on Petri Net Technologies for Modelling Communication Based Systems},
  OPTeditor = {H. Weber and H. Ehrig and W. Reisig},
  publisher = {Fraunhofer ISST},  
  year = {2001},
  pages = {1--20},  
  OPTurl = {http://fmt.isti.cnr.it/~mtbeek/berlin.ps.gz}
}

@inproceedings{BEKR01b, 
  author = {ter Beek, Maurice H. and Clarence A. Ellis and Jetty Kleijn and Grzegorz Rozenberg}, 
  title = {{Team Automata for Spatial Access Control}}, 
  booktitle = {Proceedings of the 7th European Conference on Computer-Supported Cooperative Work (ECSCW 2001)},
  editor = {Wolfgang Prinz and  Matthias Jarke and Yvonne Rogers and Kjeld Schmidt and Volker Wulf},
  publisher = {Kluwer},
  year = {2001},
  pages = {59--77},  
  OPTurl = {http://fmt.isti.cnr.it/~mtbeek/ecscw.ps.gz},
  doi = {10.1007/0-306-48019-0_4}
}

@incollection{EG02,
  author = {Gregor Engels and Luuk P. J. Groenewegen}, 
  title = {{Towards Team-Automata-Driven Object-Oriented Collaborative Work}},
  booktitle = {Formal and Natural Computing},
  OPTbooktitle = {Formal and Natural Computing---Essays Dedicated to Grzegorz Rozenberg},
  series = {LNCS},
  volume = {2300},
  editor = {Wilfried Brauer and Hartmut Ehrig and Juhani Karhum{\"{a}}ki and Arto Salomaa}, 
  publisher = {Springer},
  year = {2002},
  pages = {257--276},  
  OPTurl = {http://www.link.springer.de/link/service/series/0558/bibs/2300/23000257.htm},
  doi = {10.1007/3-540-45711-9_15}
}

@article{BEKR03,
  author = {ter Beek, Maurice H. and Clarence A. Ellis and Jetty Kleijn and Grzegorz Rozenberg},
  title = {{Synchronizations in Team Automata for Groupware Systems}},
  journal = {Comput. Sup. Coop. Work},
  OPTjournal = {Computer Supported Cooperative Work---The Journal of Collaborative Computing},
  volume = {12},
  number = {1},  
  year = {2003},
  pages = {21--69},  
  doi = {10.1023/A:1022407907596}
}

@inproceedings{BLP03,
  author = {ter Beek, Maurice and Gabriele Lenzini and Marinella Petrocchi},
  title = {{Team Automata for Security Analysis of Multicast/Broadcast Communication}},
  booktitle = {Proceedings of the ICATPN Workshop on Issues in Security and Petri Nets (WISP 2003)},
  publisher = {Eindhoven University of Technology},
  editor = {Nadia Busi and Roberto Gorrieri and Fabio Martinelli},
  year = {2003},
  pages = {57--71},  
  OPTurl = {http://fmt.isti.cnr.it/~mtbeek/wisp.ps.gz}
}

@inproceedings{BK03,
  author = {ter Beek, Maurice H. and Jetty Kleijn},
  title = {{Team Automata Satisfying Compositionality}},
  OPTbooktitle = {FME 2003},
  booktitle = {Proceedings of the 12th International Symposium of Formal Methods Europe (FME 2003)},
  series = {LNCS},
  volume = {2805},
  editor = {Keijiro Araki and Stefania Gnesi and Dino Mandrioli},
  publisher = {Springer},
  year = {2003},
  pages = {381--400},  
  OPTurl = {http://www.link.springer.de/link/service/series/0558/bibs/2805/28050381.htm},
  doi = {10.1007/978-3-540-45236-2_22}
}

@phdthesis{Bee03,
  author = {ter Beek, Maurice H.},
  title = {{Team Automata---{A} Formal Approach to the Modeling of Collaboration Between System Components}},
  school = {Leiden University},
  OPTschool = {Leiden Institute of Advanced Computer Science, Leiden University},
  year = {2003}, 
  pages = {348},
  OPTurl = {http://fmt.isti.cnr.it/~mtbeek/webphdA4.ps.gz},
  url = {https://hdl.handle.net/1887/29570}
}

@incollection{Kle03,
  author = {Jetty Kleijn},
  title = {{Team Automata for {CSCW} --~{A} Survey~--}},
  booktitle = {Petri Net Technology for Communication-Based Systems: Advances in Petri Nets},
  OPTbooktitle = {Petri Net Technology for Communication-Based Systems---Advances in Petri Nets},
  series = {LNCS},
  volume = {2472},
  editor = {Hartmut Ehrig and Wolfgang Reisig and Grzegorz Rozenberg and Herbert Weber},
  publisher = {Springer},
  year = {2003},
  pages = {295--320},  
  OPTurl = {http://www.link.springer.de/link/service/series/0558/bibs/2472/24720295.htm},
  doi = {10.1007/978-3-540-40022-6_15}
}

@inproceedings{BCM03,
  author = {ter Beek, Maurice H. and Erzs{\'{e}}bet Csuhaj-Varj{\'{u}} and Victor Mitrana},
  title = {{Teams of Pushdown Automata}},
  OPTbooktitle = {PSI 2003},
  booktitle = {Revised papers of the 5th International Andrei Ershov Memorial Conference on Perspectives of System Informatics (PSI 2003)},
  series = {LNCS},
  volume = {2890},
  editor = {Manfred Broy and Alexandre V. Zamulin},
  publisher = {Springer},
  year = {2003},
  pages = {329--337},  
  OPTurl = {http://www.link.springer.de/link/service/series/0558/bibs/2890/28900329.htm},
  doi = {10.1007/978-3-540-39866-0_32}
}

@article{BCM04,
  author = {ter Beek, Maurice H. and Erzs{\'{e}}bet Csuhaj-Varj{\'{u}} and Victor Mitrana},
  title = {{Teams of Pushdown Automata}},
  journal = {Int. J. Comput. Math.},
  OPTjournal = {International Journal of Computer Mathematics},
  volume = {81},
  number = {2},  
  year = {2004},
  pages = {141--156},  
  OPTurl = {http://taylorandfrancis.metapress.com/link.asp?id=24tnev0wfgma6uvm},
  doi = {10.1080/00207160310001650099}
}

@inproceedings{CK04b,
  author = {Josep Carmona and Jetty Kleijn}, 
  title = {{Interactive Behaviour of Multi-Component Systems}},
  booktitle = {Proceedings of the ICATPN Workshop on Token-Based Computing (ToBaCo 2004)},
  editor = {Jordi Cortadella and Alex Yakovlev},
  publisher = {University of Bologna},
  year = {2004}, 
  pages = {27--31},
  OPTurl = {http://www.liacs.nl/~kleijn/papers/tobacofinalsubmisweb.pdf}
}

@phdthesis{Pet05,
  author = {Marinella Petrocchi},
  title = {{Aspects of Modeling and Verifying Secure Procedures}}, 
  school = {University of Pisa},
  year = {2005}
}

@article{BLP05,
  author = {ter Beek, Maurice H. and Gabriele Lenzini and Marinella Petrocchi},
  title = {{Team Automata for Security --~{A} Survey~--}},
  OPTbooktitle = {Proceedings of the 2nd International Workshop on Security Issues in Coordination Models, Languages, and Systems (SecCo 2004)},
  OPTeditor = {Riccardo Focardi and Gianluigi Zavattaro},
  OPTseries = {Electronic Notes in Theoretical Computer Science},
  journal = {Electron. Notes Theor. Comput. Sci.},
  OPTjournal = {Electronic Notes in Theoretical Computer Science},
  volume = {128},
  issue = {5},
  OPTpublisher = {Elsevier Science Publishers, Amsterdam},
  year = {2005},
  pages = {105--119},  
  OPTurl = {http://fmt.isti.cnr.it/~mtbeek/secco05.pdf},
  doi = {10.1016/j.entcs.2004.11.044}
}

@article{BK05,
  author = {ter Beek, Maurice H. and Jetty Kleijn},
  title = {{Modularity for Teams of {I}/{O} Automata}},
  journal = {Inf. Process. Lett.},
  OPTjournal = {Information Processing Letters},
  volume = {95},
  number = {5},
  year = {2005},
  pages = {487--495},
  OPTurl = {http://dx.doi.org/doi:10.1016/j.ipl.2005.05.012},
  doi = {10.1016/j.ipl.2005.05.012}
}

@phdthesis{Len05,
  author = {Gabriele Lenzini},
  title = {{Integration of Analysis Techniques in Security and Fault-Tolerance}}, 
  school = {University of Twente},
  series = {CTIT Ph.D.\ Thesis Series},
  number = {05-70},
  year = {2005},
  ISBN = {90-365-2200-5}
}

@article{EP06,
  author = {Lavinia Egidi and Marinella Petrocchi},
  title = {{Modelling a Secure Agent with Team Automata}},
  OPTbooktitle = {Proceedings of the 1st International Workshop on Views On Designing Complex Architectures (VODCA 2004)},
  OPTeditor = {ter Beek, Maurice H. and Fabio Gadducci},
  OPTseries = {Electronic Notes in Theoretical Computer Science},
  journal = {Electron. Notes Theor. Comput. Sci.},
  OPTjournal = {Electronic Notes in Theoretical Computer Science},
  volume = {142},
  OPTpublisher = {Elsevier Science Publishers, Amsterdam},
  year = {2006},
  pages = {111--127},
  OPTurl = {http://dx.doi.org/10.1016/j.entcs.2004.12.046},
  doi = {10.1016/j.entcs.2004.12.046}
}

@inproceedings{BGJ06,
  author = {ter Beek, Maurice H. and Fabio Gadducci and Dirk Janssens},
  title = {A calculus for team automata},
  booktitle = {Proceedings of the 9th Brazilian Symposium on Formal Methods (SBMF 2006)},
  editor = {Leila Ribeiro and Martins Moreira, Anamaria}, 
  publisher = {Istituto de Informatica da UFRGS, Porto Alegre}, 
  year = {2006},
  pages = {59--72},
  OPTurl = {http://fmt.isti.cnr.it/~mtbeek/terBeekGadducciJanssens.pdf}
}

@article{BLP06,
  author = {ter Beek, Maurice H. and Gabriele Lenzini and Marinella Petrocchi},
  title = {{A Team Automaton Scenario for the Analysis of Security Properties in Communication Protocols}},
  journal = {J. Autom. Lang. Comb.},
  OPTjournal = {Journal of Automata, Languages and Combinatorics},
  volume = {11}, 
  number = {4},
  year = {2006},
  pages = {345--374},  
  OPTurl = {http://fmt.isti.cnr.it/~mtbeek/jalc.pdf},
  doi = {10.25596/jalc-2006-345}
}

@inproceedings{SSM07,
  author = {Mehran Sharafi and Shams Aliee, Fereidoon and Ali Movaghar},
  title = {{A Review on Specifying Software Architectures Using Extended Automata-Based Models}},
  OPTbooktitle = {FSEN 2007},
  booktitle = {Proceedings of the 2nd International Symposium on Fundamentals of Software Engineering (FSEN 2007)},
  editor = {Farhad Arbab and Marjan Sirjani},
  year = {2007},
  pages = {423--431},
  publisher = {Springer},
  series = {LNCS},
  volume = {4767},  
  OPTurl = {http://dx.doi.org/10.1007/978-3-540-75698-9_30},
  doi = {10.1007/978-3-540-75698-9_30}
}

@article{BGJ08,
  author = {ter Beek, Maurice H. and Fabio Gadducci and Dirk Janssens},
  title = {A calculus for team automata},
  OPTbooktitle = {Proceedings of the 9th Brazilian Symposium on Formal Methods (SBMF 2006)},
  OPTeditor = {A. Martins Moreira and L. Ribeiro}, 
  OPTseries = {Electronic Notes in Theoretical Computer Science},
  journal = {Electron. Notes Theor. Comput. Sci.},
  OPTjournal = {Electronic Notes in Theoretical Computer Science},
  volume = {195},
  OPTpublisher = {Elsevier Science Publishers, Amsterdam},
  year = {2008},
  pages = {41--55},
  OPTurl = {http://fmt.isti.cnr.it/~mtbeek/SBMF08.pdf},
  doi = {10.1016/j.entcs.2007.08.022}
}

@inproceedings{Sha08,
  author = {Mehran Sharafi},
  title = {{Extending Team Automata to Evaluate Software Architectural Design}},
  booktitle = {Proceedings of the 32nd IEEE International Computer Software and Applications Conference (COMPSAC 2008)},
  OPTbooktitle = {Proceedings of the 32nd Annual IEEE International Computer Software and Applications Conference (COMPSAC 2008)},
  year = {2008},
  pages = {393--400},
  publisher = {IEEE},  
  OPTurl = {http://doi.ieeecomputersociety.org/10.1109/COMPSAC.2008.57},
  doi = {10.1109/COMPSAC.2008.57}
}

@article{BK09,
  author = {ter Beek, Maurice H. and Jetty Kleijn},
  title = {{Associativity of Infinite Synchronized Shuffles and Team Automata}},
  journal = {Fundam. Inform.},
  OPTjournal = {Fundamenta Informaticae},
  volume = {91},
  number = {3-4},
  year = {2009},
  pages = {437--461},
  OPTurl = {http://dx.doi.org/10.3233/FI-2009-0051},
  doi = {10.3233/FI-2009-0051}
}

@inproceedings{JVK10,
  author = {Jaisankar Narayanasamy and Sankaradass Veeramalai and Arputharaj Kannan},
  title = {{Team Automata Based Framework for Spatio-Temporal {RBAC} Model}},
  OPTbooktitle = {BAIP 2010},
  booktitle = {Proceedings of the International Conference on Recent Trends in Business Administration and Information Processing (BAIP 2010)},
  editor = {Vinu V. Das and R. Vijayakumar and Narayan C. Debnath and Janahanlal Stephen and Natarajan Meghanathan and Suresh Sankaranarayanan and P. M. Thankachan and Ford Lumban Gaol and Nessy Thankachan},
  year = {2010},
  pages = {586--591},
  publisher = {Springer},
  series = {CCIS},
  OPTseries = {Communications in Computer and Information Science},
  volume = {70},  
  OPTurl = {http://www.springerlink.com/content/p4t727635r063qkq/},
  doi = {10.1007/978-3-642-12214-9_106}
}

@article{BK12,
  author = {ter Beek, Maurice H. and Jetty Kleijn},
  title = {{Vector Team Automata}},
  journal = {Theor. Comput. Sci.},
  OPTjournal = {Theoretical Computer Science},
  volume = {429},
  number = {},
  year = {2012},
  pages = {21--29},
  OPTurl = {http://dx.doi.org/10.1016/j.tcs.2011.12.020},
  doi = {10.1016/j.tcs.2011.12.020}
}

@article{CK13,
  author = {Josep Carmona and Jetty Kleijn},
  title = {Compatibility in a multi-component environment},
  journal = {Theor. Comput. Sci.},
  OPTjournal = {Theoretical Computer Science},
  volume = {484},
  number = {},
  year = {2013},
  pages = {1--15},
  OPTurl = {http://dx.doi.org/10.1016/j.tcs.2013.03.006},
  doi = {10.1016/j.tcs.2013.03.006}
}

@article{BK14,
  author = {ter Beek, Maurice H. and Jetty Kleijn},
  title = {{On Distributed Cooperation and Synchronised Collaboration}},
  journal = {J. Autom. Lang. Comb.},
  OPTjournal = {Journal of Automata, Languages and Combinatorics},
  volume = {19}, 
  number = {1-4},
  year = {2014},
  pages = {17--32},  
  OPTurl = {http://www.jalc.de/search/j19_i.html},
  doi = {10.25596/jalc-2014-017}
}

@inproceedings{BCK16,
  author = {ter Beek, Maurice H. and Josep Carmona and Jetty Kleijn},
  title = {{Conditions for Compatibility of Components: {T}he Case of Masters and Slaves}},
  OPTbooktitle = {ISoLA 2016},
  booktitle = {Proceedings of the 7th International Symposium on Leveraging Applications of Formal Methods, Verification and Validation: Foundational Techniques (ISoLA 2016)},
  editor = {Tiziana Margaria and Bernhard Steffen},
  year = {2016},
  pages = {784--805},
  publisher = {Springer},
  series = {LNCS},
  volume = {9952},  
  OPTurl = {http://dx.doi.org/10.1007/978-3-319-47166-2_55},
  doi = {10.1007/978-3-319-47166-2_55}
}

@inproceedings{BCHK17,
  author = {ter Beek, Maurice H. and Josep Carmona and Rolf Hennicker and Jetty Kleijn},
  title = {{Communication Requirements for Team Automata}},
  OPTbooktitle = {COORDINATION 2017},
  booktitle = {Proceedings of the 19th IFIP WG 6.1 International Conference on Coordination Models and Languages (COORDINATION 2017)},
  editor = {Jean-Marie Jacquet and Mieke Massink},
  publisher = {Springer},
  series = {LNCS},
  volume = {10319},
  year = {2017},
  pages = {256--277},
  OPTurl = {http://dx.doi.org/10.1007/978-3-319-59746-1_14},
  doi = {10.1007/978-3-319-59746-1_14}
}

@inproceedings{BHK20a,
  author = {ter Beek, Maurice H. and Rolf Hennicker and Jetty Kleijn},
  title = {{Team Automata@Work: {O}n Safe Communication}},
  OPTbooktitle = {COORDINATION 2020},
  booktitle = {Proceedings of the 22nd IFIP WG 6.1 International Conference on Coordination Models and Languages (COORDINATION 2020)},
  editor = {Simon Bliudze and Laura Bocchi},
  publisher = {Springer},
  series = {LNCS},
  volume = {12134},
  year = {2020},
  pages = {77--85},
  OPTurl = {http://dx.doi.org/10.1007/978-3-030-50029-0_5},
  doi = {10.1007/978-3-030-50029-0_5}
}

@inproceedings{BHK20b,
  author = {ter Beek, Maurice H. and Rolf Hennicker and Jetty Kleijn},
  title = {{Compositionality of Safe Communication in Systems of Team Automata}},
  OPTbooktitle = {ICTAC 2020},
  booktitle = {Proceedings of the 17th International Colloquium on Theoretical Aspects of Computing (ICTAC 2020)},
  editor = {Ka I Pun, Violet and Volker Stolz and Adenilso Sim{\~{a}}o},
  publisher = {Springer},
  series = {LNCS},
  volume = {12545},
  year = {2020},
  pages = {200--220},
  OPTurl = {http://dx.doi.org/10.1007/978-3-030-64276-1_11},
  doi = {10.1007/978-3-030-64276-1_11}
}

@techreport{BHK20c,
  author = {ter Beek, Maurice H. and Rolf Hennicker and Jetty Kleijn},
  title = {{Compositionality of Safe Communication in Systems of Team Automata}},
  institution = {Zenodo},
  day = {25},
  month = {Sep},
  year = {2020},
  OPTurl = {http://dx.doi.org/10.5281/zenodo.4050293},
  doi = {10.5281/zenodo.4050293}
}

@inproceedings{BCHP21,
  author = {ter Beek, Maurice H. and Cledou, Guillermina and Hennicker, Rolf and Proen{\c{c}}a, Jos{\'{e}}},
  title = {{Featured Team Automata}},
  OPTbooktitle = {FM 2021},
  booktitle = {Proceedings of the 24th International Symposium on Formal Methods (FM 2021)},
  editor = {Marieke Huisman and Corina P{\v{a}}s{\v{a}}reanu and Naijun Zhan},
  publisher = {Springer},
  series = {LNCS},
  volume = {13047},
  year = {2021},
  pages = {483--502},
  OPTurl = {http://dx.doi.org/10.1007/978-3-030-90870-6_26},
  doi = {10.1007/978-3-030-90870-6_26}
}

@techreport{BCHP21x,
  author = {ter Beek, Maurice H. and Cledou, Guillermina and Hennicker, Rolf and Proen{\c{c}}a, Jos{\'{e}}},
  title = {{Featured Team Automata}},
  institution = {arXiv},
  day = {3},
  month = {Aug},
  year = {2021},
  OPTurl = {https://doi.org/10.48550/arXiv.2108.01784},
  doi = {10.48550/arXiv.2108.01784}
}

@techreport{BCHP22,
  author = {ter Beek, Maurice H. and Cledou, Guillermina and Hennicker, Rolf and Proen{\c{c}}a, Jos{\'{e}}},
  title = {{Can we Communicate? Using Dynamic Logic to Verify Team Automata (Extended Version)}},
  institution = {Zenodo},
  day = {9},
  month = {Dec},
  year = {2022},
  OPTurl = {http://dx.doi.org/10.5281/zenodo.7418074},
  doi = {10.5281/zenodo.7418074}
}

@inproceedings{BCHP23,
  author = {ter Beek, Maurice H. and Cledou, Guillermina and Hennicker, Rolf and Proen{\c{c}}a, Jos{\'{e}}},
  title = {{Can we Communicate? Using Dynamic Logic to Verify Team Automata}},
  OPTbooktitle = {FM 2023},
  booktitle = {Proceedings of the 25th International Symposium on Formal Methods (FM 2023)},
  editor = {Marsha Chechik and Joost-Pieter Katoen and Martin Leucker},
  publisher = {Springer},
  series = {LNCS},
  volume = {14000},
  year = {2023},
  pages = {122--141},
  OPTurl = {http://dx.doi.org/10.1007/978-3-031-27481-7_9},
  doi = {10.1007/978-3-031-27481-7_9}
}

@misc{BCHP23-tool,
  author = {ter Beek, Maurice H. and Cledou, Guillermina and Hennicker, Rolf and Proen{\c{c}}a, Jos{\'{e}}},
  title = {{Can we Communicate? Using Dynamic Logic to Verify Team Automata (Software Artefact)}},
  year = {2022},
  OPTday = {19},
  OPTmonth = {Nov},
  publisher = {Zenodo},
  doi = {10.5281/zenodo.7338440},
  OPTurl = {https://doi.org/10.5281/zenodo.7338440},
  noUrlBecauseOfDoi = {http://arcatools.org/feta}
}

@techreport{BHP23z,
  author = {ter Beek, Maurice H. and Rolf Hennicker and Jos{\'{e}} Proen{\c{c}}a},
  title = {{Realisability of Global Models of Interaction (Extended Version)}},
  institution = {Zenodo},
  day = {25},
  month = {Sep},
  year = {2023},
  OPTurl = {http://dx.doi.org/10.5281/zenodo.8377188},
  doi = {10.5281/zenodo.8377188}
}

@inproceedings{BHP23,
  author = {ter Beek, Maurice H. and Rolf Hennicker and Jos{\'{e}} Proen{\c{c}}a},
  title = {{Realisability of Global Models of Interaction}},
  OPTbooktitle = {ICTAC 2023},
  booktitle = {Proceedings of the 20th International Colloquium on Theoretical Aspects of Computing (ICTAC 2023)},
  editor = {Erika {\'{A}}brah{\'{a}}m and Clemens Dubslaff and Tapia Tarifa, Silvia Lizeth},
  publisher = {Springer},
  series = {LNCS},
  volume = {14446},
  year = {2023},
  pages = {236--255},
  OPTurl = {https://doi.org/10.1007/978-3-031-47963-2_15},
  doi = {10.1007/978-3-031-47963-2_15}
}

@inproceedings{P23,
  author = {Proen{\c{c}}a, Jos\'{e}},
  title = {{Overview on Constrained Multiparty Synchronisation in Team Automata}},
  OPTbooktitle = {FACS 2023},
  booktitle = {Revised Selected Papers of the 19th International Conference on Formal Aspects of Component Software (FACS 2023)},
  editor = {C\'{a}mara, Javier and Jongmans, Sung-Shik},
  publisher = {Springer},
  series = {LNCS},
  volume = {14485},
  year = {2023},
  pages = {194--205},
  OPTurl = {https://doi.org/10.1007/978-3-031-52183-6_10},
  doi = {10.1007/978-3-031-52183-6_10}
}

@inproceedings{BHP24,
  author = {ter Beek, Maurice H. and Rolf Hennicker and Jos{\'{e}} Proen{\c{c}}a},
  title = {{Team Automata: {O}verview and Roadmap}},
  booktitle = {Proceedings of the 26th IFIP WG 6.1 International Conference on Coordination Models and Languages (COORDINATION 2024)},
  editor = {Ilaria Castellani and Francesco Tiezzi},
  publisher = {Springer},
  series = {LNCS},
  volume = {14676},
  year = {2024},
  pages = {161--198},
  OPTurl = {https://doi.org/10.1007/978-3-031-62697-5_10},
  doi = {10.1007/978-3-031-62697-5_10}
}

@inproceedings{BBP26,
  author = {Basile, Davide and ter Beek, Maurice H. and Proen{\c{c}}a, Jos{\'{e}}},
  title = {{Asynchronous Team Automata}},
  OPTbooktitle = {FM 2026},
  booktitle = {Proceedings of the 27th International Symposium on Formal Methods (FM 2026)},
  editor = {Augusto Sampaio and Mari{\"{e}}lle Stoelinga},
  publisher = {Springer},
  series = {LNCS},
  volume = {16557},
  year = {2026},
  pages = {27--49},
  OPTurl = {},
  doi = {10.1007/978-3-032-26220-2_2},
  OPTnote = {preprint in A-Team artefact, \url{https://doi.org/10.5281/zenodo.18601856}}
}

@misc{BBP26-artefact,
  author = {Proen{\c{c}}a, Jos{\'{e}} and ter Beek, Maurice H. and Basile, Davide},
  title = {{A-Team artefact}},
  howpublished = {Zenodo},
  OPTday = {26},
  OPTmonth = {Feb},
  year = {2026},
  OPTurl = {https://zenodo.org/records/18784823},
  doi = {10.5281/zenodo.18784823}
}

@inproceedings{TCV22,
  author = {Viktor Teren and Jordi Cortadella and Tiziano Villa},
  title = {{Decomposition of transition systems into sets of synchronizing Free-choice {P}etri Nets}},
  OPTbooktitle = {DSD 2022},
  booktitle = {Proceedings of the 25th Euromicro Conference on Digital System Design (DSD 2022)},
  publisher = {IEEE},
  year = {2022},
  pages = {165--173},
  doi = {10.1109/DSD57027.2022.00031}
}

@inproceedings{TCV21,
  author = {Viktor Teren and Jordi Cortadella and Tiziano Villa},
  title = {Decomposition of transition systems into sets of synchronizing state machines},
  OPTbooktitle = {DSD 2021},
  booktitle = {Proceedings of the 24th Euromicro Conference on Digital System Design (DSD 2021)},  
  OPTeditor = {Francesco Leporati and Salvatore Vitabile and Amund Skavhaug},
  publisher = {IEEE},
  year = {2021},
  pages = {77--81},
  doi = {10.1109/DSD53832.2021.00021}
}

@article{Lut16,
  author = {Bas Luttik},
  title = {Unique parallel decomposition in branching and weak bisimulation semantics},
  journal = {Theor. Comput. Sci.},
  volume = {612},
  year = {2016},
  pages = {29--44},
  doi = {10.1016/j.tcs.2015.10.013}
}

@article{CGM98,
  author = {Flavio Corradini and Roberto Gorrieri and Davide Marchignoli},
  title = {Towards parallelization of concurrent systems},
  journal = {RAIRO Theor. Informatics Appl.},
  volume = {32},
  number = {4-6},
  year = {1998},
  pages = {99--125},
  doi = {10.1051/ita/1998324-600991}
}

@inproceedings{GM92,
  author = {Groote, Jan Friso and Faron Moller},
  title = {{Verification of Parallel Systems via Decomposition}},
  OPTbooktitle = {CONCUR 1992},
  booktitle = {Proceedings of the 3rd International Conference on Concurrency Theory (CONCUR 1992)},
  editor = {Rance Cleaveland},
  series = {LNCS},
  volume = {630},
  publisher = {Springer},
  year = {1992},
  pages = {62--76},
  doi = {10.1007/BFb0084783}
}

@article{MM93,
  author = {Robin Milner and Faron Moller},
  title = {{Unique Decomposition of Processes}},
  journal = {Theor. Comput. Sci.},
  volume = {107},
  number = {2},
  year = {1993},
  pages = {357--363},
  doi = {10.1016/0304-3975(93)90176-T}
}

@article{BS08,
  author = {Simon Bliudze and Joseph Sifakis},
  title = {{The Algebra of Connectors: Structuring Interaction in {BIP}}},
  journal = {IEEE Trans. Comput.},
  volume = {57},
  number = {10},
  year = {2008},
  pages = {1315--1330},
  doi = {10.1109/TC.2008.26}
}

@article{DBLP:journals/corr/abs-1203-0780,
  author = {Giuseppe Castagna and Mariangiola Dezani{-}Ciancaglini and Luca Padovani},
  title = {{On Global Types and Multi-Party Sessions}},
  journal = {Log. Meth. Comp. Sci.},
  volume = {8},
  number = {1},
  year = {2012},
  pages = {24:1--24:45},
  doi = {10.2168/LMCS-8(1:24)2012}
}

@article{BY09,
  author = {Andi Bejleri and Nobuko Yoshida},
  title = {{Synchronous Multiparty Session Types}},
  OPTjournal = {ENTCS},
  journal = {Electr. Notes Theor. Comput. Sci.},
  OPTjournal = {Electronic Notes in Theoretical Computer Science},
  volume = {241},
  year = {2009},
  pages = {3--33},
  doi = {10.1016/j.entcs.2009.06.002}
}

@inproceedings{HYC08,
  author = {Kohei Honda and Nobuko Yoshida and Marco Carbone},
  title = {Multiparty asynchronous session types},
  OPTbooktitle = {POPL 2008},
  booktitle = {Proceedings of the 35th ACM SIGPLAN-SIGACT Symposium on Principles of Programming Languages (POPL 2008)},
  OPTeditor = {George C. Necula and Philip Wadler},
  publisher = {ACM},
  year = {2008},
  pages = {273--284},
  doi = {10.1145/1328438.1328472}
}

@article{BES94,
  author = {van Benthem, Johan and van Eijck, Jan and Vera Stebletsova},
  title = {{Modal Logic, Transition Systems and Processes}},
  journal = {J. Log. Comput.},
  volume = {4},
  number = {5},
  year = {1994},
  pages = {811--855},
  doi = {10.1093/logcom/4.5.811}
}

@incollection{HT03,
  author = {David Harel and P. S. Thiagarajan},
  title = {{Message Sequence Charts}},
  booktitle = {{UML} for Real: {D}esign of Embedded Real-Time Systems},
  editor = {Luciano Lavagno and Grant Martin and Bran Selic},
  publisher = {Kluwer},
  year = {2003},
  pages = {77--105},
  doi = {10.1007/0-306-48738-1_4}
}

@misc{ITU11,
  author = {{ITU (International Telecommunication Union)}},
  title = {{Message Sequence Chart~({MSC})}},
  howpublished = {Recommendation ITU-T Z.120},
  year = {2011},
  month = {Feb},
  url = {http://www.itu.int/rec/T-REC-Z.120},
  OPTurl = {https://www.itu.int/rec/dologin_pub.asp?lang=e&id=T-REC-Z.120-201102-I!!PDF-E&type=items}
}

@inproceedings{BLT20,
  author = {Franco Barbanera and Ivan Lanese and Emilio Tuosto},
  title = {{Choreography Automata}},
  OPTbooktitle = {COORDINATION 2020},
  booktitle = {Proceedings of the 22nd IFIP WG 6.1 International Conference on Coordination Models and Languages (COORDINATION 2020)},
  editor = {Simon Bliudze and Laura Bocchi},
  series = {LNCS},
  volume = {12134},
  publisher = {Springer},
  year = {2020},
  pages = {86--106},
  doi = {10.1007/978-3-030-50029-0_6}
}

@inproceedings{BLT22,
  author = {Franco Barbanera and Ivan Lanese and Emilio Tuosto},
  title = {{Formal Choreographic Languages}},
  OPTbooktitle = {COORDINATION 2022},
  booktitle = {Proceedings of the 24th IFIP WG 6.1 International Conference on Coordination Models and Languages (COORDINATION 2022)},
  editor = {ter Beek, Maurice H. and Marjan Sirjani},
  series = {LNCS},
  volume = {13271},
  publisher = {Springer},
  year = {2022},
  pages = {121--139},
  doi = {10.1007/978-3-031-08143-9_8}
}

@article{BLT23,
  author = {Franco Barbanera and Ivan Lanese and Emilio Tuosto},
  title = {{A Theory of Formal Choreographic Languages}},
  journal = {Log. Meth. Comp. Sci.},
  volume = {19},
  number = {3},
  year = {2023},
  pages = {9:1--9:36},
  doi = {10.46298/LMCS-19(3:9)2023}
}

@inproceedings{CMT99,
  author = {Ilaria Castellani and Madhavan Mukund and P. S. Thiagarajan},
  title = {{Synthesizing Distributed Transition Systems from Global Specification}},
  OPTbooktitle = {FSTTCS 1999},
  booktitle = {Proceedings of the 19th Conference on Foundations of Software Technology and Theoretical Computer Science (FSTTCS 1999)},
  editor = {C. Pandu Rangan and Venkatesh Raman and Ramaswamy Ramanujam},
  series = {LNCS},
  volume = {1738},
  publisher = {Springer},
  year = {1999},
  pages = {219--231},
  doi = {10.1007/3-540-46691-6_17}
}

@inproceedings{PM19,
  author = {Jos{\'{e}} Proen{\c{c}}a and Alexandre Madeira},
  title = {{Taming Hierarchical Connectors}},
  OPTbooktitle = {FSEN 2019},
  booktitle = {Revised Selected Papers of the 8th International Conference on Fundamentals of Software Engineering (FSEN 2019)},
  editor = {Hossein Hojjat and Mieke Massink},
  series = {LNCS},
  volume = {11761},
  publisher = {Springer},
  year = {2019},
  pages = {186--193},
  doi = {10.1007/978-3-030-31517-7_13}
}

@article{BOV23,
  author = {Laura Bocchi and Dominic Orchard and A. Laura Voinea},  
  title = {{A Theory of Composing Protocols}},
  journal = {Art Sci. Eng. Program.},
  volume = {7},
  number = {2},
  artno = {6},
  year = {2023},
  pages = {6:1--6:76},
  doi = {10.22152/programming-journal.org/2023/7/6}
}

@article{BB22,
  author = {Davide Basile and ter Beek, Maurice H.},
  title = {{Contract Automata Library}},
  journal = {Sci. Comput. Program.},
  volume = {221},
  year = {2022},
  doi = {10.1016/j.scico.2022.102841}
}

@article{BB24,
  author = {Davide Basile and ter Beek, Maurice H.},
  title = {Advancing Orchestration Synthesis for Contract Automata},
  journal = {J. Log. Algebr. Methods Program.},
  volume = {},
  number = {},
  year = {2024},
  pages = {},
  doi = {}
}

@book{HKT00,
  author = {David Harel and Dexter Kozen and Jerzy Tiuryn},
  title = {Dynamic Logic},
  series = {Foundations of Computing},
  publisher = {MIT Press},
  year = {2000},
  doi = {10.7551/mitpress/2516.001.0001}
}

@inproceedings{BDL04,
  author = {Gerd Behrmann and Alexandre David and Kim Guldstrand Larsen},
  title = {{A Tutorial on {Uppaal}}},
  OPTbooktitle = {SFM-RT 2004},
  booktitle = {Revised Lectures of the 4th International School on Formal Methods for the Design of Computer, Communication and Software Systems: Formal Methods for the Design of Real-Time Systems (SFM-RT 2004)},
  editor = {Marco Bernardo and Flavio Corradini},
  series = {LNCS},
  volume = {3185},
  publisher = {Springer},
  year = {2004},
  pages = {200--236},
  doi = {10.1007/978-3-540-30080-9_7}
}

@inproceedings{LLN18,
  author = {Kim Guldstrand Larsen and Florian Lorber and Brian Nielsen},
  title = {{20 Years of \emph{Real} Real Time Model Validation}},
  OPTbooktitle = {FM 2018},
  booktitle = {Proceedings of the 22nd International Symposium on Formal Methods (FM 2018)},
  editor = {Klaus Havelund and Jan Peleska and Bill Roscoe and Erik P. de Vink},
  series = {LNCS},
  volume = {10951},
  publisher = {Springer},
  year = {2018},
  pages = {22--36},
  doi = {10.1007/978-3-319-95582-7_2}
}

@book{GM14,
  author = {Groote, Jan Friso and Mohammad Reza Mousavi},
  title = {Modeling and Analysis of Communicating Systems},
  publisher = {MIT Press},
  OPTaddress = {Cambridge}, 
  year = {2014}
}

@book{AG23,
  author = {Muhammad Atif and Groote, Jan Friso},
  title = {Understanding Behaviour of Distributed Systems Using {mCRL2}},
  OPTseries = {SSDC},
  OPTseries = {Studies in Systems, Decision and Control},
  OPTvolume = {458},
  publisher = {Springer},
  year = {2023},
  doi = {10.1007/978-3-031-23008-0}
}

@book{ABKS13,
  author = {Sven Apel and Don Batory and Christian K{\"{a}}stner and Gunter Saake},
  title = {Feature-Oriented Software Product Lines: Concepts and Implementation},
  publisher = {Springer},
  OPTaddress = {Berlin, Germany},
  year = {2013},
  doi = {10.1007/978-3-642-37521-7}
}

@article{TAKSS14,
  author = {Th{\"{u}}m, Thomas and Apel, Sven and K{\"{a}}stner, Christian and Schaefer, Ina and Saake, Gunter},
  title = {{A Classification and Survey of Analysis Strategies for Software Product Lines}},
  journal = {ACM Comput. Surv.},
  volume = {47},
  number = {1},
  year = {2014},
  OPTeid = {6},
  pages = {6:1--6:45},
  doi = {10.1145/2580950}
}

@inproceedings{CHSLR10,
  author = {Andreas Classen and Patrick Heymans and Pierre-Yves Schobbens and Axel Legay and Jean-Fran{\c{c}}ois Raskin},
  title = {{Model Checking \underline{Lots} of Systems: Efficient Verification of Temporal Properties in Software Product Lines}},
  OPTbooktitle = {ICSE 2010},
  booktitle = {Proceedings of the 32nd International Conference on Software Engineering (ICSE 2010)},
  publisher = {ACM},
  year = {2010},
  pages = {335--344},
  doi = {10.1145/1806799.1806850},
}

@article{CCSHLR13,
  author = {Classen, Andreas and Cordy, Maxime and Schobbens, Pierre-Yves and Heymans, Patrick and Legay, Axel and Raskin, Jean-Fran\c{c}ois},
  title = {{Featured Transition Systems: {F}oundations for Verifying Variability-Intensive Systems and Their Application to LTL Model Checking}},
  journal = {IEEE Trans. Softw. Eng.},
  volume = {39},
  number = {8},
  year = {2013},
  pages = {1069--1089},
  doi = {10.1109/TSE.2012.86}
}

@article{CDKB18,
  author = {Chrszon, Philipp and Dubslaff, Clemens and Kl{\"u}ppelholz, Sascha and Baier, Christel},
  title = {{{ProFeat}: {F}eature-Oriented Engineering for Family-Based Probabilistic Model Checking}},
  journal = {Form. Asp. Comput.},
  volume = {30},
  number = {1},
  year = {2018},
  pages = {45--75},
  doi = {10.1007/s00165-017-0432-4}
}

@article{BDLMP22,
  author = {ter Beek, Maurice H. and Damiani, Ferruccio and Lienhardt, Michael and Mazzanti, Franco and Paolini, Luca},
  title = {{Efficient Static Analysis and Verification of Featured Transition Systems}},
  journal = {Empirical Software Engineering},
  volume = {22},
  number = {1},
  year = {2022},
  pages = {10:1--10:43},
  doi = {10.1007/s10664-020-09930-8}
}

@inproceedings{BDFT14,
  author = {Davide Basile and Pierpaolo Degano and Gian{-}Luigi Ferrari and Emilio Tuosto},
  title = {{From Orchestration to Choreography through Contract Automata}},
  booktitle = {Proceedings of the 7th Interaction and Concurrency Experience (ICE 2014)},
  editor = {Ivan Lanese and Lluch Lafuente, Alberto and Ana Sokolova and Torres Vieira, Hugo},
  series = {EPTCS},
  volume = {166},
  year = {2014},
  pages = {67--85},
  doi = {10.4204/EPTCS.166.8}
}

@inproceedings{LTY15,
 author = {Julien Lange and Emilio Tuosto and Nobuko Yoshida},
 title = {{From Communicating Machines to Graphical Choreographies}},
 booktitle = {Proceedings of the 42nd ACM SIGPLAN-SIGACT Symposium on Principles of Programming Languages (POPL 2015)},
 OPTbooktitle = {Proceedings of the 42nd Annual ACM SIGPLAN-SIGACT Symposium on Principles of Programming Languages (POPL 2015)},
 publisher = {ACM},
 year = {2015},
 pages = {221--232},
 doi = {10.1145/2676726.2676964},
}

@article{BDF16,
  author = {Davide Basile and Pierpaolo Degano and Gian-Luigi Ferrari},
  title = {{Automata for Specifying and Orchestrating Service Contracts}},
  journal = {Log. Meth. Comp. Sci.},
  volume = {12},
  number = {4:6},
  year = {2016},
  pages = {1--51},
  doi = {10.2168/LMCS-12(4:6)2016}
}

@article{BZ83,
  author = {Daniel Brand and Pitro Zafiropulo},
  title = {{On Communicating Finite-State Machines}},
  journal = {J. ACM},
  volume = {30},
  number = {2},
  year = {1983},
  pages = {323--342},
  doi = {10.1145/322374.322380}
}

@article{BdLH19,
  author = {Franco Barbanera and Ugo de'Liguoro and Rolf Hennicker},
  title = {Connecting open systems of communicating finite state machines},
  journal = {J. Log. Algebr. Methods Program.},
  volume = {109},
  year = {2019},
  doi = {10.1016/j.jlamp.2019.07.004}
}

@article{BDFT16,
  author = {Davide Basile and Pierpaolo Degano and Gian-Luigi Ferrari and Emilio Tuosto},
  title = {Relating two automata-based models of orchestration and choreography},
  journal = {J. Log. Algebr. Methods Program.},
  volume = {85},
  number = {3},
  year = {2016},
  pages = {425--446},
  doi = {10.1016/J.JLAMP.2015.09.011}
}

@inproceedings{BGGDF17,
  author = {Davide Basile and Di Giandomenico, Felicita and Stefania Gnesi and Pierpaolo Degano and Gian-Luigi Ferrari},
  title = {{Specifying Variability in Service Contracts}},
  booktitle = {Proceedings of the 11th International Workshop on Variability Modelling of Software-intensive Systems (VaMoS 2017)},
  OPTeditor = {ter Beek, Maurice H. and N. Siegmund and I. Schaefer},
  publisher = {ACM},
  year = {2017},
  pages = {20--27},
  doi = {10.1145/3023956.3023965}
}

@inproceedings{BBDG17,
  author = {Davide Basile and ter Beek, Maurice H. and Di Giandomenico, Felicita and Stefania Gnesi},
  title = {{Orchestration of Dynamic Service Product Lines with Featured Modal Contract Automata}},
  booktitle = {Proceedings of the 21st International Systems and Software Product Line Conference (SPLC 2017)},
  volume = {2},
  OPTeditor = {M. ter Beek and W. Cazzola and O. D\'{\i}az and M. La Rosa and R. E. L\'{o}pez-Herrej\'{o}n and T. Th\"{u}m and J. Troya and A. Ruiz-Cort\'{e}s and D. Benavides},
  publisher = {ACM},
  year = {2017},
  pages = {117--122},
  doi = {10.1145/3109729.3109741}
}

@article{BBDLFGD20,
  author = {Davide Basile and ter Beek, Maurice H. and Pierpaolo Degano and Axel Legay and Gian-Luigi Ferrari and Stefania Gnesi and Di Giandomenico, Felicita},
  title = {{Controller synthesis of service contracts with variability}},
  journal = {Sci. Comput. Program.},
  volume = {187},
  year = {2020},
  pages = {},
  doi = {10.1016/j.scico.2019.102344}
}

@article{BBL20,
  author = {Davide Basile and ter Beek, Maurice H. and Axel Legay},
  title = {{Timed service contract automata}},
  journal = {Innov. Syst. Softw. Eng.},
  OPTjournal = {Innovations in Systems and Software Engineering},
  number = {16},
  volume = {2},
  year = {2020},
  pages = {199--214},
  doi = {10.1007/s11334-019-00353-3}
}

@article{CGP09,
  author = {Giuseppe Castagna and Nils Gesbert and Luca Padovani},
  title = {{A Theory of Contracts for Web Services}},
  journal = {ACM Trans. Program. Lang. Syst.},
  volume = {31},
  number = {5},
  year = {2009},
  pages = {19:1--19:61},
  doi = {10.1145/1538917.1538920}
}

@inproceedings{BCZ15,
  author = {Massimo Bartoletti and Tiziana Cimoli and Roberto Zunino},
  title = {{Compliance in Behavioural Contracts: {A} Brief Survey}},
  booktitle = {Programming Languages with Applications to Biology and Security},
  editor = {Chiara Bodei and Gian-Luigi Ferrari and Corrado Priami},
  series = {LNCS},
  volume = {9465},
  publisher = {Springer},
  year = {2015},
  pages = {103--121},
  doi = {10.1007/978-3-319-25527-9_9}
}

@article{HB18,
  author = {Rolf Hennicker and Michel Bidoit},
  title = {{Compatibility Properties of Synchronously and Asynchronously Communicating Components}},
  journal = {Log. Meth. Comp. Sci.},
  volume = {14},
  number = {1},
  year = {2018},
  pages = {1--31},
  doi = {10.23638/LMCS-14(1:1)2018}
}

@inproceedings{BSBM04,
  author = {Bordeaux, Lucas and Sala{\"u}n, Gwen and Berardi, Daniela and Mecella, Massimo},
  title = {{When are Two Web Services Compatible?}},
  OPTbooktitle = {TES 2004},
  booktitle = {Proceedings of the 5th International Workshop on Technologies for E-Services (TES 2004)},
  editor = {Shan, Ming-Chien and Dayal, Umeshwar and Hsu, Meichun},
  publisher = {Springer},
  series = {LNCS},
  volume = {3324},
  year = {2005},
  pages = {15--28}, 
  doi = {10.1007/978-3-540-31811-8_2}
}

@article{DOS12,
  author = {Francisco Dur{\'{a}}n and Meriem Ouederni and Gwen Sala{\"{u}}n},
  title = {{A generic framework for $n$-protocol compatibility checking}},
  journal = {Sci. Comput. Program.},
  volume = {77},
  number = {7-8},
  pages = {870--886},
  year = {2012},
  doi = {10.1016/j.scico.2011.03.009}
}

@incollection{DH05,
  author = {de Alfaro, Luca and Thomas A. Henzinger},
  title = {{Interface-Based Design}},
  booktitle = {Engineering Theories of Software Intensive Systems},
  OPTeditor = {M. Broy and J. Gr\"{u}nbauer and D. Harel and T. Hoare},
  series = {NATO Science Series},
  volume = {195},
  publisher = {Springer},
  year = {2005},
  pages = {83--104},
  doi = {10.1007/1-4020-3532-2_3}
}

@inproceedings{CC02,
  author = {Josep Carmona and Jordi Cortadella},
  title = {{Input/{O}utput Compatibility of Reactive Systems}},
  OPTbooktitle = {FMCAD 2002},
  booktitle = {Proceedings of the 4th International Conference on Formal Methods in Computer-Aided Design (FMCAD 2002)},
  editor = {Mark Aagaard and John W. O'Leary},
  series = {LNCS},
  volume = {2517},
  publisher = {Springer},
  year = {2002},
  pages = {360--377},
  doi = {10.1007/3-540-36126-X_22}
}

@inproceedings{LNW07,
  author = {Kim Guldstrand Larsen and Ulrik Nyman and Andrzej W\k{a}sowski},
  title = {{Modal I/O Automata for Interface and Product Line Theories}},
  OPTbooktitle = {ESOP 2007},
  booktitle = {Proceedings of the 16th European Symposium on Programming (ESOP 2007)},
  editor = {De Nicola, Rocco},
  series = {LNCS},
  volume = {4421},
  publisher = {Springer},
  year = {2007},
  pages = {64--79},
  doi = {10.1007/978-3-540-71316-6_6}
}

@inproceedings{LPT09,
  author = {Kim Lauenroth and Klaus Pohl and Simon T{\"o}hning},
  title = {{Model Checking of Domain Artifacts in Product Line Engineering}},
  booktitle = {Proceedings of the 24th International Conference on Automated Software Engineering (ASE 2009)},
  publisher = {IEEE},
  year = {2009},
  pages = {269--280},
  doi = {10.1109/ASE.2009.16}
}

@inproceedings{MPC11,
  author = {Radu Muschevici and Jos{\'{e}} Proen{\c{c}}a and Dave Clarke},
  title = {{Modular Modelling of Software Product Lines with Feature Nets}},
  OPTbooktitle = {SEFM 2011},
  booktitle = {Proceedings of the 9th International Conference on Software Engineering and Formal Methods (SEFM 2011)},
  editor = {Gilles Barthe and Alberto Pardo and Gerardo Schneider},  
  publisher = {Springer},
  series = {LNCS},
  volume = {7041},
  year = {2011},
  pages = {318--333},
  doi = {10.1007/978-3-642-24690-6_22}
}

@article{MPC16,
  author = {Radu Muschevici and Jos{\'{e}} Proen{\c{c}}a and Dave Clarke},
  title = {{Feature Nets: behavioural modelling of software product lines}},
  journal = {Softw. Sys. Model.},
  volume = {15},
  number = {4},
  year = {2016},
  pages = {1181--1206},
  doi = {10.1007/s10270-015-0475-z}
}

@inproceedings{BMSH10,
  author = {Sebastian S. Bauer and Philip Mayer and Andreas Schroeder and Rolf Hennicker},
  title = {{On Weak Modal Compatibility, Refinement, and the {MIO} Workbench}},
  OPTbooktitle = {TACAS 2010},
  booktitle = {Proceedings of the 16th International Conference on Tools and Algorithms for the Construction and Analysis of Systems (TACAS 2010)},
  editor = {Javier Esparza and Rupak Majumdar},
  series = {LNCS},
  volume = {6015},
  publisher = {Springer},
  year = {2010},
  pages = {175--189},
  doi = {10.1007/978-3-642-12002-2_15}
}

@book{KLSV10,
  author = {Dilsun K. Kaynar and Nancy Lynch and Roberto Segala and Frits Vaandrager},
  title = {The Theory of Timed {I/O} Automata},
  series = {Synthesis Lectures on Distributed Computing Theory},
  edition = {2nd},
  publisher = {Springer},
  year = {2010},
  doi = {10.1007/978-3-031-02003-2}
}

@article{LT89,
  author = {Nancy A. Lynch and Mark R. Tuttle},
  title = {{An Introduction to {I}nput/{O}utput Automata}},
  journal = {CWI Q.},
  OPTjournal = {CWI Quarterly},
  volume = {2},
  number = {3},
  year = {1989},
  pages = {219--246},
  note = {\url{https://ir.cwi.nl/pub/18164}}
}

@article{LT26,
  author = {Nancy A. Lynch and Mark R. Tuttle},
  title = {{An Introduction to {I}nput/{O}utput Automata}},
  journal = {Form. Asp. Comput.},
  OPTjournal = {Formal Aspects of Computing},
  volume = {38},
  number = {3},
  year = {2026},
  pages = {25:1--25:22},
  doi = {10.1145/3788687}
}

@book{Arn94,
  author = {A. Arnold},
  title = {Finite Transition Systems: Semantics of Communicating Systems},
  publisher = {Prentice Hall},
  year = {1994}
}

@inproceedings{DH01,
  author = {de Alfaro, Luca and Thomas A. Henzinger},
  title = {{Interface Automata}},
  OPTbooktitle = {ESEC/FSE 2001},
  booktitle = {Proceedings of the 8th European Software Engineering Conference held jointly with 9th {ACM} {SIGSOFT} International Symposium on Foundations of Software Engineering (ESEC/FSE 2001)},
  publisher = {ACM},
  year = {2001},
  pages = {109--120},
  doi = {10.1145/503209.503226}
}

@article{GS05,
  author = {Gregor G{\"{o}}ssler and Joseph Sifakis},
  title = {Composition for component-based modeling},
  journal = {Sci. Comput. Program.},
  volume = {55},
  year = {2005},
  pages = {161--183},
  doi = {10.1016/j.scico.2004.05.014}
}

@inproceedings{BBS06,
  author = {Ananda Basu and Marius Bozga and Joseph Sifakis},
  title = {{Modeling Heterogeneous Real-time Components in BIP}},
  OPTbooktitle = {SEFM 2006},
  booktitle = {Proceedings of the 4th IEEE International Conference on Software Engineering and Formal Methods (SEFM 2006)},
  publisher = {IEEE},
  year = {2006},
  pages = {3--12},
  doi = {10.1109/SEFM.2006.27}
}

@phdthesis{Jon87,
  author = {Bengt Jonsson},
  title = {Compositional Verification of Distributed Systems},
  school = {Uppsala University},
  year = {1987}
}

@inproceedings{Ohe02,
  author = {David von Oheimb},
  title = {{Interacting State Machines: {A} Stateful Approach to Proving Security}},
  OPTbooktitle = {FASec 2002},
  booktitle = {Revised Papers of the 1st International Conference on Formal Aspects of Security (FASec 2002)},
  editor = {Ali E. Abdallah and Peter Y. A. Ryan and Steve A. Schneider},
  series = {LNCS},
  volume = {2629},
  publisher = {Springer},
  year = {2002},
  pages = {15--32},
  doi = {10.1007/978-3-540-40981-6_4}
}

@article{BCVZ06,
  author = {Lubo{\v{s}} Brim and Ivana {\v{C}}ern{\'{a}} and Pavl{\'{\i}}na Va{\v{r}}ekov{\'{a}} and Barbora Zimmerov{\'{a}}},
  title = {{Component-Interaction Automata as a Verification-Oriented Component-Based System Specification}},
  journal = {ACM Softw. Eng. Notes},
  volume = {31},
  number = {2},
  year = {2006},
  doi = {10.1145/1118537.1123063}
}

@article{RW87,
  author = {Peter J. Ramadge and Walter M. Wonham},
  title = {{Supervisory Control of a Class of Discrete Event Processes}},
  journal = {SIAM J. Control Optim.},
  volume = {25},
  number = {1},
  year = {1987},
  pages = {206--230},
  doi = {10.1137/0325013}
}

@article{BBP20,
  author = {Davide Basile and ter Beek, Maurice H. and Rosario Pugliese},
  title = {{Synthesis of Orchestrations and Choreographies: {B}ridging the Gap between Supervisory Control and Coordination of Services}},
  journal = {Log. Meth. Comp. Sci.},
  volume = {16},
  number = {2},
  year = {2020},
  pages = {9:1--9:29},
  doi = {10.23638/LMCS-16(2:9)2020},
  OPTnote = {{S}pecial Issue with Selected Papers of COORDINATION 2019}
}

@inproceedings{KKWVBS16,
  author = {Igor V. Konnov and Tomer Kotek and Qiang Wang and Helmut Veith and Simon Bliudze and Joseph Sifakis},
  title = {{Parameterized Systems in {BIP}: Design and Model Checking}},
  OPTbooktitle = {CONCUR 2016},
  booktitle = {Proceedings of the 27th International Conference on Concurrency Theory (CONCUR 2016)},
  editor = {Jos{\'{e}}e Desharnais and Radha Jagadeesan},
  series = {LIPIcs},
  volume = {59},
  publisher = {Schloss Dagstuhl - Leibniz-Zentrum f{\"{u}}r Informatik},
  year = {2016},
  pages = {30:1--30:16},
  doi = {10.4230/LIPICS.CONCUR.2016.30}
}

@inproceedings{C07,
  author = {Dave Clarke},
  title = {{Coordination: {Reo}, Nets, and Logic}},
  OPTbooktitle = {FMCO 2007},
  booktitle = {Revised Lectures of the 6th International Symposium on Formal Methods for Components and Objects (FMCO 2007)},
  editor = {Frank S. de Boer and Marcello M. Bonsangue and Susanne Graf and Willem P. de Roever},
  series = {LNCS},
  volume = {5382},
  publisher = {Springer},
  year = {2007},
  pages = {226--256},
  doi = {10.1007/978-3-540-92188-2_10}
}

@article{SD19,
  author = {Paula Severi and Mariangiola Dezani{-}Ciancaglini},
  title = {{Observational Equivalence for Multiparty Sessions}},
  journal = {Fundam. Inform.},
  volume = {170},
  number = {1-3},
  year = {2019},
  pages = {267--305},
  doi = {10.3233/FI-2019-1863}
}

@article{SY19,
  author = {Alceste Scalas and Nobuko Yoshida},
  title = {{Less Is More: Multiparty Session Types Revisited}},
  journal = {Proc. ACM Program. Lang.},
  volume = {3},
  OPTnumber = {POPL},
  year = {2019},
  pages = {30:1--30:29},
  doi = {10.1145/3290343}
}

@article{Arb04,
  author = {Farhad Arbab},
  title = {Reo: a channel-based coordination model for component composition},
  journal = {Math. Struct. Comput. Sci.},
  volume = {14},
  number = {3},
  year = {2004},
  pages = {329--366},
  doi = {10.1017/S0960129504004153}
}

@article{BSAR06,
  author = {Christel Baier and Marjan Sirjani and Farhad Arbab and Jan J. M. M. Rutten},
  title = {Modeling component connectors in {Reo} by constraint automata},
  journal = {Sci. Comput. Program.},
  volume = {61},
  number = {2},
  year = {2006},
  pages = {75--113},
  doi = {10.1016/J.SCICO.2005.10.008}
}

@article{JA12,
  author = {Sung-Shik T. Q. Jongmans and Farhad Arbab},
  title = {{Overview of Thirty Semantic Formalisms for {Reo}}},
  journal = {Sci. Ann. Comput. Sci.},
  volume = {22},
  number = {1},
  year = {2012},
  pages = {201--251},
  doi = {10.7561/SACS.2012.1.201}
}

@inproceedings{KC09,
  author = {Christian Koehler and Dave Clarke},
  title = {{Decomposing Port Automata}},
  booktitle = {Proceedings of the 24th ACM Symposium on Applied Computing (SAC 2009)},
  editor = {Sung Y. Shin and Sascha Ossowski},
  publisher = {ACM},
  year = {2009},
  pages = {1369--1373},
  doi = {10.1145/1529282.1529587}
}

@book{Rei13,
  author = {Wolfgang Reisig},
  title = {Understanding Petri Nets: Modeling Techniques, Analysis Methods, Case Studies},
  publisher = {Springer},
  year = {2013},
  doi = {10.1007/978-3-642-33278-4}
}

@book{BD24,
  author = {Eike Best and Raymond Devillers},
  title = {Petri Net Primer: A Compendium on the Core Model, Analysis, and Synthesis},
  OPTseries = {CSFAL},
  OPTseries = {Computer Science Foundations and Applied Logic},
  publisher = {Springer},
  year = {2024},
  doi = {10.1007/978-3-031-48278-6}
}

@article{NPW81,
  author = {Mogens Nielsen and Gordon D. Plotkin and Glynn Winskel},
  title = {{{Petri} Nets, Event Structures and Domains, Part~{I}}},
  journal = {Theor. Comput. Sci.},
  volume = {13},
  year = {1981},
  pages = {85--108},
  doi = {10.1016/0304-3975(81)90112-2}
}

@inproceedings{Win88,
  author = {Glynn Winskel},
  title = {An introduction to event structures},
  OPTbooktitle = {REX 1988},
  booktitle = {Linear Time, Branching Time and Partial Order in Logics and Models for Concurrency: Proceedings of the Research and Education in Concurrent Systems Workshop (REX 1988)},
  editor = {Jaco W. de Bakker and Willem P. de Roever and Grzegorz Rozenberg},
  series = {LNCS},
  volume = {354},
  publisher = {Springer},
  OPTaddress = {Germany},
  year = {1988},
  pages = {364--397},
  doi = {10.1007/BFB0013026}
}

@article{BarBliu20-express,
  author = {Eduard Baranov and Simon Bliudze},
  title = {Expressiveness of component-based frameworks: {A} study of the expressiveness of {BIP}},
  journal = {Acta Inform.},
  volume = {57},
  year = {2020},
  pages = {761--800},
  doi = {10.1007/s00236-019-00337-7},
  OPTpdf = {http://www.bliudze.me/simon/articles/Baranov-Bliudze2019_Article_ExpressivenessOfComponent-base.pdf},
}

@article{JavaBIP-spe,
  author = {Simon Bliudze and Anastasia Mavridou and Radoslaw Szymanek and Alina Zolotukhina},
  title = {Exogenous coordination of concurrent software components with {JavaBIP}},
  journal = {Softw. Pract. Exper.},
  OPTjournal = {Software: Practice and Experience},
  volume = {47},
  number = {11},
  year = {2017},
  pages = {1801--1836},
  doi = {10.1002/spe.2495},
  OPTpdf = {http://www.bliudze.me/simon/articles/javabip-spe.pdf}
}

@inproceedings{MBBS16,
  author = {Anastasia Mavridou and Eduard Baranov and Simon Bliudze and Joseph Sifakis},
  title = {{Architecture Diagrams: {A} Graphical Language for Architecture Style Specification}},
  OPTbooktitle = {ICE 2016},
  booktitle = {Proceedings of the 9th Interaction and Concurrency Experience (ICE 2016)},
  editor = {Massimo Bartoletti and Ludovic Henrio and Sophia Knight and Torres Vieira, Hugo},
  series = {EPTCS},
  volume = {223},
  year = {2016},
  pages = {83--97},
  doi = {10.4204/EPTCS.223.6}
}

@inproceedings{OPBLT21,
  author = {Simone Orlando and Di Pasquale, Vairo and Franco Barbanera and Ivan Lanese and Emilio Tuosto},
  title = {{Corinne, a Tool for Choreography Automata}},
  OPTbooktitle = {FACS 2021},
  booktitle = {Proceedings of the 17th International Conference on Formal Aspects of Component Software (FACS 2021)},
  editor = {Gwen Sala{\"{u}}n and Anton Wijs},
  series = {LNCS},
  volume = {13077},
  publisher = {Springer},
  year = {2021},
  pages = {82--92},
  doi = {10.1007/978-3-030-90636-8_5}
}

@inproceedings{KL98,
  author = {Joost-Pieter Katoen and Lennard Lambert},
  title = {{Pomsets for {M}essage {S}equence {C}harts}},
  OPTbooktitle = {SAM 1998},
  booktitle = {Proceedings of the 1st Workshop of the SDL Forum Society on SDL and MSC (SAM 1998)},
  OPTbooktitle = {Formale Beschreibungstechniken f{\"{u}}r verteilte Systeme},
  editor = {Yair Lahav and Adam Wolisz and Joachim Fischer and Eckhardt Holz},
  OPTeditor = {H. K{\"{o}}nig and P. Langend{\"{o}}rfer},
  publisher = {Humboldt-Universit{\"{a}}t zu Berlin},
  OPTpublisher = {Shaker},
  year = {1998},
  pages = {197--207}
}

@article{EJPI24,
  author = {Luc Edixhoven and Sung-Shik Jongmans and Jos\'{e} Proen\c{c}a and
Ilaria Castellani},
  title = {Branching pomsets: design, expressiveness and applications to choreographies},
  journal = {J. Log. Algebr. Methods Program.},
  volume = {136},
  year = {2024},
  OPTpages = {100919},
  doi = {10.1016/j.jlamp.2023.100919},
  OPTurl = {https://www.sciencedirect.com/science/article/pii/S2352220823000731}
}

@inproceedings{KKSB11,
  author = {Joachim Klein and Sascha Kl{\"{u}}ppelholz and Andries Stam and Christel Baier},
  title = {{Hierarchical Modeling and Formal Verification: {A}n Industrial Case Study Using {Reo} and {Vereofy}}},
  OPTbooktitle = {FMICS 2011},
  booktitle = {Proceedings of the 16th International Workshop on Formal Methods for Industrial Critical Systems (FMICS 2011)},
  editor = {Gwen Sala{\"{u}}n and Bernhard Sch{\"{a}}tz},
  series = {LNCS},
  volume = {6959},
  publisher = {Springer},
  year = {2011},
  pages = {228--243},
  doi = {10.1007/978-3-642-24431-5_17}
}

@inproceedings{DA18,
  author = {Kasper Dokter and Farhad Arbab},
  title = {{Treo: {T}extual Syntax for {Reo} Connectors}},
  booktitle = {Proceedings of the 1st International Workshop on Methods and Tools for Rigorous System Design (MeTRiD 2018)},
  editor = {Simon Bliudze and Saddek Bensalem},
  series = {EPTCS},
  volume = {272},
  year = {2018},
  pages = {121--135},
  doi = {10.4204/EPTCS.272.10}
}

@article{PC17,
  author = {Proen{\c{c}}a, Jos{\'{e}} and Clarke, Dave},
  title = {Typed connector families and their semantics},
  journal = {Sci. Comput. Program.},
  volume = {146},
  year = {2017},
  pages = {28--49},
  doi = {10.1016/j.scico.2017.03.002}
}

@inproceedings{CP18,
  author = {Cruz, Rúben and Proen{\c{c}}a, Jos{\'{e}}},
  title = {{{ReoLive}: {A}nalysing Connectors in Your Browser}},
  OPTbooktitle = {STAF 2018},
  booktitle = {Revised Selected Papers of the STAF 2018 Collocated Workshops},
  editor = {Mazzara, Manuel and Ober, Iulian and Salaün, Gwen},
  series = {LNCS},
  volume = {11176},
  publisher = {Springer},
  year = {2018},
  pages = {336--350},
  doi = {10.1007/978-3-030-04771-9_25}
}

@inproceedings{PC20,
  author = {Proen{\c{c}}a, Jos{\'{e}} and Cledou, Guillermina},
  title = {{{ARx}: {R}eactive Programming for Synchronous Connectors}},
  OPTbooktitle = {COORDINATION 2020},
  booktitle = {Proceedings of the 22nd IFIP WG 6.1 International Conference on Coordination Models and Languages (COORDINATION 2020)},
  editor = {Bliudze, Simon and Bocchi, Laura},
  publisher = {Springer},
  series = {LNCS},
  volume = {12134},
  year = {2020},
  pages = {39--56},
  doi = {10.1007/978-3-030-50029-0_3}
}

@mastersthesis{S18,
  author = {Smeyers, Maarten},
  title = {{A Browser-Based Graphical Editor for {Reo} Networks}},
  school = {Leiden University},
  year = {2018},
  url = {https://theses.liacs.nl/1536}
}

@inproceedings{KKV10,
  author = {Natallia Kokash and Christian Krause and de Vink, Erik P.},
  title = {{Data-Aware Design and Verification of Service Compositions with {Reo} and {mCRL2}}},
  OPTboooktitle = {SAC 2010},
  booktitle = {Proceedings of the 25th ACM Symposium on Applied Computing (SAC 2010)},
  OPTeditor = {Sung Y. Shin and Sascha Ossowski and Michael Schumacher and Mathew J. Palakal and Chih{-}Cheng Hung},
  publisher = {ACM},
  year = {2010},
  pages = {2406--2413},
  doi = {10.1145/1774088.1774590}
}

@inproceedings{PCVA12,
  author = {Jos{\'{e}} Proen{\c{c}}a and Dave Clarke and de Vink, Erik and Farhad Arbab},
  title = {Dreams: a framework for distributed synchronous coordination},
  OPTboooktitle = {SAC 2012},
  booktitle = {Proceedings of the 27th ACM Symposium on Applied Computing (SAC 2012)},
  OPTeditor = {Sascha Ossowski and Paola Lecca},
  publisher = {ACM},
  year = {2012},
  pages = {1510--1515},
  doi = {10.1145/2245276.2232017}
}

@misc{FKMMP08,
  author = {Arbab, Farhad and Krause, Christian and Maraikar, Ziyan and Moon, Young-Joo and Proen{\c{c}}a, Jos{\'{e}}},
  title = {{Modeling, Testing and Executing {Reo} Connectors with the {Eclipse} Coordination Tools}},
  howpublished = {Tool demo session of FACS 2008},
  year = {2008}
}

@phdthesis{P11,
  author = {Proen\c{c}a, Jos\'{e}},
  school = {Leiden University},
  title = {Synchronous Coordination of Distributed Components},
  year = {2011},
  OPTmonth = {May},
  url = {https://hdl.handle.net/1887/17624}
}

@inproceedings{BCJMRSW15,
  author = {Simon Bliudze and Alessandro Cimatti and Mohamad Jaber and Sergio Mover and Marco Roveri and Wajeb Saab and Qiang Wang},
  title = {{Formal Verification of Infinite-State {BIP} Models}},
  booktitle = {ATVA 2015},
  OPTbooktitle = {Proceedings of the 13th International Symposium on Automated Technology for Verification and Analysis (ATVA 2015)},
  editor = {Bernd Finkbeiner and Geguang Pu and Lijun Zhang},
  series = {LNCS},
  volume = {9364},
  publisher = {Springer},
  year = {2015},
  pages = {326--343},
  doi = {10.1007/978-3-319-24953-7_25}
}

@inproceedings{BBHRS23,
  author = {Simon Bliudze and van den Bos, Petra and Marieke Huisman and Robert Rubbens and Larisa Safina},
  editor = {Leen Lambers and Sebasti{\'{a}}n Uchitel},
  title = {{{JavaBIP} meets {VerCors}: {T}owards the Safety of Concurrent Software Systems in {Java}}},
  OPTbooktitle = {FASE 2023},
  booktitle = {Proceedings of the 26th International Conference on Fundamental Approaches to Software Engineering (FASE 2023)},
  series = {LNCS},
  volume = {13991},
  publisher = {Springer},
  year = {2023},
  pages = {143--150},
  doi = {10.1007/978-3-031-30826-0_8}
}

@article{BBBCJNS11,
  author = {Ananda Basu and Saddek Bensalem and Marius Bozga and Jacques Combaz and Mohamad Jaber and Thanh{-}Hung Nguyen and Joseph Sifakis},
  title = {{Rigorous Component-Based System Design Using the {BIP} Framework}},
  journal = {IEEE Softw.},
  volume = {28},
  number = {3},
  year = {2011},
  OPTmonth = {May/June},
  pages = {41--48},
  doi = {10.1109/MS.2011.27}
}

@incollection{Y24,
  author = {Nobuko Yoshida},
  title = {{Programming Language Implementations with Multiparty Session Types}},
  booktitle = {Active Object Languages: Current Research Trends},
  editor = {Frank S. de Boer and Ferruccio Damiani and Reiner H{\"{a}}hnle and Einar Broch Johnsen and Eduard Kamburjan},
  series = {LNCS},
  volume = {14360},
  publisher = {Springer},
  year = {2024},
  pages = {147--165},
  doi = {10.1007/978-3-031-51060-1_6}
}

@article{GT19,
  author = {Roberto Guanciale and Emilio Tuosto},
  title = {Realisability of pomsets},
  journal = {J. Log. Algebr. Methods Program.},
  volume = {108},
  year = {2019},
  pages = {69--89},
  doi = {10.1016/J.JLAMP.2019.06.003}
}

@article{ABBJSZ18,
  author = {Paul C. Attie and Saddek Bensalem and Marius Bozga and Mohamad Jaber and Joseph Sifakis and Fadi A. Zaraket},
  title = {{Global and Local Deadlock Freedom in {BIP}}},
  journal = {ACM Trans. Softw. Eng. Methodol.},
  volume = {26},
  number = {3},
  year = {2018},
  pages = {9:1--9:48},
  doi = {10.1145/3152910}
}

@inproceedings{AKM08,
  author = {Farhad Arbab and Natallia Kokash and Sun Meng},
  title = {{Towards Using {Reo} for Compliance-Aware Business Process Modeling}},
  booktitle = {ISoLA 2008},
  OPTbooktitle = {Proceedings of the 3rd International Symposium on Leveraging Applications of Formal Methods, Verification and Validation (ISoLA 2008)},
  editor = {Tiziana Margaria and Bernhard Steffen},
  series = {CCIS},
  OPTseries = {Communications in Computer and Information Science},
  volume = {17},
  publisher = {Springer},
  year = {2008},
  pages = {108--123},
  doi = {10.1007/978-3-540-88479-8_9},
}

@article{BH26,
  author = {Barbanera, Franco and Hennicker, Rolf},
  title = {{Safe orchestrated multicomposition of systems of communicating finite state machines}},
  journal = {J. Log. Algebr. Methods Program.},
  volume = {150},
  year = {2026},
  pages = {101109},
  doi = {10.1016/J.JLAMP.2026.101109}
}

@inproceedings{SK26,
  author = {Saeedloei, Neda and Klu{\'{z}}niak, Feliks},
  title = {{Timed Scenario Expressions and Realisability}},
  booktitle = {Proceedings of the 28th IFIP WG 6.1 International Conference on Coordination Models and Languages (COORDINATION 2026)},
  editor = {Roberto Casadei and Fatemeh Ghassemi},
  publisher = {Springer},
  series = {LNCS},
  volume = {16590},
  year = {2026},
  pages = {3--25},
  doi = {10.1007/978-3-032-28358-0_1}
}

@inproceedings{DGJPY15,
  author = {Mariangiola Dezani{-}Ciancaglini and Silvia Ghilezan and Svetlana Jak{\v{s}}i{\'{c}} and Jovanka Pantovi{\'{c}} and Nobuko Yoshida},
  title = {Precise subtyping for synchronous multiparty sessions},
  booktitle = {Proceedings of the 8th International Workshop on Programming Language Approaches to Concurrency- and Communication-cEntric Software (PLACES 2015)},
  editor = {Simon Gay and Jade Alglave},
  series = {EPTCS},
  volume = {203},
  year = {2015},
  pages = {29--43},
  doi = {10.4204/EPTCS.203.3}
}

@inproceedings{BCADDY08,
  author = {Lorenzo Bettini and Mario Coppo and D'Antoni, Loris and De Luca, Marco and Mariangiola Dezani{-}Ciancaglini and Nobuko Yoshida},
  title = {{Global Progress in Dynamically Interleaved Multiparty Sessions}},
  booktitle = {Proceedings of the 19th International Conference Concurrency Theory (CONCUR 2008)},
  editor = {van Breugel, Franck and Marsha Chechik},
  series = {LNCS},
  volume = {5201},
  publisher = {Springer},
  year = {2008},
  pages = {418--433},
  doi = {10.1007/978-3-540-85361-9_33}
}

@article{CDYP16,
  author = {Mario Coppo and Mariangiola Dezani{-}Ciancaglini and Nobuko Yoshida and Luca Padovani},
  title = {Global progress for dynamically interleaved multiparty sessions},
  journal = {Math. Struct. Comput. Sci.},
  volume = {26},
  number = {2},
  year = {2016},
  pages = {238--302},
  doi = {10.1017/S0960129514000188}
}

\end{document}